\documentclass[review=false]{siamart}
\usepackage[numbers]{natbib}
\usepackage{multicol}  
\usepackage{color}
\usepackage{amssymb}
\usepackage{mathrsfs}
\usepackage{epstopdf}
\usepackage{makeidx}
\usepackage{bm}
\usepackage{epsf,amsfonts,amsmath}
\usepackage{amssymb,amsmath}
\usepackage{appendix}
\usepackage{graphicx}
\usepackage{multirow}
\usepackage{array}
\usepackage{caption}
\usepackage{times}
\usepackage{float}
\usepackage{grffile}

\usepackage{tikz}
\usetikzlibrary{shapes,arrows,snakes,backgrounds}
\usetikzlibrary{mindmap,trees}
\usetikzlibrary{decorations.pathreplacing}
\usetikzlibrary{plotmarks}
\usetikzlibrary{calc}
\usetikzlibrary{fadings}
\usepackage{pgflibraryshapes}  
\usetikzlibrary{arrows}
\usetikzlibrary{positioning}

\usetikzlibrary{calc,patterns,angles,quotes}  

\usepackage[version-1-compatibility]{siunitx}
\tikzstyle{pblock} = [rectangle, draw, fill=yellow!20, text width=5.5em, text centered, minimum height=5em, node distance=7.5cm]
\tikzstyle{void} = [fill=white, node distance=5cm]
\tikzstyle{connector} = [->,thick]

\newcommand{\sobol}{Sobol'}
\newcommand{\hfm}{high-fidelity model}
\newcommand{\sm}{surrogate model}

\newcommand{\opal}{\textsc{OPAL}}

\newcommand{\opalcycl}{\textsc{OPAL-cycl}}

\renewcommand{\epsilon}{\varepsilon}

\newcommand{\figref}[1]{Figure~\ref{#1}}
\newcommand{\tabref}[1]{Table~\ref{#1}}

\newcommand{\secref}[1]{Section~\ref{#1}}

\renewcommand {\Re}{{\rm I \kern-2pt R}}

\newcommand {\RM}[1]{\mathrm{#1}}

\newcommand {\bs}[1]{\mathbf #1}

\newcommand{\xivec}{\mathbf{\xi}}

\newcommand {\htp}{$\text{H}_2^+$~}

\usepackage{xspace}

\newcommand{\DAD}{DAE$\delta$ALUS\xspace}

\newcommand{\nue}{\ensuremath{\nu_e}\xspace}

\newcommand{\numu}{\ensuremath{\nu_{\mu}}\xspace}

\newcommand{\nutau}{\ensuremath{\nu_{\tau}}\xspace}

\newcommand{\TheTitle}{On Non-Intrusive Uncertainty Quantification and Surrogate Model Construction in Particle Accelerator Modelling} 
\newcommand{\TheAuthors}{Andreas Adelmann}
\headers{\TheTitle}{\TheAuthors}

\title{{\TheTitle}}

\author{
  Andreas Adelmann \thanks{Paul Scherrer Institut, Switzerland
    (\email{andreas.adelmann@psi.ch}).}
 }

\begin{document}
\maketitle

\begin{abstract}
\label{Abstract}
Using a cyclotron based model problem, we demonstrate for the first time the applicability and usefulness of an uncertainty quantification (UQ) approach in order to construct surrogate models.\ The surrogate model quantities for example emittance, energy spread, or the halo parameter, can be used to construct a global sensitivity model along with error propagation and error analysis.\ The model problem is chosen such that it represents a template for general high-intensity particle accelerator modelling tasks.\ The usefulness and 
applicability of the presented UQ approach is then demonstrated on an ongoing research project, aiming at the design of a compact high-intensity cyclotron.
The proposed UQ approach is based on polynomial chaos expansions and relies on a well defined number of high fidelity particle accelerator simulations.\ Important uncertainty sources are identified using \sobol\ indices within the global sensitivity analysis.

\end{abstract}



\section{INTRODUCTION}
\label{sec:introduction}

Uncertainty Quantification (UQ) describes the
{\em origin}, {\em propagation},  and {\em interplay}
of different sources of uncertainties in the analysis and behavioural prediction of generally complex and high dimensional systems, such as particle accelerators.\ With uncertainty, one might question how accurately a mathematical model can describe the true physics and what impact the model uncertainty
(structural or parametric) has on the outputs from the model.\ Given a mathematical model, we need to estimate the error.\ ``How accurately is
a specified output approximated by a given numerical method?\ Can the error in the numerical solutions and the specified outputs be reliably estimated and controlled by adapting resources?''\ For example, in beam dynamics simulations with space charge, grid sizes would be such a resource.

UQ techniques allow one to quantify output variability in the presence of uncertainty. These techniques can generally tackle all sources 
of uncertainties, including structural ones. However, in this paper we focus on parametric uncertainty of input parameters.\ The moments of the output distributions are sampled using Monte Carlo~\cite{McQMc} or Quasi-Monte Carlo~\cite{QMC} methods, or newer approaches such as Multi-Level Monte Carlo \cite{adelmannml2015}.\ Other approaches exist and are known as non-sampling based methods.\
For an introduction to response surface methods see ~\cite{Response1:book,Response2:book}.\ The most popular method these days, which is used in this paper, is the Polynomial Chaos (PC) based method~\cite{Wiener}. Strictly speaking, PC also requires sampling, but it is not random sampling as in Monte-Carlo type approaches.

Polynomial chaos based techniques for propagating uncertainty and model reduction have been used in the past in almost all important scientific areas. An incomplete list consists of: climate modelling \cite{Sargsyan:14}, transport in heterogeneous media~\cite{Ghanem1998}, Ising models~\cite{TuhinPoly}, combustion~\cite{Najm2009}, fluid flow~\cite{ISI:000262972800003,Xiu2003}, materials models~\cite{Cit:Materials}, battery design \cite{Hadigol2015507}, and Hamiltonian systems \cite{Jose2013}.

In probabilistic UQ approaches, one represents uncertain model parameters as random variables or processes.
Among these methods, stochastic spectral methods \cite{Ghanem91a,Xiu10a} based on PC expansions \cite{Wiener,Cameron47} have received special attention due to their advantages over traditional UQ techniques. For a
more detailed discussion on that subject, consult the introduction of Hadigol et.al.\ \cite{Hadigol2015507}, or alternatively, the book of Smith \cite{Smith2014}.\

In the field of particle accelerator science, non-intrusive methods are far more attractive than intrusive methods.\ The complexity of
the physics model would most likely require a total rewrite of the existing simulation packages, in order to facilitate intrusive methods.\ Because non-intrusive methods
allow the use of existing beam dynamics codes as black boxes, they are the methods of choice. 
A non-intrusive method to solve an inverse was proposed in \cite{lee2006}.\ A proton beam from a linear charged particle accelerator, is focused through the use of successive quadrupoles. The goal of the inverse problem is to find the unknown initial state of the beam, in terms of particle position and momentum. 
Measurement data on the projection of the phase space was used where available beyond the focusing region. This setup is that of an inverse problem, in which a computer simulator is used to link an initial state configuration to observable values, and then inference is performed for the distribution of the initial state. 
The used Bayesian approach allows estimation of uncertainty in the initial distributions and beam predictions.

In this paper, we use \opal\ \cite{YanAdeHum10,bi:cyclotron_sim} as the black-box solver.\ As we will see later, only independent solution realisations are needed, hence embarrassingly parallel implementation is straightforward.

The proposed PC approach, first introduced in \cite{Ghanem91a, Xiu02}, computes the statistics for Quantity of Interest (QoI) with a small number of accelerator simulations.\ However, in contrast to \cite{Ghanem91a, Xiu02} we do not exploit the sparsity of expansion coefficients, this is subject to further research.\ Additionally, the presented UQ framework enables one to perform a global sensitivity analysis (SA) to identify the most important uncertain parameters affecting the variability of the output quantities. 

To avoid confusion, we firstly point out a misnomer by mentioning that polynomial chaos~\cite{Wiener} and chaos theory~\cite{Cit:Stro} are unrelated areas. Originally proposed by Norbert Wiener~\cite{Wiener} in 1938 (prior to the development of chaos theory---hence the unfortunate usage of the term \emph{chaos}), polynomial chaos expansions are a popular method for propagating uncertainty through low dimensional systems with smooth dynamics.

This work presents a sampling-based PC approach to study the effects of uncertainty in various model parameters of accelerators.\ As a model problem, we use the central region of a ``PSI Injector 2 like'' high-intensity cyclotron, where we only consider the first 10 turns of the cyclotron. While this paper's focus is mainly to introduce UQ to the field of particle accelerator science we add a realistic example of an ongoing design effort.

\subsection{Motivation in lieu of an actual research project} 
Searches for CP violation in the neutrino sector, and ``sterile'' neutrinos, respectively need a lot of statistics i.e.\ events. This translates, 
in the Decay-At-rest Experiment for $\delta_{\textrm{CP}}$ violation At a Laboratory for Underground Science (\DAD) \cite{Abs:2012wp} and the Isotope Decay-At-Rest experiment (IsoDAR)  \cite{PhysRevLett.109.141802}, into high fluxes of protons and compact accelerators in our example cyclotrons. The detailed exposition how UQ and PC Expansion (PCE) is used in
an ongoing research project is given in Section \ref{sec:realwordprob}, here we want to motivate this approach, fix language and notation. 

\begin{figure}[h!] \centering
\begin{tikzpicture}[every node/.style={anchor=south west,inner sep=0pt},x=1mm, y=1mm, z=1mm,]   
    \node (fig1) at (-10,0)
       {\includegraphics[width=0.4\textwidth,angle=0]{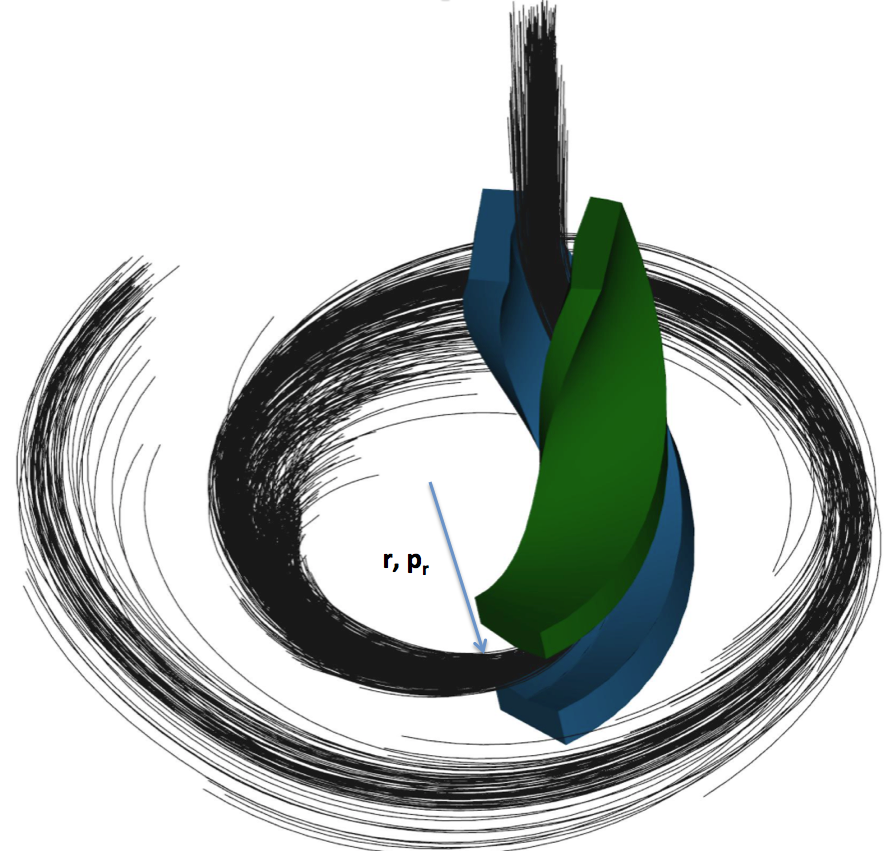}};

\coordinate (O) at (-10, 1 -1);
\draw[->] (O) -- +(10, 0,  0) node [right] {$x$};
\draw[->] (O) -- +(0,  10, 0) node [above] {$z$};
\draw[->] (O) -- +(0,  0, 5) node [above] {$y$};

\coordinate[label=below right:$$] (o) at (25.0,40);
\coordinate[label=below left: $$] (x) at (1.5,0);
\pic[draw,thick,"$\theta$",angle radius=5mm,angle eccentricity=1.3,->]{angle=x--o--o};

\coordinate[label=below right:$\color{red}\circ$] (o) at (35.0,18.0);
\coordinate[label=below right:$\color{red}\circ$] (o) at (11.0,22.0);
\coordinate[label=below right:$\color{red}\circ$] (o) at (11.0,24.0);

\coordinate[label=below right:$\color{red}\circ$] (o) at (18.0,27.5);

\end{tikzpicture}
\caption{Spiral inflector with particle trajectories, from the IsoDAR example presented in \secref{sec:realwordprob}.\label{fig:sensIsodar0}}
\end{figure}
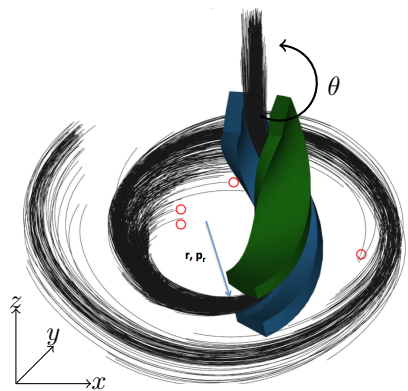

In high intensity accelerators the working point is to a large extend defined with setting the flux of particles per time, i.e.\ the intensity $I$ (c.f.\ Section \ref{sec:appl}). In the compact accelerator studied in Section \ref{sec:realwordprob} the angle  $\theta$ of the spiral inflector defines also a working point.\  The spiral inflector and part of the central region (magnets and cavities not shown) is depicted in \figref{fig:sensIsodar0}. The spiral inflector can be rotated around the $z$ axes by an angle $\theta$. 

We consider these as {\em design or controllable} parameters.  The machine is operated at only a few distinct different values. For example in high intensity i.e.\ production mode, or for machine development in lower intensities mode to prevent damage or activate of the accelerator. Similar arguments can be made for $\theta$ c.f.\ Section \ref{sec:realwordprob}. 

The other category of parameters are the {\em model parameters}. The model parameters are either quantities that are not measurable or measurable with an associated uncertainty.  
In the problem of Section \ref{sec:realwordprob}, the radius $r$ of the injected particles and the associated radial momenta $p_r$ (c.f.\ \figref{fig:sensIsodar0}) can not measured in-situ, hence empirical values
or values from simple models together with a meaningful PDF is used. Other quantities can be measured, for example the phase of the cavity $\phi$, but we want to find optimal values.  In the works, all design parameters are i.d.d., bounded and uniformly distributed.

Maximising performance in high intensity accelerators has two main dimensions: 1.\ maximize the number of transmitted particles throughout  the accelerator and at the 
same time 2.\ minimize particle losses.\ In \figref{fig:sensIsodar0} you can already see by eye that particle tracks are terminating at the not shown walls. A few are marked as red
circles for illustration purposes.\ The tolerable particle losses have to be at levels of $3$ to $4$ standard deviations of the particle density. Particle losses are associated
with "halo", i.e. particles that are sufficiently far away from the core of the distribution, such that they have a high probability to be lost.\
This all translates into the necessity
to solve large N-body problems, taking into account the non-linear particle particle interaction, together with complicated boundary conditions. Furthermore, as hinted above, with the design
of such complex scientific instruments, large scale multi-objective optimisation must be worked out and correlations and sensitivities identified. This motivated the search for inexpensive to evaluate surrogate models and is one of the main motivation behind this works.

In \secref{sec:PCE} we present our stochastic modelling approach which is based on non-intrusive PC expansions.\ After the derivation of the surrogate model, we then continue with reviewing a global sensitivity analysis approach using \sobol\ indices.\
\secref{sec:amodel} introduces the simulation model and a model problem.\ \secref{sec:appl} applies the UQ to the
stated problem, and shows the main features of this approach.\ The features are general in nature and {\em not} restricted to cyclotrons.\  \secref{sec:realwordprob} reports on an ongoing design effort using UQ. Conclusions are presented in \secref{sec:concl}.

\section{UQ VIA POLYNOMIAL CHAOS EXPANSION}
\label{sec:PCE}

Wiener in 1938 \citep{Wiener} introduced polynomial chaos expansion.\ In 1991, Ghanem and Spanos \citep{Ghanem91a} reintroduced this technique to the field of engineering.\ They first studied problems  with Gaussian input uncertainties and extended their method to non-Gaussian random inputs. In their studies,  orthogonal polynomials of the Askey scheme were used. This is known as a generalised polynomial chaos (gPC) expansion \citep{Xiu02}. The method of gPC expansion provides a framework to approximate the solution of a stochastic system by projecting it onto a basis of polynomials of the random inputs. 

An overview and some details on the correspondence between distributions and polynomials
can be found in \cite{Ogura}.\ A framework to generate polynomials
for arbitrary distributions has been developed
in~\cite{BeyondWienerAskey}.\ The advantage of using polynomial chaos is that it provides exponential convergence for smooth models.\ However, the approach suffers from the curse of dimensionality, making them challenging for problems with number of parameters in the range $10 \ldots 50$.\ To mitigate the curse of dimensionality, sparse grid techniques have traditionally been used~\cite{Nobile2008, Zabaras2008}.\ More recently, iterative methods to propagate uncertainty in complex networks have also been developed~\cite{surana_uq, Sahai2012, Klus2011}.

\subsection{The surrogate model}

\begin{figure}[h!]
\centering
\begin{tikzpicture}[scale=0.70]
\draw [rounded corners=15pt, thick ] (4,1) -- (.5,3) -- (1,5) -- (1.2,7) -- (5.3,7.5) -- (7,6) -- (6,4) -- (6.7,2.72)  --cycle;

\foreach \x in {1,...,7}
\foreach \y in {1,...,7}
{
       \draw[red] (\x,\y) circle (2pt) ;
}

\fill[blue!20!white] (4,4) rectangle (5,5);
\fill[blue!20!white] (3,3) rectangle (4,4);
\fill[blue!20!white] (4,3) rectangle (5,4);

\node at (3.6,3.7) {$\mathbf{\nu}$};
\node at (3.3,3.2) {$\circ$};
\node at (5,5.4) {$\mathbf{D}^*  \subset \mathbf{D}$};
\node at (3,7.7) {$ \mathbf{D} \subset \mathbb{R}^{d}$};
\end{tikzpicture}
\caption{Admissible design parameter search space $\mathbf D$, and one of the many possible ideal configurations $\mathbf{\nu}$ (working point) of the accelerator.\
The red circles depicts the training points, from which the surrogate model will be constructed.\ The equidistance of these points is not necessary, however it is sufficient 
to introduce the overall concept. We furthermore assume that subspace $ \mathbf{D}^*$ is much smaller than $\mathbf{D}$. \label{fig:modelred}}
\end{figure}
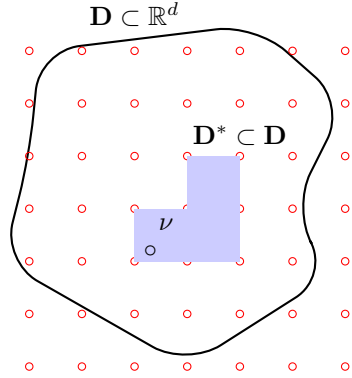

Suppose you are designing or optimising complex systems such as particle accelerators.\  As a particular example, consider the case of a high-intensity hadron machine.\ In such a machine one needs  
to characterise and minimise some QoI's (for example halo, and at the same time increase the beam quality).\ In order to accomplish this task, usually a large number of design and model parameters, in the search space
$\mathbf{D}$ (c.f.\ \figref{fig:modelred}), have to be considered.\  Let us furthermore assume that $\mathbf{D}$ is the admissible space, i.e.\ where the accelerator is working. The goal is to find a desired (optimal) working point $\mathbf{\nu}$, such that  properties of the QoI's are met.\ The restriction to one point is arbitrary, but allows a more focussed 
discussion.This endeavour is usually accompanied with large and
extensive multi-objective optimizations. 

In an ideal world you would run a large number of high fidelity simulations (in some proportion to the size of $\mathbf{D}$) to solve the problem. However, even with state-of-the-art tools,
and in cases of practical interest, it is impossible to accomplish this task due to the prohibitive time to solution.

With the help of adequate surrogate models, there are at least two ways to tackle the problem.\ Firstly, with a high fidelity simulator we build a surrogate model from a coarser, discrete search space, depicted by the red points in \figref{fig:modelred}.\ With this surrogate model we then predict $\mathbf{\nu}$ which yields eventually an optimal solution.

In the second option we would first find the smaller domain $\mathbf{D}^*$, with the help of the surrogate model constructed from $\mathbf{D}$. Because $\mathbf{D}^*$ is much smaller than $\mathbf{D}$, it is feasible to use the expensive \hfm\  to obtain $\mathbf{\nu} \in \mathbf{D}^*$. 

It is important to mention that the surrogate model does not really reduce the search space. Rather, it is an approximation to the full model over the area of the search space where one believes that the model matters the most.\ The goal of the surrogate model is to create a cheap-to-sample approximation of the full model.

\subsection{Mathematical bases of UQ}
We briefly introduce the mathematical bases in the style and the notation of \cite{Smith2014,Ghanem91a, Xiu02,Xiu10a,Hadigol2015507}, more details can be found in 
Appendix \ref{apptheory}.

%

All square integrable, second-order random variables with finite variance output, $u(\bm{\xi}) \in L_2\left(\Omega,\mathcal{F},\mathcal{P} \right)$, can be written as
\begin{equation*} 
u(\bm{\xi}) = \sum_{|\bm{i}| = 0}^\infty \alpha_{\bm{i}} \Psi_{\bm{i}}(\bm{\xi}).
\end{equation*}
Hence $\alpha_{\bm{i}}$ denotes the deterministic coefficients and $\Psi_{\bm{i}}(\bm{\xi})$ are the multivariate PC basis 
functions \cite[10.1.1]{Smith2014} \cite{Ghanem91a} and $\bm{i}$ is a mult-index.\ Note that the uncertain QoI, $u$, is represented by a vector of deterministic parameters $\alpha_{\bm{i}}$.\
Input uncertainties of the system have been discretised and approximated by the random vector 
\begin{equation} \label{eq:xi}
\bm{\xi} = \left(\xi_1,\cdots, \xi_d\right):\Omega \rightarrow \mathbb{R}^{d}, 
\end{equation}
$d \in \mathbb{N}$. The probability density function (pdf) of the random variable, $\xi_k$, is denoted by $\rho_k(\bm{\xi})$.\ Similarly, $\rho(\bm{\xi})$ represents the joint pdf of $\bm{\xi}$.\
For the truncated PCE to order $p$ in $d$ dimensions of \eqref{appeq:rvpce0} we get
\begin{equation} 
\label{eq:rvpce}
\hat{u}(\bm{\xi}) = \sum_{\bm{i} \in \mathcal{I}_{d,p}} \alpha_{\bm{i}} \Psi_{\bm{i}}(\bm{\xi}),
\end{equation}
with  $\Psi_{\bm i}(\bm\xi)$ certain orthogonal bases functions, and $ \mathcal{I}_{d,p} $ a set of multi-indices. 


The number $K$ of PC basis functions of total order  $p$  in dimension $d$ can be calculated to 
\begin{displaymath} 
\label{eq:pplo}
K= \frac{(p+d)!}{p!d!}.
\end{displaymath}
Because of the orthogonality of $\Psi_{{i}_k}(\xi_k)$ and the independence of $\xi_k$, as $ p \rightarrow \infty$, the truncated PC expansion in \eqref{eq:rvpce} converges in the mean-square sense, if and only if  the following two conditions are fulfilled: 1)
$u(\bm\xi)$ has finite variance and 2) the coefficients $\alpha_{\bm{i}}$ are computed from the projection equation \citep{Xiu02}
\begin{equation}
\alpha_{\bm{i}} = \frac{\mathbb{E}[\hat{u} \Psi_{\bm i}]}{\mathbb{E}[\Psi_{\bm i}^2]}.
\end{equation}
\subsection{Non-intrusive polynomial chaos expansion}
\label{sec:Non-intrusive PCE}
In PC-based methods, one obtains the coefficients of the solution expansion either intrusively \citep{Debusschere04} 
or non-intrusively \citep{Eldred09}.\ An intrusive approach requires significant modification of the deterministic solvers and increases the number of equations to solve.\ 

Non-intrusive methods on the other hand can make use of existing deterministic solvers ($\mathcal{M}$) as black boxes.\  First, one needs to generate a set of $N$ deterministic or random samples of $\bm\xi$, denoted by $\{{\bm{\xi}}^{(i)}\}_{i=1}^{N}$. The second step is to generate $N$ realisations of the output QoI, $\{u({\bm{\xi}}^{(i)})\}_{i=1}^{N}$, with the available deterministic solver $\mathcal{M}$ and without any solver modifications.\ The third and final step is to solve for the PC coefficients using the obtained realisations.\ Methods such as least-squares regression \citep{Hosder06}, pseudo-spectral collocation \citep{Xiu10a}, Monte Carlo sampling \citep{LeMaitre10}, and compressive sampling \citep{Doostan11a} are available.\ Along these lines an in-depth discussion on least-squares regression and compressive sampling
can be found in \cite[3.1.1,3.1.2]{Hadigol2015507}.\

The mean, $\mathbb{E}[\cdot]$, and variance, $\mathrm{Var}[\cdot]$, of $u(\bm{\xi})$ can be directly approximated from the PC coefficients because of polynomial basis orthogonality given by 
\begin{equation}
\mathbb{E}[\hat{u}] =  \alpha_{\bm{0}},
\end{equation}
and                 
\begin{equation}
\label{eq:vpce}
\mathrm{Var}[\hat{u}] =  \sum_{\substack{\bm{i} \in \mathcal{I}_{d,p}\\ \bm{i} \neq \bm{0}}} \alpha_{\bm{i}}^2 ~
\mathbb{E}[\Psi^2({\xivec}_{\bm{i}})].\
\end{equation}
A more complete description will be shown later in \secref{sec:UQFrame}.

\subsection{Global sensitivity analysis}
\label{sec:GSA}
The expensive, deterministic high-fidelity particle accelerator model, $\mathcal{M}$, is described by a function $\vec{u}=\mathcal{M}(\vec{x})$, where the input $\vec{x}$ is a point inside 
$\mathbf{D}$ (c.f. \figref {fig:modelred}) and $\vec{u}$ is a vector of QoI's.\ Finding correlations in these high dimensional spaces is nontrivial, however it is vital for a deep understanding of the underlying  physics.\ For example, reducing the search space is of great interest in the modelling and optimization process.\ In the spirit of Sobol' \citep{Sobol01}, let $\vec{u}^{*}=\mathcal{M}(\vec{x}^{*})$ be the sought  (true) solution.\ The local sensitivity of the solution $\vec{u}^{*}$ with respect to $x_{k}$ is estimated by $(\partial \vec{u} /\partial x_{k})_{\vec{x}=\vec{x}^{*}}$.\ On the contrary, the global sensitivity approach does not specify the input $\vec{x}=\vec{u}^{*}$, it only considers the model $\mathcal{M}(\vec{x})$.\ Therefore, global sensitivity analysis should be regarded as a tool for studying the mathematical model rather than a specific solution ($\vec{x}=\vec{x}^{*}$).\
For details we refer to Appendix \ref{appGSA}.

\subsection{The UQTk based framework} 
\label{sec:UQFrame}
In this section a detailed description is provided on how the particle accelerator UQ framework is constructed.\ The framework is based on the Uncertainty Quantification Toolkit (UQTk) \cite{uqtk}, a lightweight C++/Python library that helps perform basic UQ tasks including intrusive and non-intrusive forward propagation.\ UQTk can also be used for inverse modelling via Bayesian or optimization techniques.\ The corresponding tools used from UQTk are indicated in typewriter style in the following algorithm.\

Let's denote  $\mathcal{M}$ as the black-box solver, $\vec{\lambda}$ as the model parameters and $\vec{x}$ as the design or controllable parameter, with $l$ distinct values.\ \footnote{For a fixed value of the design parameter, the surrogate construction algorithm is described in \cite{ISI:000262972800003}.}\ The nonintrusive propagation of uncertainty from the $d$-dimensional model parameter $\vec{\lambda}$ to the output $\vec{u}_{i} = \mathcal{M}(\vec{\lambda},x_{i})$ follows a 
collocation procedure, given a $K$-dimensional basis $\vec{\Psi} = (\Psi_1, \ldots, \Psi_K)$ and $K=\frac{(d+p)!}{d!p!}$ multivariate basis terms with $p$ being the maximal polynomial order. \\

{\bf Algorithm 1:} generate for each $x_{i}$ (design or controllable), a PC surrogate model
\begin{enumerate}
  \item generate $N = (p+1)^{d}$ quadrature point-weight pairs  $(\vec{\xi}^n, w_n)$ \\ ({\tt generate\_quad})
 \item  for each of quadrature point $\vec{\xi}^n$ compute corresponding model input $\vec{\lambda}^n$ by
  \begin{eqnarray}\label{eq:modelevel}
 \vec{\lambda}^n =  \lambda_j^{n} &=& \displaystyle \sum_{k=0}^{K-1} \lambda_{jk} \Psi_k(\vec{\xi}^n) \qquad j=1,\dots,d \label{eq:pcin}.
  \end{eqnarray}
  
  \item create the training points with high fidelity simulations (\opal)
  \begin{eqnarray} 
  	u^n_i = \mathcal{M}(  \vec{\lambda}^n, x_i) \qquad\qquad i=1,\dots,l. \label{eq:blackbox}
  \end{eqnarray}
  \item calculate the expectation
              via orthogonal projection ({\tt pce\_resp}) using quadrature
  \begin{equation}
  \label{eq:proj}
   \alpha_{ki} = \frac{\langle u \Psi_k \rangle} {\langle \Psi^2_k \rangle} = \frac{1}{{\langle \Psi^2_k \rangle}}\displaystyle \sum_{n=1}^N u^n_i \Psi_k(\vec{\xi}^n)w_n, \quad k=0,\ldots,K-1. 
  \end{equation}
  \item ~~~ Given the computed $\alpha_{ki}$ values for each $i$ and $k$, one assembles the PCE
  \begin{equation} 
  \label{eq:surmodel}
    \hat{u}_{i} =  \displaystyle \sum_{k=0}^{K-1} \alpha_{ki} \Psi_k(\vec{\xi}), \quad k=0,\ldots,K-1.
  \end{equation} 
\end{enumerate}

\emph{Remark 1}: The input PC in Eq.~(\ref{eq:pcin}) is assumed to be given by an expert. For example, often only bounds for the inputs are known, in which case, Eq.~(\ref{eq:pcin}) is simply a linear PC or just scaling from $\xi_j\in[-1,1]$ to $\lambda_j\in[a_j,b_j]$ for each $j=1,\dots,d$.\ More explicitly stated, in Eq.~(\ref{eq:pcin}) $\lambda_{j0}=\frac{a_j+b_j}{2}$, and $\lambda_{jk}=\delta_{jk} \frac{b_j-a_j}{2}$. Thus, Eq.~(\ref{eq:pcin}) becomes 

\begin{equation}\label{eq:lambdaj}
\lambda_j^n=\frac{b_j+a_j}{2}+\frac{b_j-a_j}{2}\xi_j^n.\
\end{equation}

\emph{Remark 2}: If samples $\vec{\xi}^n$ are randomly selected from the distribution of $\vec{\xi}$, then the projection formula~Eq.\ (\ref{eq:proj}) still holds, as long as one sets $w_n=1/N$ for all $n$, and it becomes an importance sampling Monte-Carlo.\

\begin{figure}[h!]
  \begin{center}
  \begin{tikzpicture}[scale=0.8, transform shape]
    \begin{scope}[shape=rectangle,rounded corners,text centered]
 
 \node at (-1,5.5) [shape=rectangle,draw] {
            \begin{tabular}{c | c c}
		$\vec{\lambda}$  & l-bound & u-bound\\
		\hline
		\hline
		$\lambda_{1}$ & $a_{1}$ & $b_{1}$ \\
		$\lambda_{2}$ & $a_{2}$ & $b_{2}$ \\
		$\vdots$ & $\vdots$ & $\vdots$ \\
		$\lambda_{d}$ & $a_{d}$ & $b_{d}$ 
            \end{tabular}
        };

 \node at (-1.,2.5) [shape=rectangle,draw] {
            \begin{tabular}{c  c c}
		 $\vec{x} = $ & $(x_1, \ldots,  x_l)$ &  \quad \quad \quad \quad \quad\\
            \end{tabular}
        };

     \draw [ultra thick] [<-] (3.5, 2.5) -- (2.2,2.5);	

     \draw [ultra thick] [<-] (6, 3.6) -- (6., 4.4);

     \draw  [ultra thick]  [->] (2.2,5.5) -- (3.5,5.5);

      \draw [fill=orange] (3.7,4.5) rectangle (8.2,6.5);
      \node at (6, 5.5) [align=center] {\textbf{$N$ Quadrature Points}\\ Eq. (6)};

      \draw [fill=yellow] (3.7,1.5) rectangle (8.3,3.5);
      \node at (6, 2.5) [align=center] {\textbf{Model evaluations} Eq. (7) \\ $u_m=\mathcal{M}(\vec{\lambda}^m,\vec{x})$ \\ $m= 1 \ldots lN $ };

      \draw  [ultra thick]  [->] (6.0,1.4) -- (4,0.0);

      \node at (1.7, 0.) [align=center] {\textbf{Using $N$ samples} (Eq. 9)};
      \node at (2,-1) [shape=rectangle,draw] {
   
	$\hat{u}_{i}  =\displaystyle \sum_{k=0}^{K-1} \alpha_{ki} \Psi_k(\vec{\xi}), \quad i= 1 \ldots l$ .
        };
  
   \draw  [ultra thick]  [<-] (5.5,-4.5) -- (3.9,-1.9);
   \draw  [ultra thick]  [<-] (-1,  -4.5) -- (0.1,-1.9);
    \draw [fill=orange] (-3.5,-7) rectangle (1.5,-5);
    \node at (-1, -6) {\textbf{Surrogate Model $\hat{u}_{i}$}};

     \draw [fill=orange] (3.2,-7) rectangle (8.8,-5);
     \node at (6, -6) {\textbf{Global Sensitivity Analysis}};

   \node at (-1.8, 7.5) {\textbf{Model Parameters}};
   \node at (-0.8, 3.25) {\textbf{One design (or controllable) Parameter $\vec{x}$}};
    
    \end{scope}
  \end{tikzpicture}
  \caption{The Uncertainty Quantification Framework, with the discretised input uncertainties of the system are denoted by $\vec{\xi}$ c.f.\ Equation \ref{eq:xi}.
  In case of different design (or controllable) parameters $\vec{x}$ we would build $l$ separate response surfaces. Details can be found in Algorithm 1 and Appendix A.
  }
  \label{fig:dksconcept}
  \end{center}
\end{figure}
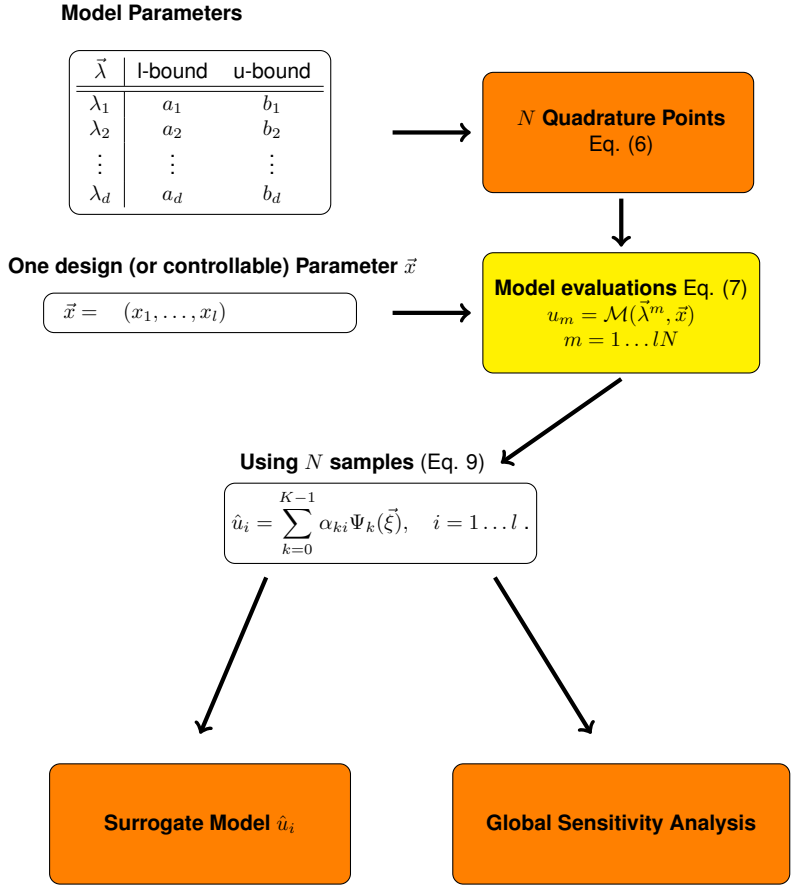

\emph{Remark 3}: In \figref{fig:dksconcept} a design parameter $\vec{x}$ is introduced. In case of  $p+1 < l$, i.e.\ if you only have,
a few, discrete values for the design parameter, a reduced number model evaluation is obtained. Instead of sampling this parameter, you create $l$ different 
response surfaces.

\subsubsection{Evaluation of the Surrogate model}
Having constructed the PC-coefficients, according to \eqref{eq:proj} the utility {\tt pce\_eval} can be used to evaluate $\hat{u}_{i}$ \eqref{eq:surmodel}.\

\subsubsection{Sensitivity Analysis}
As shown in \secref{sec:GSA}, the same information used in the surrogate model construction can be used in the sensitivity analysis.\ In the UQTk {\tt pce\_sens} will compute the total and joint sensitivities along with the variance fraction of each PC term individually. 

\section{A GENERAL MODEL PROBLEM} \label{sec:amodel}
Charged particle accelerators are among the largest and most complex scientific instruments. The application of charged particle accelerators ranges from material science, biology to 
fundamental physics questions, currently addressed for example with the LHC or in the future maybe with experiments like DAE$\delta$ALUS/IsoDAR \cite{daedalus,isodar} (c.f.\ \secref{sec:realwordprob}).\ There exists a wide range of different accelerator types, and a commonly used classification in  linear and circular types based on the geometrical nature. Given this taxonomy, a circular accelerator with
non constant  radius of curvature is the most general accelerator and used as a template for all other conventional types. Even in the simplest incarnation of a cyclotron \cite{RevModPhys.64.795} a rich dynamic is present, periodic or near periodic orbits or in general twist maps.

The Hamiltonian that describes the
motion of a classical relativistic charged particle in a general magnetic field \cite{wied:07} is given by
\begin{equation} \label{relativisticemhamiltonian}
H = (1 + h) \sqrt{\left( \mathbf{P} - q \mathbf{A} \right)^2 c^2 + m^2 c^4} + q \phi. 
\end{equation}
We are neglecting in this discussion the spin and radiation, for the sake of simplicity. All external electromagnetic fields (magnets etc) 
are absorbed in the vector potential $\mathbf{A}$ and with $ \mathbf{P}$ we denote the generalised momenta.\
Charge and mass are denoted by $m$ and $q$ respectively and $c$ is the speed of light.  In the case of a cyclotron, all quantities are expressed in a 
Frenet-Serret coordinate system with non constant curvature $h$ (c.v. \figref{fig:inj2model}).\ The scalar potential $\phi$ represents 
the non-linear particle-particle interaction. The computation of $\phi$ and the resulting non-linear force is computationally very expensive 
and the effect of these forces is a limiting factor of high intensity particle accelerators. The limiting aspect is based on the fact that these
repulsive forces create a halo around the core of the particles. This halo has a tendency to separate from the core and contribute to 
particle losses, that in the end activates the machine to a level where maintenance is difficult or even impossible. 

The case of a cyclotron, represents a large class of accelerator topologies. For example  in case of vanishing curvature $h$ in Eq.\ \eqref{relativisticemhamiltonian} the case of a linear accelerator is recovered. We select the cyclotron, in order to demonstrate the applicability of this framework in a very general context. 

Solving such problem under relevant circumstances, is equivalent with solving a large N-body problem with non-trivial boundary conditions. This together
with the multi-scale nature of the problem -- in time and phase space -- are calling for a hierarchy of models. On the extreme end we are 
using a $1:1$ ratio between simulation and macro-particles and computational times of days on high end parallel computers. Low dimensional models on the other hand are important
to narrow a potential high dimensional search space and making the problem more accessible.

Surrogate models as introduced in the previous section are in between the two extremes. They cover some non-linearities but 
are much faster to evaluate compared to the high fidelity model. With the sensitivity analysis we will get insight into a correlated space if QoI's and 
model parameters.

\subsection{The Accelerator Simulation Model}

For this discussion we briefly introduce \opalcycl\ \cite{PhysRevSTAB.13.064201}, one of the three flavours of \opal.\ \opal\ will be used as the back-box solver denoted by $\mathcal{M}$ in \eqref{eq:blackbox}.

\subsubsection{Governing  Equation}
In the cyclotron under consideration, the collision between particles can be neglected because the typical bunch density is low.\
In time domain, the general equations of motion for charged particles in electromagnetic fields can be expressed by
\begin{equation}\label{eq:motion}
  \frac{d\bs{p}(t)}{dt}  = q\left(c\mbox{\boldmath$\beta$}\times \bs{B} + \bs{E}\right). \nonumber \\
\end{equation}
We denote $\bs{p}=m c \gamma \mbox{\boldmath$\beta$}$ as the momentum of a particle, $\mbox{\boldmath$\beta$}=(\beta_x, \beta_y, \beta_z)$ as the normalised velocity vector, 
and $\gamma$ is the relativistic factor.\ In general, the time ($t$) and he position ($\bs{x}$) dependent electric and magnetic vector fields are
written in an abbreviated form as $\bs{B} \text{ and } \bs{E}$.\

If $\bs{p}$ is normalized by $m_0c$, 
Eq.\,(\ref{eq:motion}) can be written in Cartesian coordinates as 
\begin{eqnarray}\label{eq:motion2}
  \frac{dp_x}{dt} & = & \frac{q}{m_0c}E_x + \frac{q}{\gamma m_0}(p_y B_z - p_z B_y),    \nonumber \\
  \frac{dp_y}{dt} & = & \frac{q}{m_0c}E_y + \frac{q}{\gamma m_0}(p_z B_x - p_x B_z),   \\
  \frac{dp_z}{dt} & = & \frac{q}{m_0c}E_z + \frac{q}{\gamma m_0}(p_x B_y - p_y B_x).\    \nonumber 
\end{eqnarray}
The evolution of the beam's distribution function, $ f(\bs {x},c\mbox{\boldmath$\beta$},t): (\Re^M \times \Re^M \times \Re) \rightarrow \Re$, can be expressed by a collisionless Vlasov equation:
\begin{equation}\label{eq:Vlasov}
  \frac{df}{dt}=\partial_t f + c\mbox{\boldmath$\beta$} \cdot \nabla_x f +q(\bs{E}+ c\mbox{\boldmath$\beta$}\times\bs{B})\cdot \nabla_{c\mbox{\boldmath$\beta$}} f  =  0.\
\end{equation}
Here it is assumed that $M$ particles are within the beam.\
In this particular case, $\bs{E}$ and $\bs{B}$ include both externally applied fields and space charge fields.\
\begin{eqnarray}\label{eq:Allfield}
  \bs{E} & = & \bs{E_{\RM{ext}}}+\bs{E_{\RM{sc}}}, \nonumber\\    
  \bs{B} & = & \bs{B_{\RM{ext}}}+\bs{B_{\RM{sc}}},
\end{eqnarray}
all other fields are neglected.\

\subsubsection{The Self Fields}
The space charge fields can be obtained
by a quasi-static approximation.\ In this approach, the relative motion of the particles is non-relativistic in the beam rest frame, thus the self-induced magnetic field is practically absent and the electric field can be computed by solving Poisson's equation
\begin{equation}\label{eq:Poisson}
  \nabla^{2} \phi(\bs{x}) = - \frac{\rho(\bs{x})}{\varepsilon_0},
\end{equation}
where $\phi$ and $\rho$ are the electrostatic potential and the spatial charge density in the beam rest frame.\ The electric field can then be calculated by
\begin{equation}\label{eq:Efield}
  \bs{E}_{\RM{sc}}=-\nabla\phi,
\end{equation}
and transformed back to yield both the electric and the magnetic fields, in the lab frame, as required in Eq.\,(\ref{eq:Allfield}) by means of a Lorentz transformation.\
Because of the large vertical gap in our cyclotron, the contributions from image charges and currents are minor compared to space-charge effects \cite{Baartman:1}, and hence it is a good approximation to use 
open boundary conditions.\ Details on the space charge calculation methods utilised in \opal\ can be found in \cite{PhysRevSTAB.13.064201,adai:10,Hockney:1}.

\subsubsection{External Fields}
With respect to the external magnetic field, two possible situations can be considered.\
In the first situation, the real field map is available on the median plane of the existing cyclotron machine using measurement equipment.\

In most cases concerning cyclotrons, the vertical field, $B_z$, is measured on the median plane ($z=0$) only.\
Since the magnetic field outside the median plane is required to compute trajectories with $z \neq 0$, the field needs to be expanded in the $Z$ direction.\

According to the approach given by Gordon and Taivassalo \cite{Gordon:2}, by using a magnetic potential and measured $B_z$ on the median plane
at the point $(r,\theta, z)$ in cylindrical polar coordinates, the 3rd order field can be written as    
\begin{equation}
\label{eq:Bfield}
\vec{B}_{\RM{ext}} (r,\theta, z) =\left(z\frac{\partial B_z}{\partial r}-\frac{1}{6}z^3 C_r, \frac{z}{r}\frac{\partial B_z}{\partial \theta}-\frac{1}{6}\frac{z^3}{r} C_{\theta}, B_z-\frac{1}{2}z^2 C_z\right), 
\end{equation}

where $B_z\equiv B_z(r, \theta, 0)$ and  
\begin{eqnarray}\label{eq:Bcoeff}
  C_r & = & \frac{\partial^3B_z}{\partial r^3} + \frac{1}{r}\frac{\partial^2 B_z}{\partial r^2} - \frac{1}{r^2}\frac{\partial B_z}{\partial r} 
        + \frac{1}{r^2}\frac{\partial^3 B_z}{\partial r \partial \theta^2} - 2\frac{1}{r^3}\frac{\partial^2 B_z}{\partial \theta^2}, \nonumber  \\    
  C_{\theta} & = & \frac{1}{r}\frac{\partial^2 B_z}{\partial r \partial \theta} + \frac{\partial^3 B_z}{\partial r^2 \partial \theta}
        + \frac{1}{r^2}\frac{\partial^3 B_z}{\partial \theta^3},  \\
  C_z & = & \frac{1}{r}\frac{\partial B_z}{\partial r} + \frac{\partial^2 B_z}{\partial r^2} + \frac{1}{r^2}\frac{\partial^2 B_z}{\partial \theta^2} \nonumber .\
\end{eqnarray}
All the partial differential coefficients are computed on the median plane data by interpolation, using Lagrange's 5-point formula.\

In the second situation, a 3D field map for the region of interest is calculated numerically from a 3D model of the cyclotron.\
This is generally performed during the design phase of the cyclotron and utilises commercial software.\
In this case the calculated field will be more accurate, especially at large distances from the median plane, i.e.\ a
full 3D field map can be calculated.\ For all calculations in this paper, we use the Gordon and Taivassalo \cite{Gordon:2} method.\

For the radio-frequency cavities, a radial voltage profile $V(r)$ along the radius of the cavity is used.\ The gap-width, $g$, is included in order to correct for the transit time.\ For the time-dependent field,

\begin{equation}
\label{ }
\Delta E_{\RM{rf}} = \frac{\sin\tau}{\tau} \Delta V(r) \cos(\omega_{\RM{rf}} t - \phi),
\end{equation}
with $F$ denoting the transit time factor ($F=\frac{1}{2} \omega_{\RM{rf}}  \Delta t$), and
$ \Delta t$ the transit time defined by
\begin{equation}
\Delta t = \frac{g}{\beta c}.\
\end{equation}

In addition, a voltage profile varying along the radius will give a phase compression of the bunch, which is induced by an additional magnetic field component $B_{z}$ in the gap,
\begin{equation}\label{eq:Bz}
B_{z} \simeq \frac{1}{g \omega_{\RM{rf}} } \frac{d V(r)}{dr} \sin(\omega_{\RM{rf}} t - \phi).
\end{equation}

\begin{figure}[h!]
\centering
\includegraphics[width=0.8\textwidth]{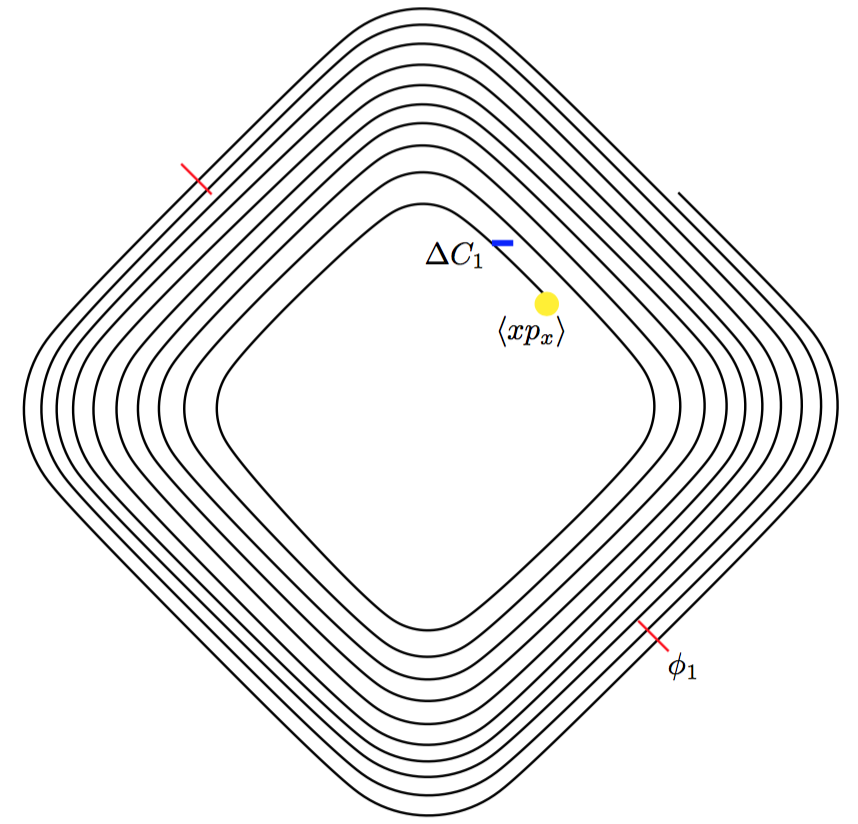}
\caption{The cyclotron model problem setup\label{fig:inj2model}.\ The two red lines indicate the 2 double gap flat-top resonators, the blue line represents a collimator, and the yellow circle stands for the initial conditions.}
\end{figure}

Finally, in this paper, both the external fields and space charge fields are used to track particles for one time step using a 4th order Runge-Kutta (RK) integrator.\ This means 
the fields are evaluated for four times in each time step.\ Space-charge fields are assumed to be constant during one time step
because their variation is typically much slower than that of external fields. 

\section{APPLICATION OF THE UQ FRAMEWORK to a MODEL PROBLEM} \label{sec:appl}
To demonstrate the usefulness and strength of UQ, consider a simplified model of the PSI Injector 2 cyclotron, which is sketched in \figref{fig:inj2model}.\  The simplifications are as follows: 1) only energies
up to 8.5 MeV (turn 10) are considered to reduce the computational burden;\ 2) a Gaussian distribution, linearly matched to the injection energy of 870 keV, is used for the initial conditions;\ 3) the magnetic field and RF structures are the same as in our full production simulation;\ 4) $P_r$ and $R$ are obtained from equilibrium orbit simulations, and 5) one collimator is introduced in order to mimic bunch shaping.\ 
Full scale high-fidelity simulations of this kind can be found in \cite{Yang201384,bi:cyclotron_sim}, where similar physics goals were pursued.

\subsection{Model parameters}
In typical design studies of high-power cyclotrons, the high number of model parameters are such that one cannot fully scan their entire range.\ For this feasibility study, one model parameter out of a family of  
three important categories (c.f.\ \figref{fig:inj2model}) was chosen:
\begin{enumerate}
  \item  initial conditions: model parameter $\langle xp_x \rangle$, correlation between the initial $x$ and $p_x$ phase-space variables,
  \item  collimator settings: model parameter  $\Delta C_{1}$ position of the collimator,
  \item rf phase settings: model parameter  $\phi_{1}$ defines the phase of the acceleration cavity.\
\end{enumerate}
From previous experience, these three categories have the most influence when designing and optimising high-precision models of high-power cyclotrons.\ The relationship of the parameters with uncertainties, $\lambda_{1}, \lambda_{2}, \lambda_{3}$, is shown in \figref{fig:dksconcept}.\

\subsection{Quantities of interest (QoI)} \label{sec:qoidef}
The phase space spanned by $M$ macro particles, in the high fidelity \opal\ model (simulation), is given by $(\vec{q}_i(t),\vec{p}_i(t)) \in \Gamma \subset \Re^{(2M + 1)}$ and $i= x,y,z$. We identify a subset of interesting QoI's such as:
\begin{enumerate}
   \item $\tilde{\epsilon}_x = \sqrt{ \langle \vec{q}_x^2 \vec{p}_x^2 \rangle - \langle \vec{q}_x \vec{p}_x \rangle^2}$ the rms projected emittance and 
   $\tilde{x}$ the rms beam size,
  \item  the kinetic energy $E$ and rms energy spread $\Delta E$,
  \item $h_t = \frac{\langle \vec{q}_x^4 \rangle}{\langle \vec{q}_x^2 \rangle ^2} - k$, the halo parameter in $x$-direction at end of turn $t$
  with $k \in \Re$, a distribution dependent normalisation constant. 
\end{enumerate}
The rms beam size $\tilde{x}$ is one of the better quantities that can be directly measured and hence among the first candidates for 
characterisation of the particle beam.\ A measure of the projected phase-space volume is the emittance $\tilde{\epsilon}_x$.\ This quantity is often
used for the estimation of the beam quality.\ The two energy related parameters $E$ and $\Delta E$ are target values to achieve.\ The first one, $E$, closely related to the experiment, where the particle beam is designed for. The energy spread, $\Delta E$, is directly related to the beam quality in the case of the presented model problem.\ Minimizing the halo of the particle beam is equal to minimizing losses, the most important quantity to optimize in high-power hadron accelerators. In the formulation
of $h_t$, this parameter is deviating from $1$ if and only if the initial choosen distribution is changing. If the initial distribution is a stationary distribution, this
measure can be attributed to the mechanism of halo generation, in case of a deviation from the value 1.

In the case of a high-intensity cyclotron model, we choose the controllable parameter $\vec{y}$ as the average current.

\subsection{UQ model setup}
The controllable parameters are not modelled with polynomials, but rather given by 10 equidistant values from $1 \dots 10$ mA.\
As a next step, the polynomial type for the model parameter is chosen according to the Wiener-Askey scheme (cf.\ Appendix \ref{app1}).\ The distribution of the three model parameters $\langle  x p_x \rangle$,  $\Delta C_1$, and the phase $\phi_1$, are modelled according to a uniform distribution using polynomials of the Legendre type. The bounds of the distribution are given in \tabref{table:designp1}.\
\begin{table}[h!]
\caption{Upper and lower bounds of the design parameters} 
\centering
\begin{tabular}[t]{ l   c  c }   
\hline \hline
v-name & l-bound & u-bound \\  
\hline \hline
$\langle xp_x \rangle$  & -0.5 & 0.5 \\  
\hline
$\Delta C_{1}$ (mm) & 0 & 5 \\  
\hline
$\phi_{1} (^{\circ})$ & -20 & 20 \\  
\hline
\end{tabular}
\label{table:designp1} 
\end{table}
Other parameters for the UQ model are listed in \tabref{table:designp2}.\
\begin{table}[h!]
\caption{Summary of UQ related parameters for the presented results. The dimension for all the experiments is $d=3$. The one controllable 
parameter $\vec{x}$ has length $l=4$.}
\centering
\begin{tabular}[t]{ l   l  c c c}   
\hline \hline
Parameter & Meaning      ~~~~~~~~~~~~~~~~~~~~~~~~~~~~~~~    Experiment & 3 & 2 & 1\\  
\hline \hline
$p$ & {\bf order of surrogate construction} & 2 & 3 &4\\
        & quadrature points per dimension $(p+1)$& 3 & 4 & 5\\
$N$ & quadrature points $N = (p+1)^{d}$ & 27& 64& 125\\
$K$ & polynomial basis terms $K=(d+p)!/d!p!$ & 10& 20& 35\\
$N\cdot l$ & number of high-fidelity runs & 108 & 256& 500 \\
\hline
\end{tabular}
\label{table:designp2} 
\end{table}

\subsection{High-Fidelity Simulations vs.\ Surrogate Model}
As a first method to determine the validity of the surrogate model, the values of the high-fidelity \opal\ simulations on the x-axis and the values of the surrogate model on the y-axis were compared.\ The distance of the corresponding point to the line $x=y$ is a measure of the surrogate model's quality.\ The QoI's, as defined in \secref{sec:qoidef}, are compared for a subset of controllable parameters: $1,5,8$, and $10$ mA, and for $3$ different orders of the surrogate model, as described in \tabref{table:designp2}.\ All data from the surrogate model and the high-fidelity model are taken at the end of turn 10 in our model problem. The  maximum training error is calculated from the dataset used to create the surrogate model.

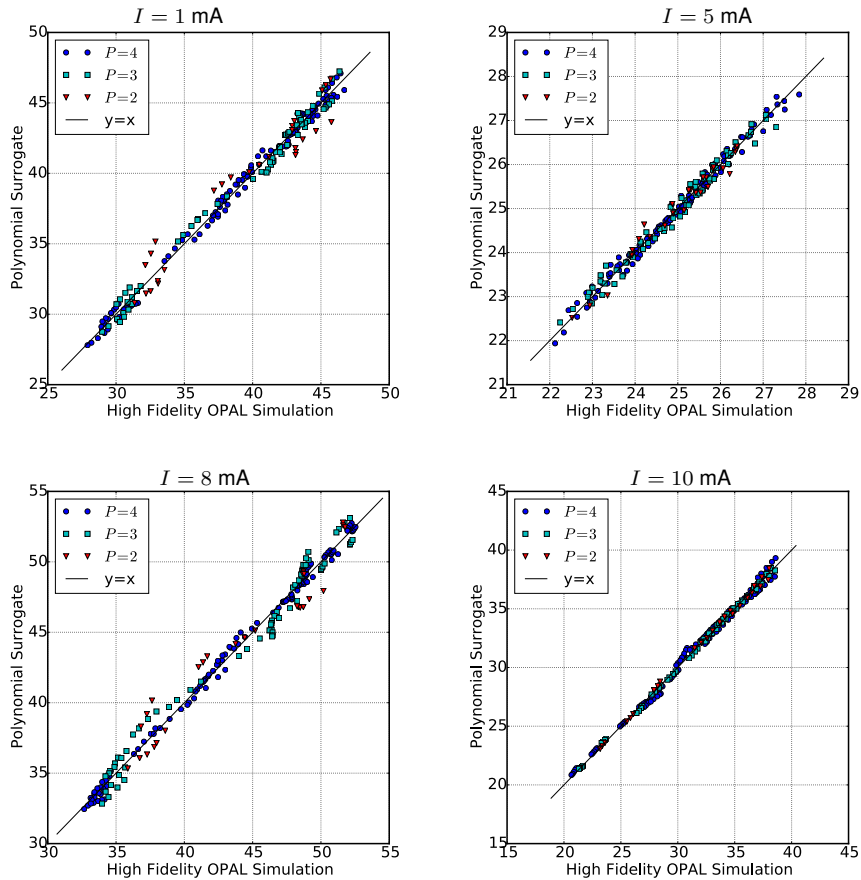
\begin{figure}[h!]
\centering
\begin{tikzpicture}
\node[above right] (img) at (0,0) {\includegraphics[width=0.45\textwidth]{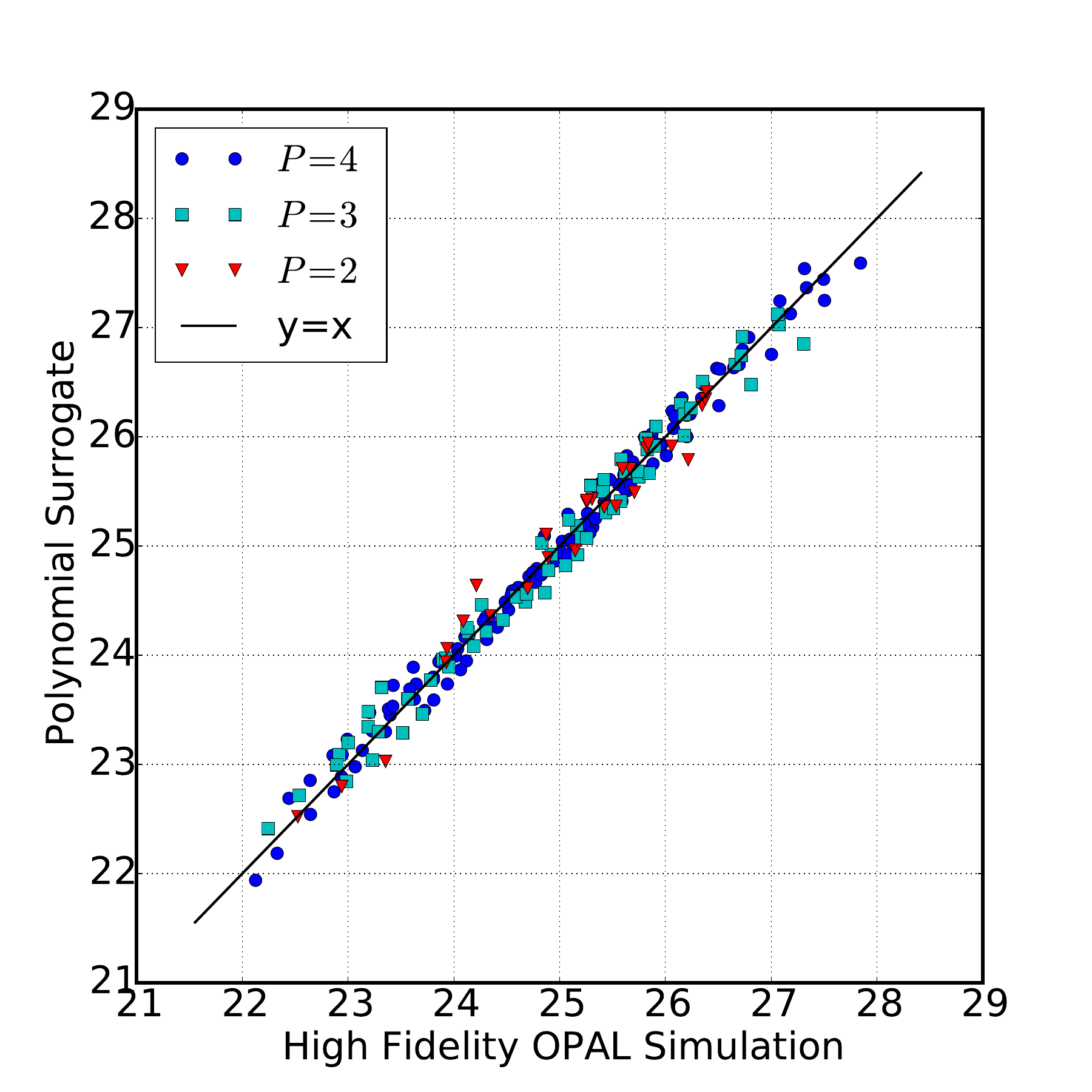}};
\node[below right] (img) at (0,0) {\includegraphics[width=0.45\textwidth]{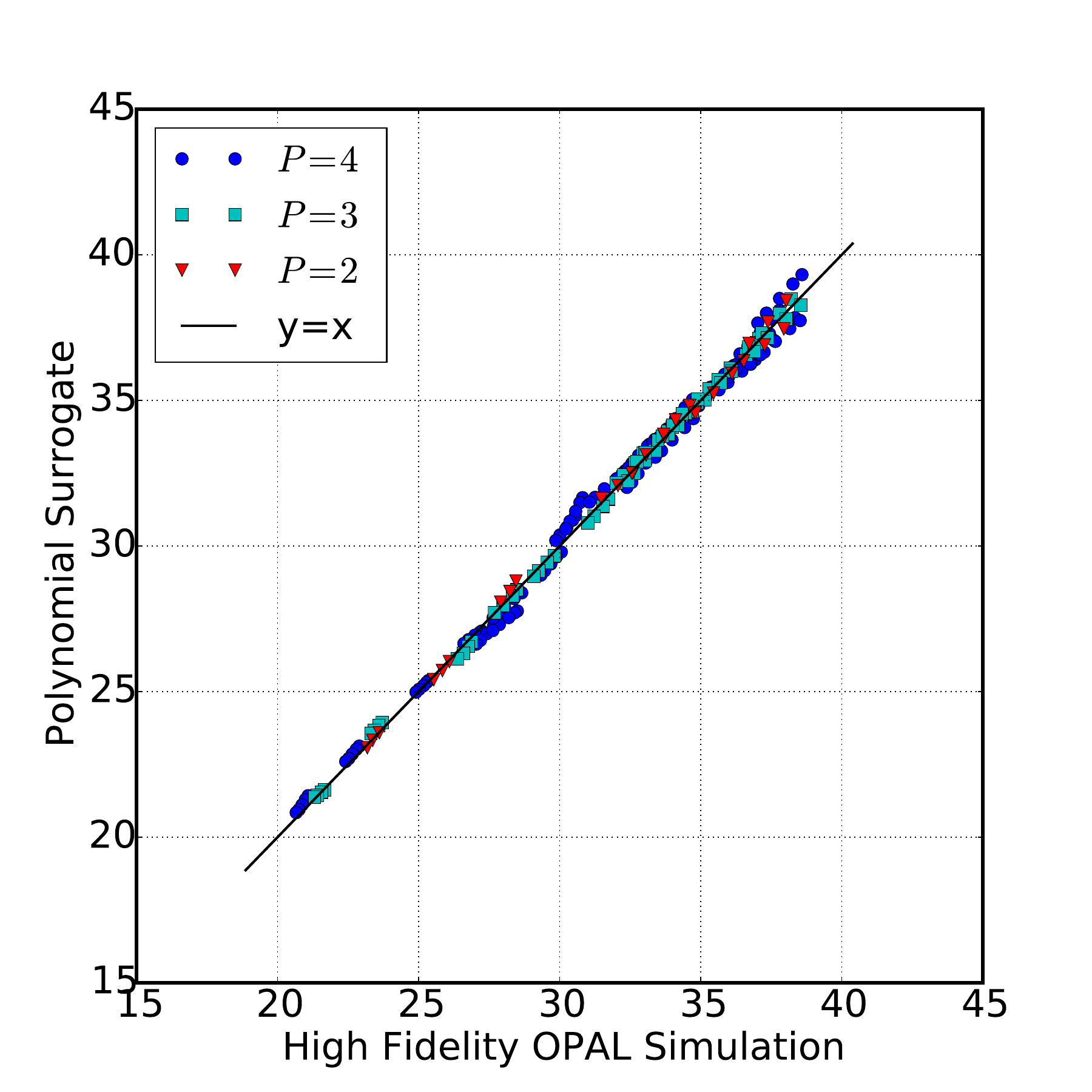}};
\node[above left] (img) at (0,0) {\includegraphics[width=0.45\textwidth]{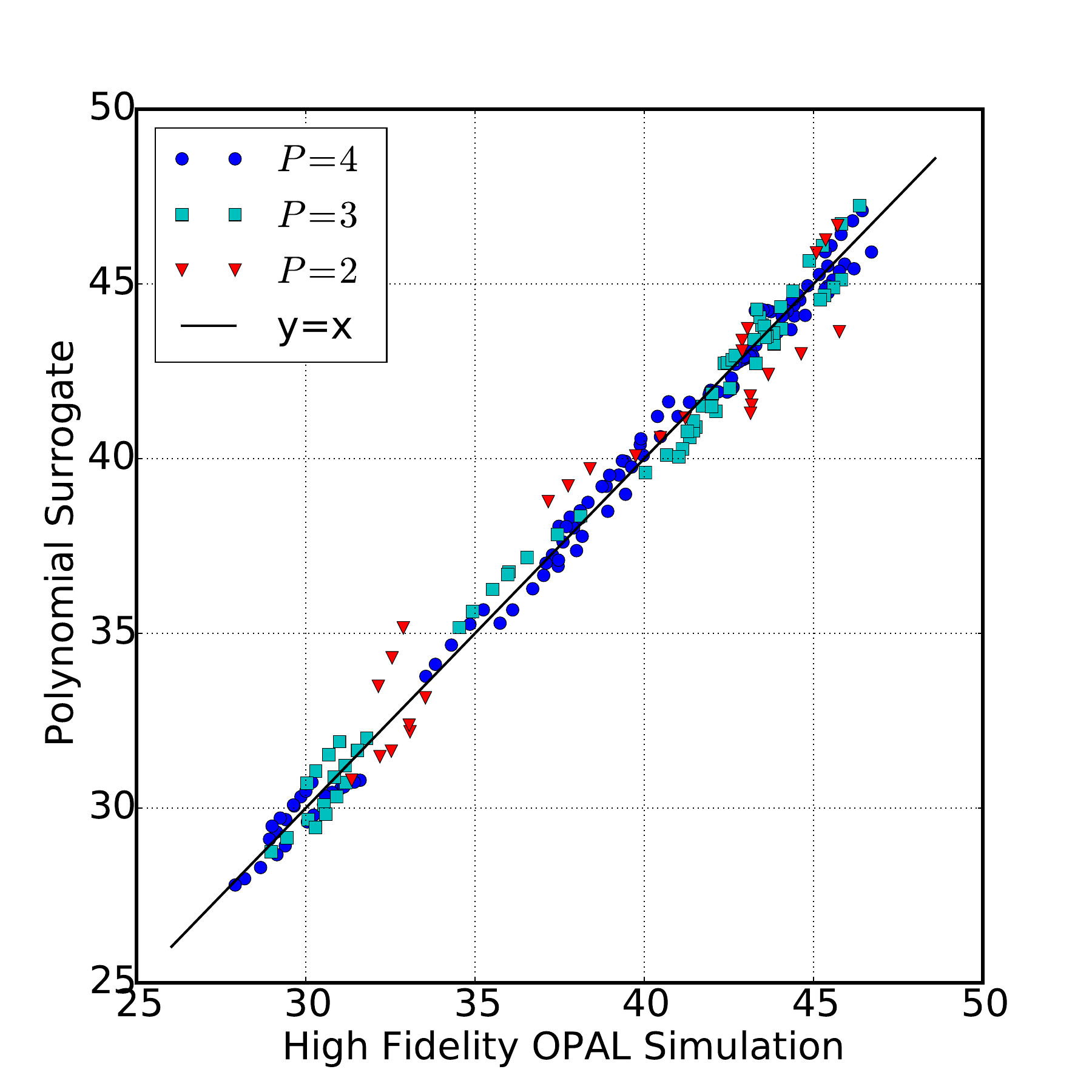}};
\node[below left] (img) at (0,0) {\includegraphics[width=0.45\textwidth]{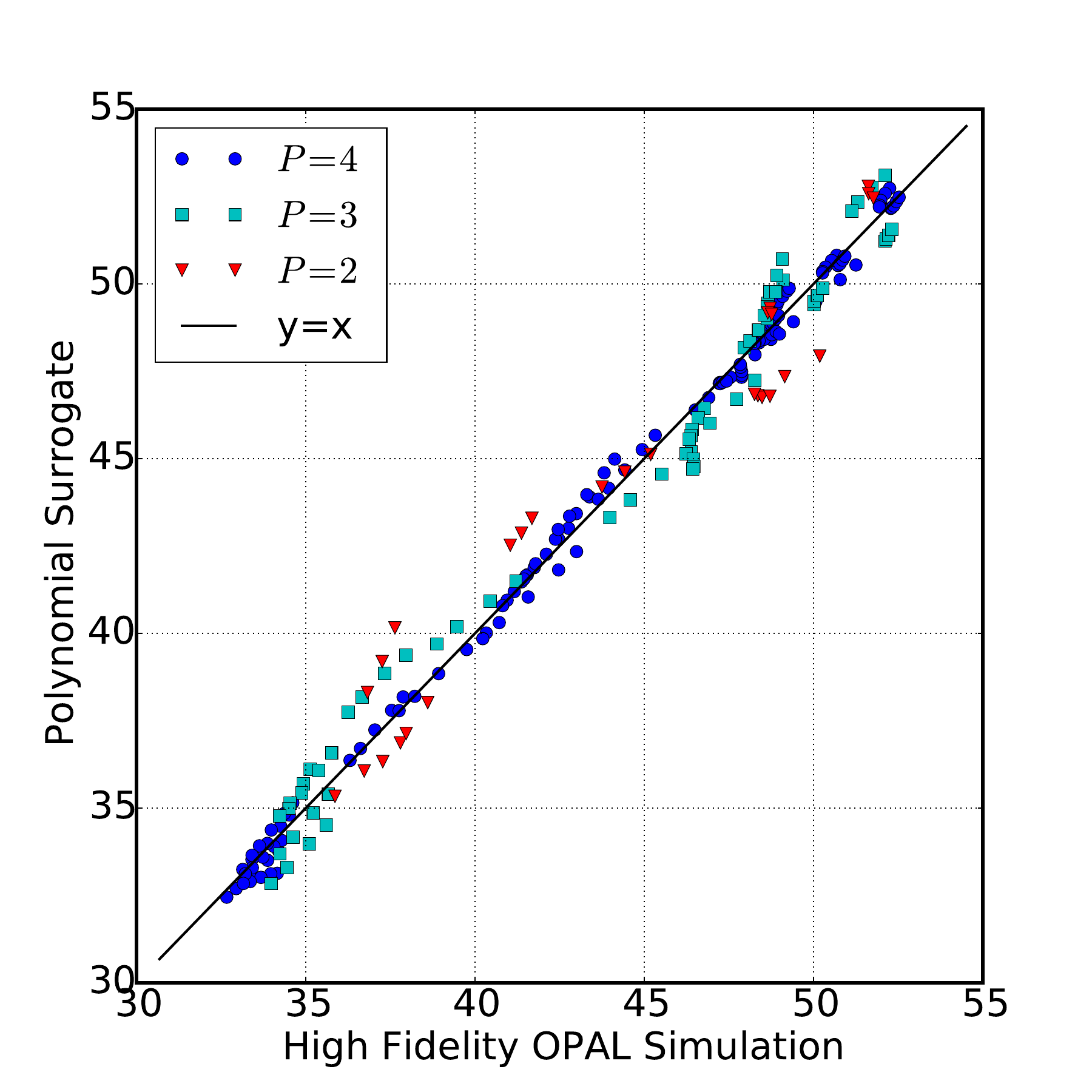}};
\node at (-100pt,160pt) {\footnotesize  $I=1$ mA};
\node at (100pt,160pt) {\footnotesize  $I=5$ mA};
\node at (-90pt,-15pt) {\footnotesize  $I=8$ mA};
\node at (90pt,-15pt) {\footnotesize $I=10$ mA};
\end{tikzpicture}
     \caption{Energy spread $\Delta E$ (keV) for all 3 experiments described in \tabref{table:designp2}.  
\label{fig:surmodel2}}
\end{figure}

Overall the expected convergence is observed when increasing $p$ as shown in \figref{fig:surmodel1} - \ref{fig:surmodel5}, and furthermore this is supported by the $L_2$ error shown in \secref{sec:l2err}.\
\clearpage
\subsubsection{Projected Emittance \& Beam size }
\begin{figure}[h!]
\centering
\begin{tikzpicture}
\node[above right] (img) at (0,0) {\includegraphics[width=0.45\textwidth]{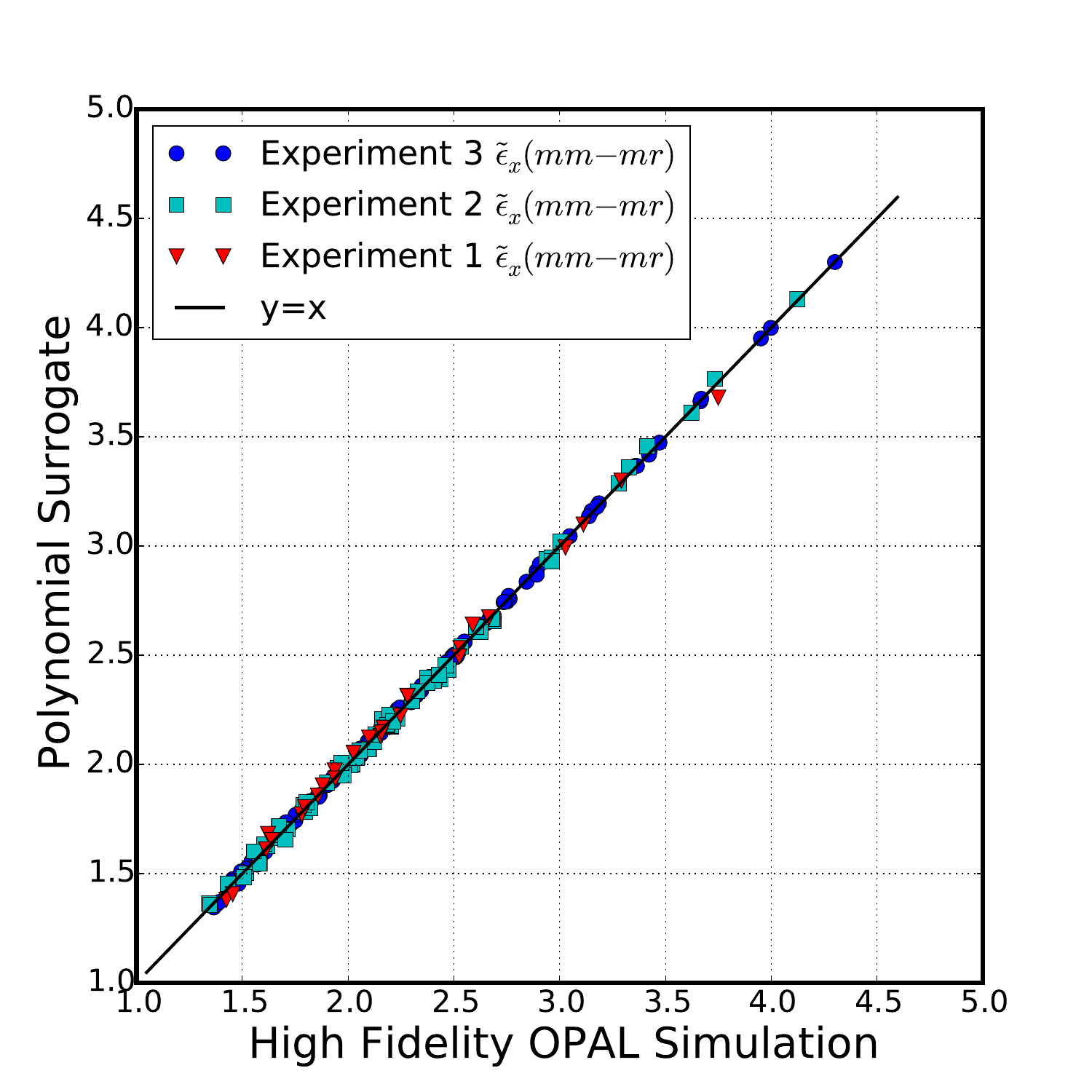}};
\node[below right] (img) at (0,0) {\includegraphics[width=0.45\textwidth]{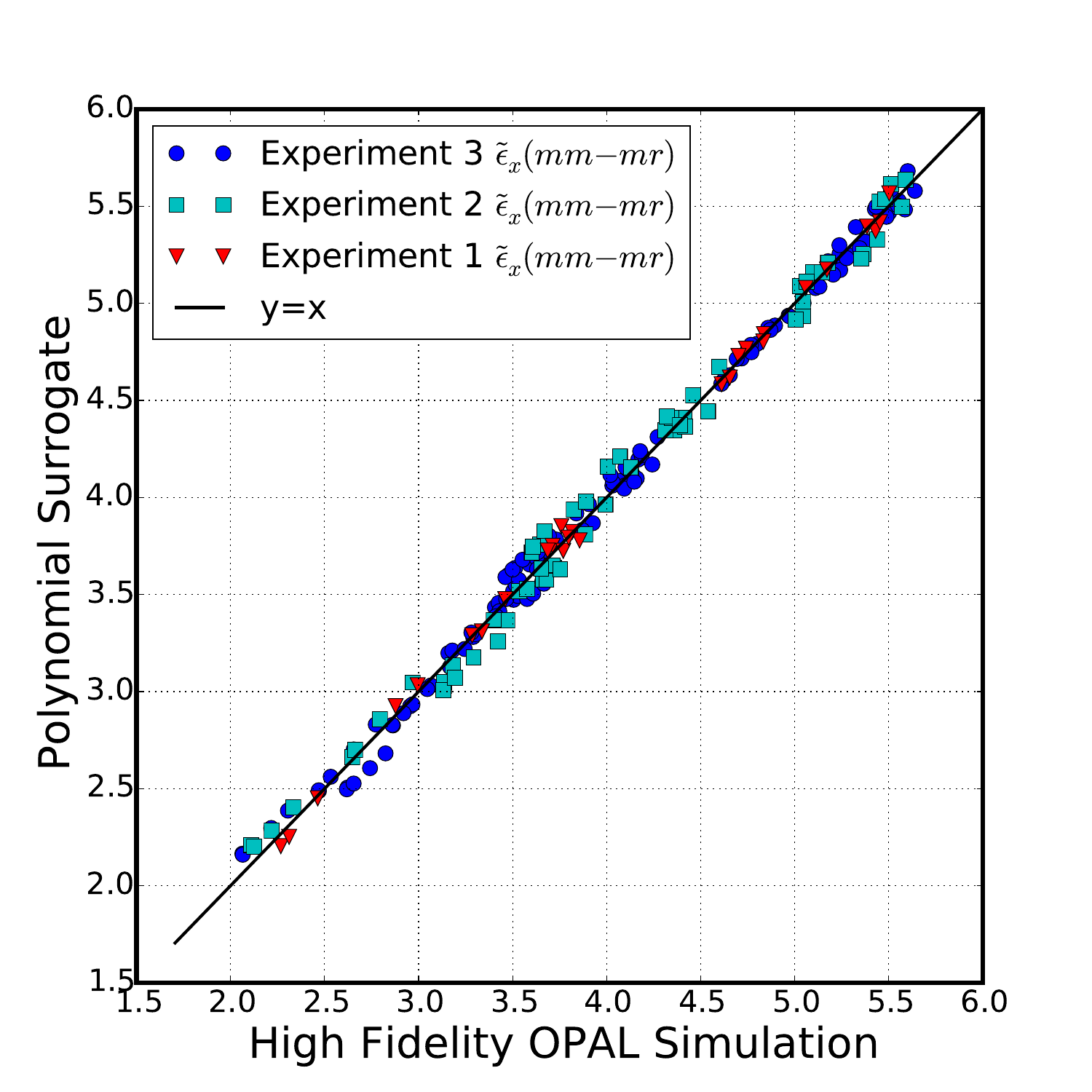}};
\node[above left] (img) at (0,0) {\includegraphics[width=0.45\textwidth]{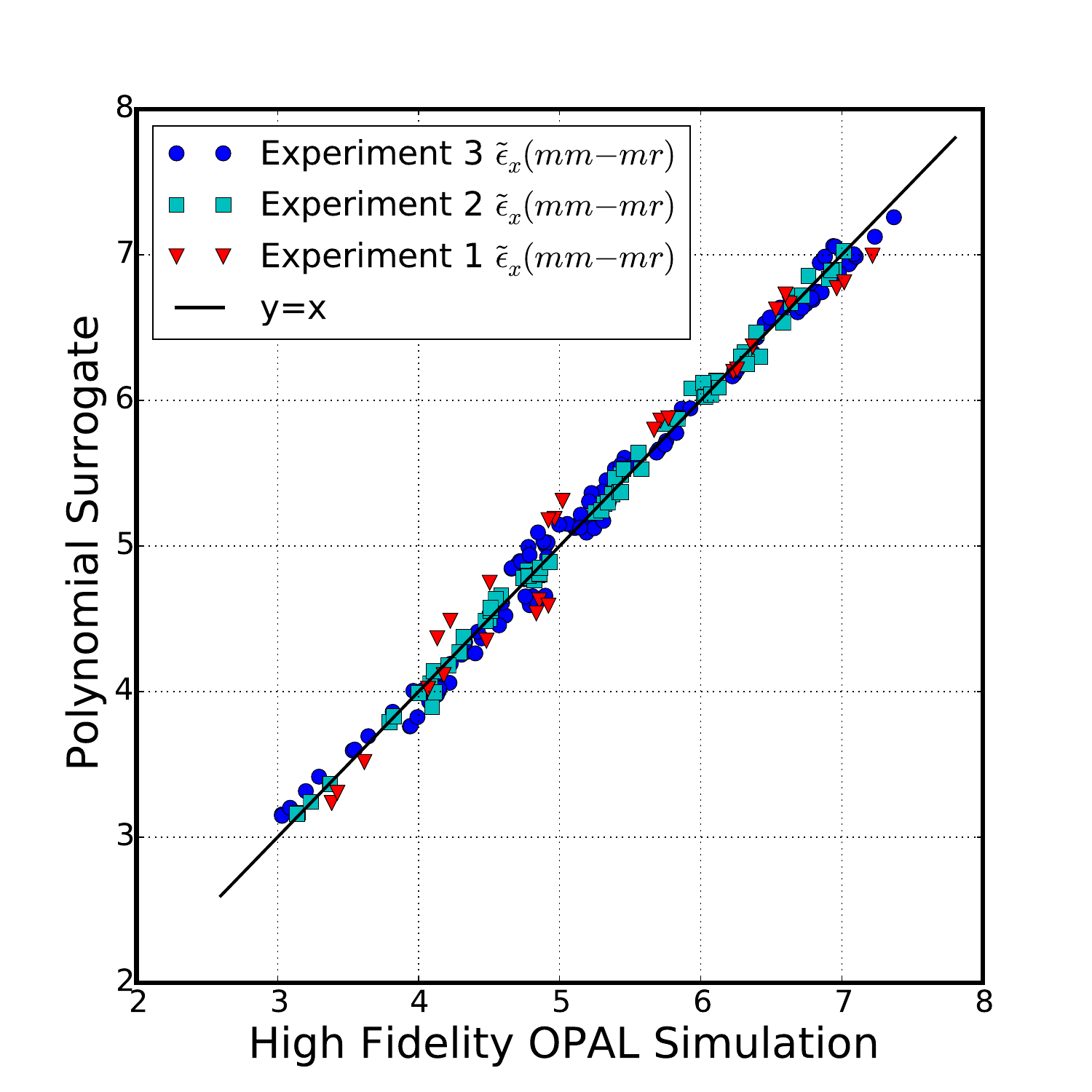}};
\node[below left] (img) at (0,0) {\includegraphics[width=0.45\textwidth]{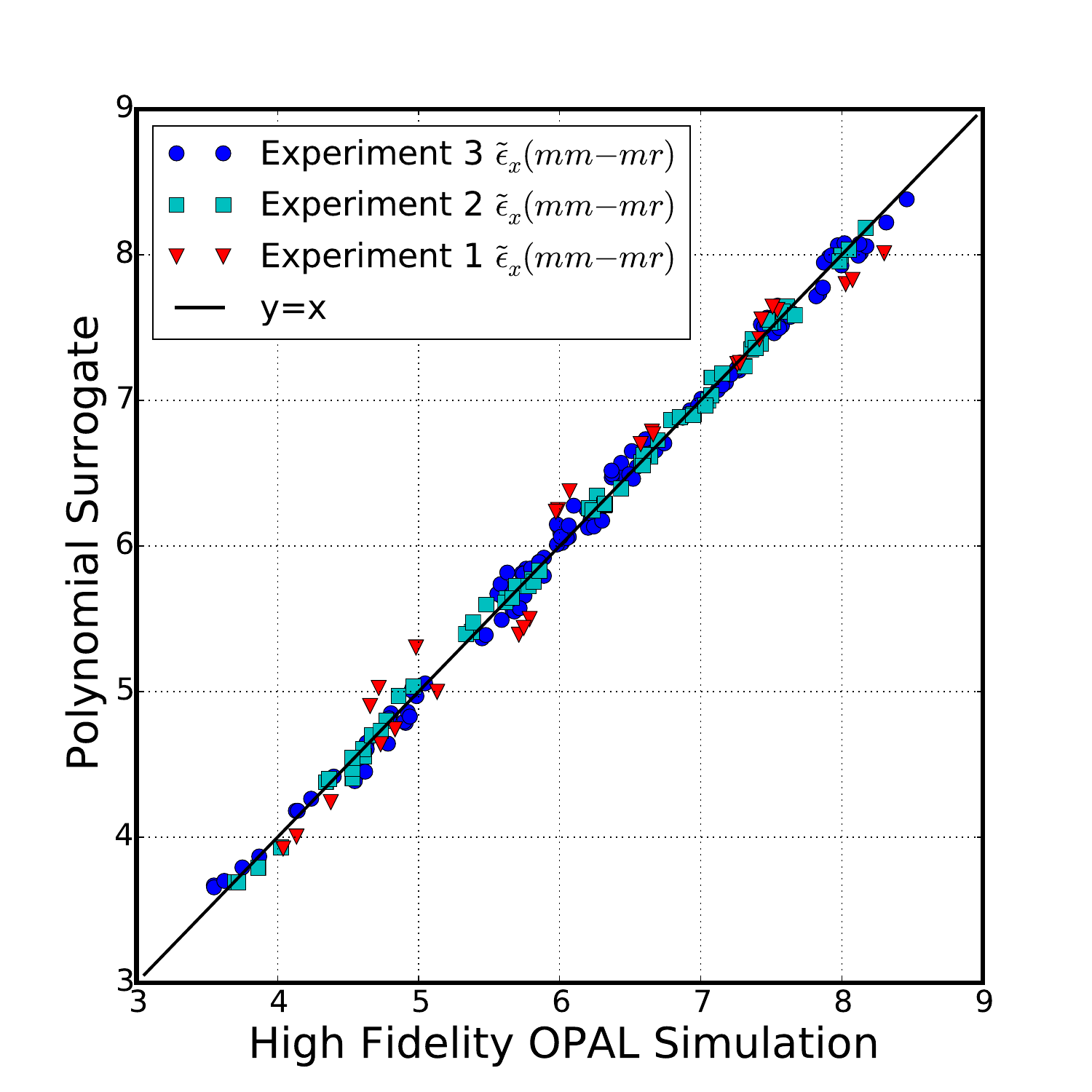}};
\node at (-100pt,160pt) {\footnotesize  $I=1$ mA};
\node at (100pt,160pt) {\footnotesize  $I=5$ mA};
\node at (-90pt,-15pt) {\footnotesize  $I=8$ mA};
\node at (90pt,-15pt) {\footnotesize $I=10$ mA};
\end{tikzpicture}
      \caption{Projected emittance $\tilde{\epsilon}_x$ (mm-mr) for all 3 experiments described in \tabref{table:designp2}. 
\label{fig:surmodel1}}
\end{figure}
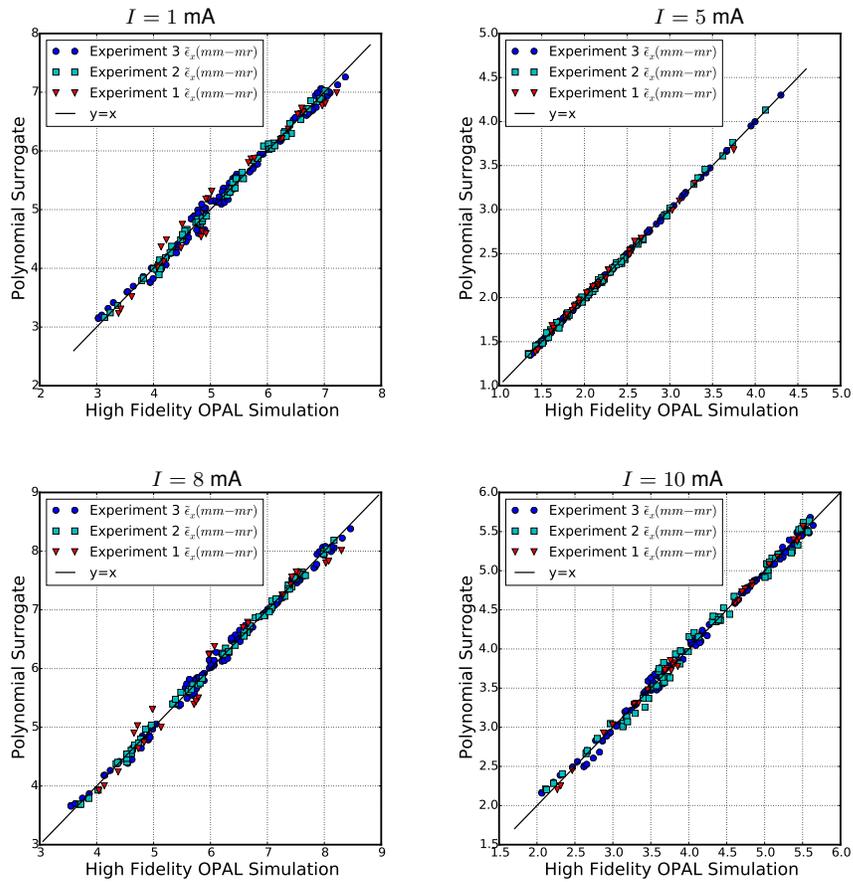

\begin{figure}[h!]
\centering
\begin{tikzpicture}
\node[above right] (img) at (0,0) {\includegraphics[width=0.45\textwidth]{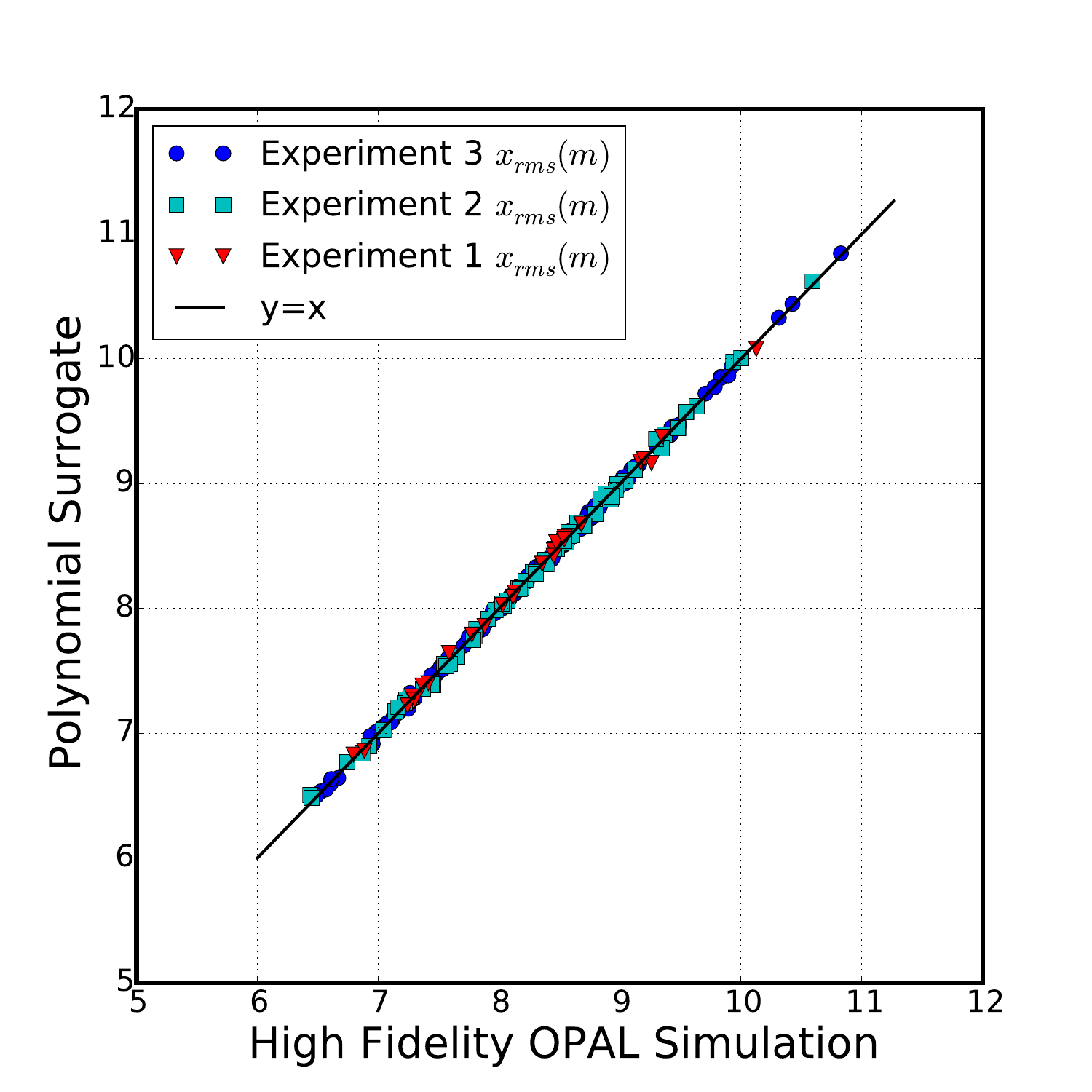}};
\node[below right] (img) at (0,0) {\includegraphics[width=0.45\textwidth]{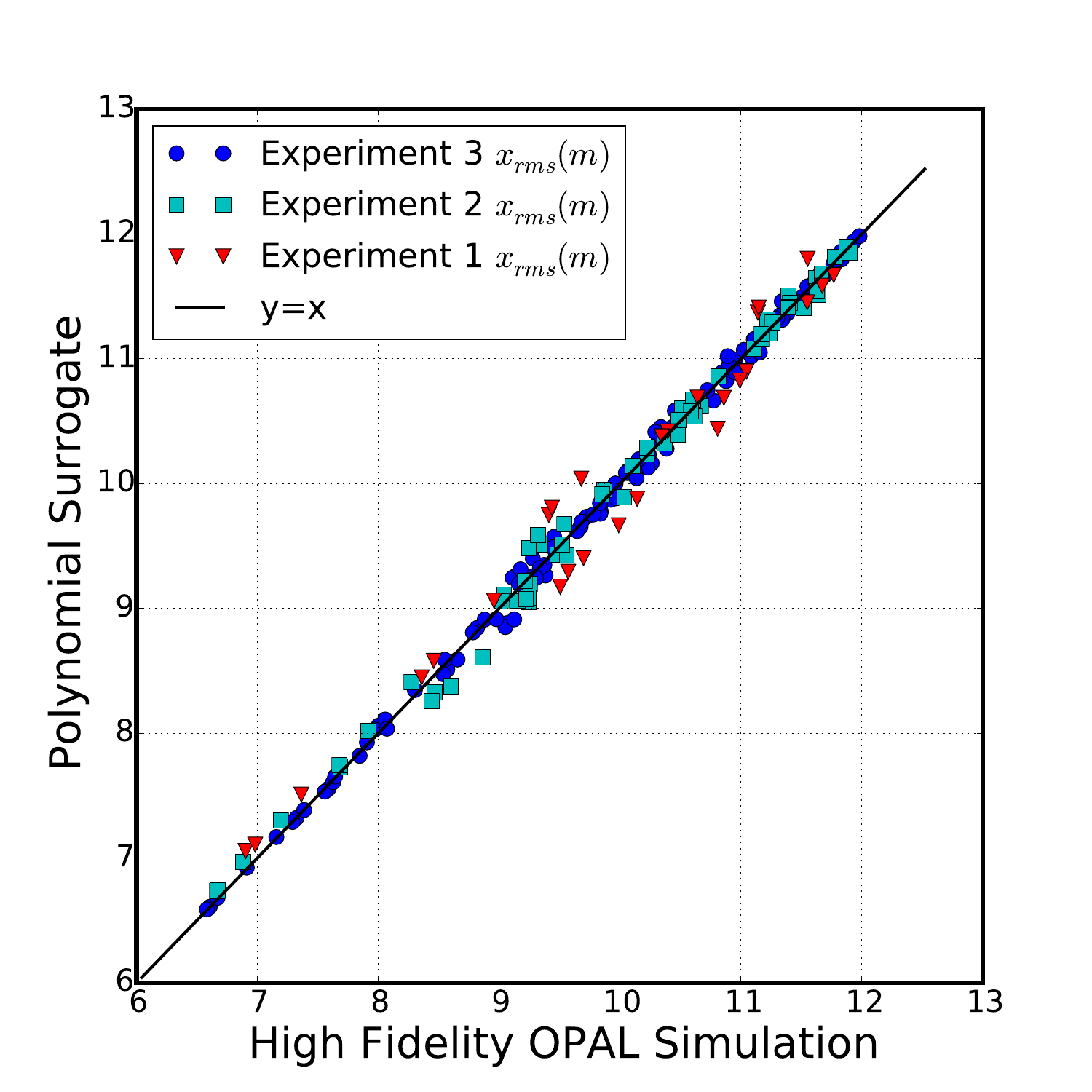}};
\node[above left] (img) at (0,0) {\includegraphics[width=0.45\textwidth]{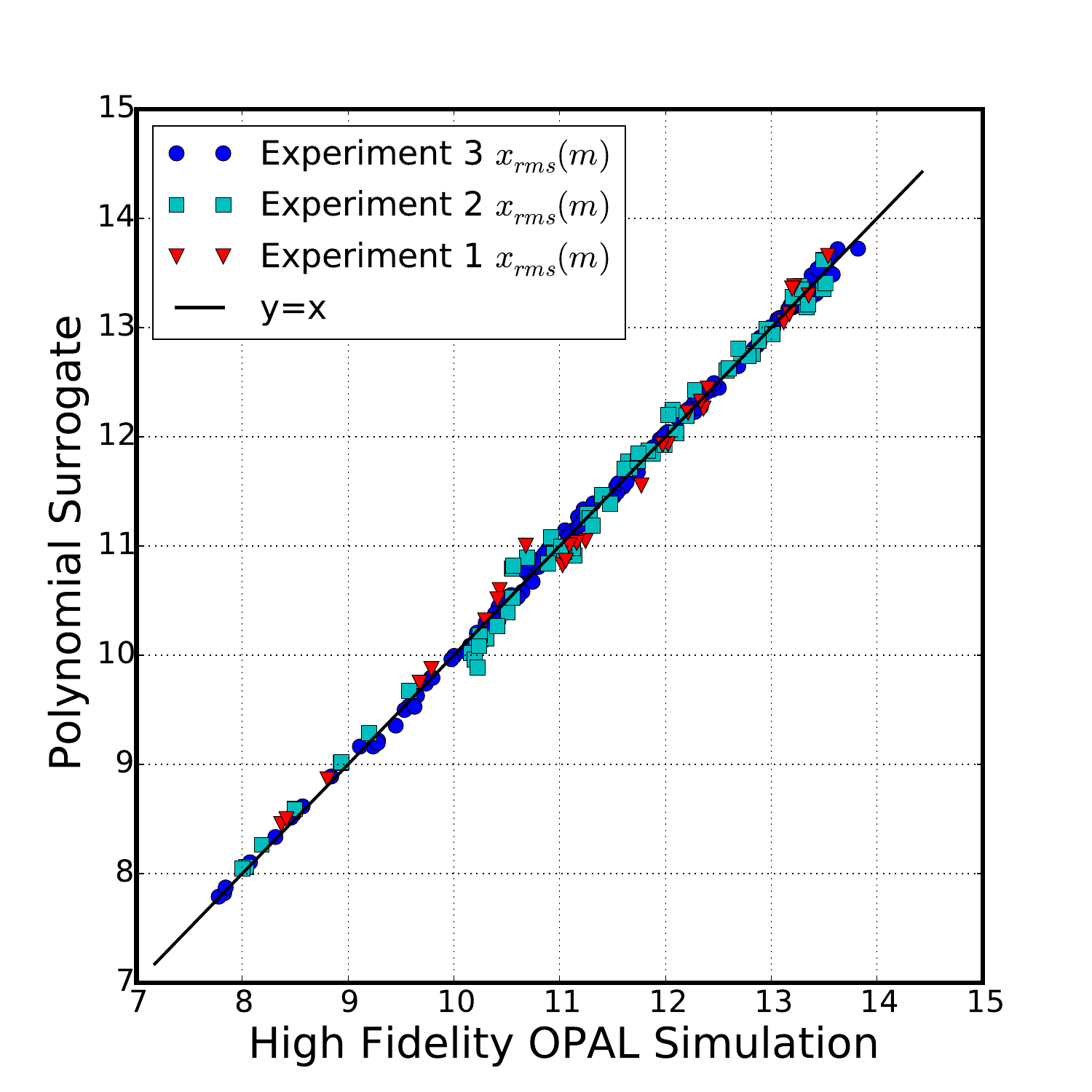}};
\node[below left] (img) at (0,0) {\includegraphics[width=0.45\textwidth]{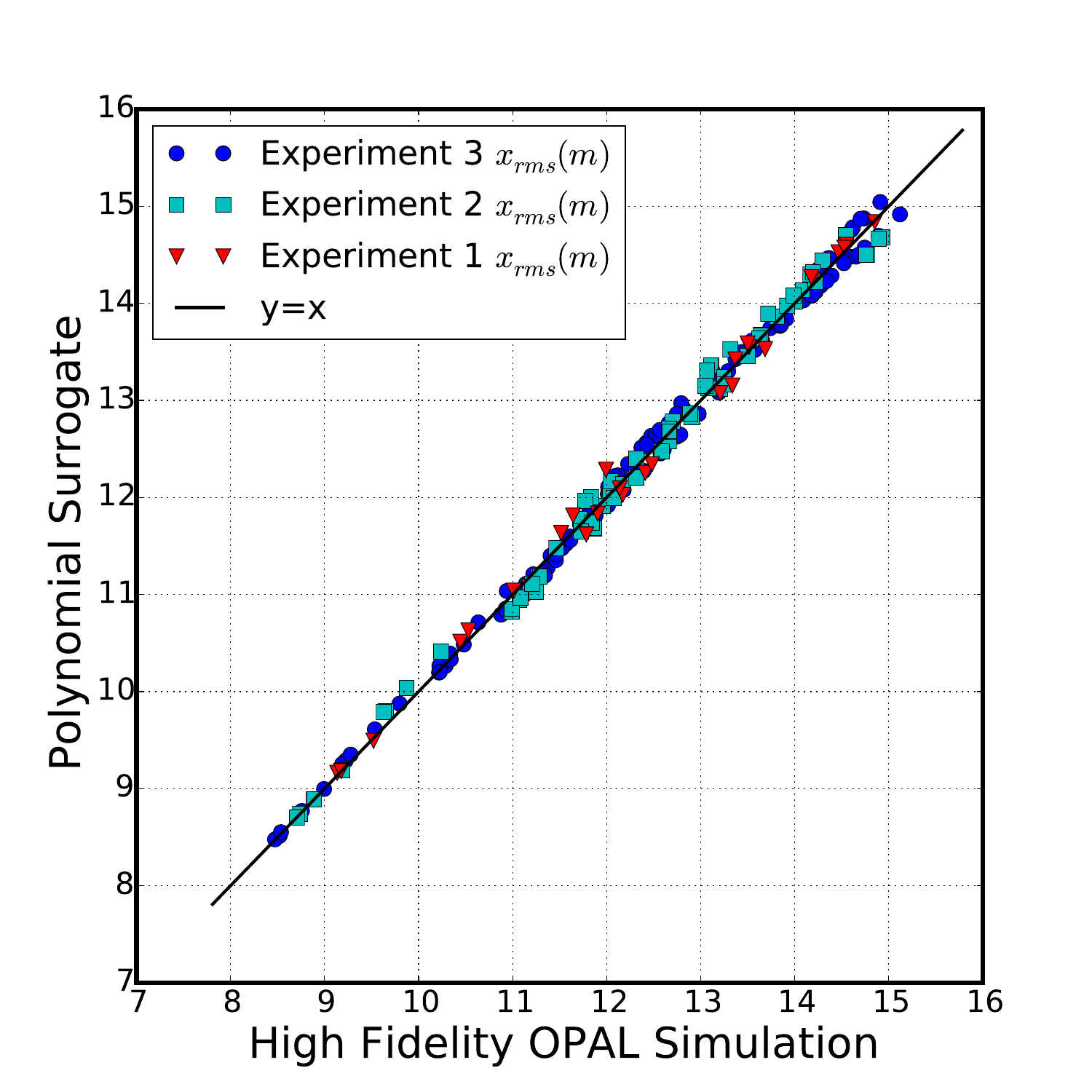}};
\node at (-100pt,160pt) {\footnotesize  $I=1$ mA};
\node at (100pt,160pt) {\footnotesize  $I=5$ mA};
\node at (-90pt,-15pt) {\footnotesize  $I=8$ mA};
\node at (90pt,-15pt) {\footnotesize $I=10$ mA};
\end{tikzpicture}
      \caption{The rms beam size $\tilde{x}$ (mm) for all 3 experiments described in \tabref{table:designp2}. 
\label{fig:surmodel22}}
\end{figure}
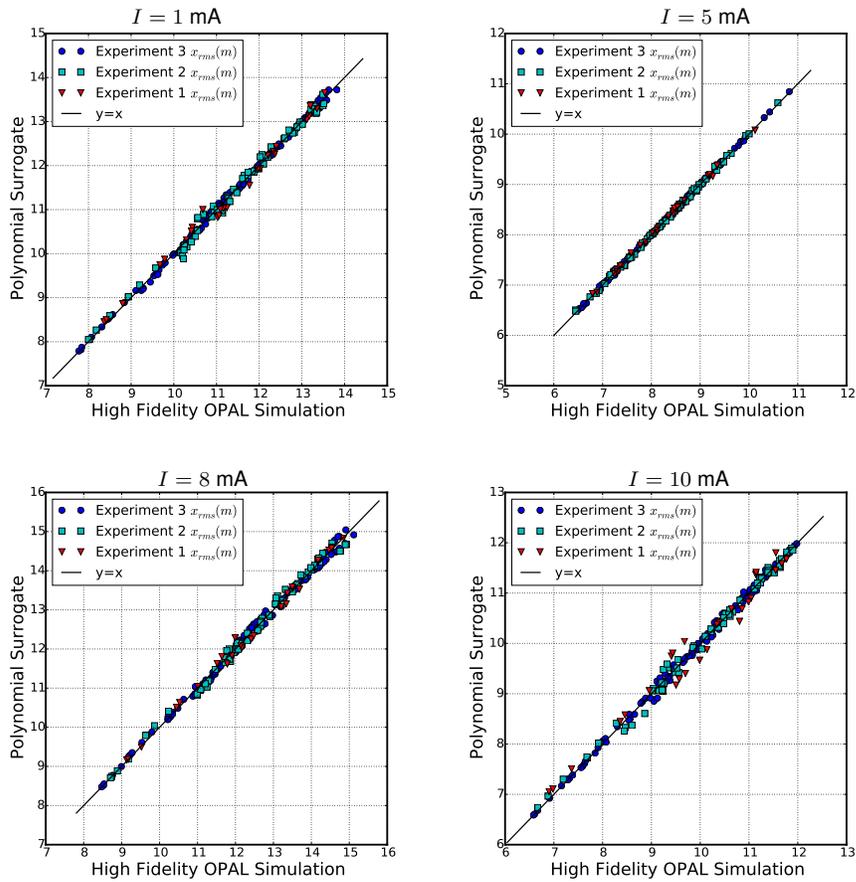

Given the fact that the emittance is a very sensitive quantity, measuring phase space volume, it is surprising, but also promising, that such a good 
agreement between the \sm\ and the \hfm\ can be achieved.\ This is graphically illustrated in \figref{fig:surmodel1} and \figref{fig:surmodel22}.\
The maximum training error in \% is given in \tabref{table:maxe1} and \tabref{table:maxe2}, and is below 7\%  for all considered cases. 
\begin{table}[h!]
\caption{Maximum training error in \% between the high-fidelity and surrogate model for the projected emittance $\tilde{\epsilon}_x$ of the beam.}
\centering
\begin{tabular}[t]{ l  l l l}   
\hline \hline
	 & $P=4$ & $P=3$ & $P=2$ \\
\hline \hline
$I=1$ mA &1.94  & 2.81  & 3.35 \\
$I=5$ mA & 5.04  & 4.77  & 2.79 \\
$I=8$ mA & 4.89  & 4.95  & 6.70 \\
$I=10$ mA &  3.6  & 2.78  & 5.60 \\
\hline
\end{tabular}
\label{table:maxe1} 
\end{table}

\begin{table}[h!]
\caption{Maximum training error in \% between the high fidelity and surrogate model for the rms beam size $\tilde{x}$ of the beam.}
\centering
\begin{tabular}[t]{ l  l l l}   
\hline \hline
	 & $P=4$ & $P=3$ & $P=2$ \\
\hline \hline
$I=1$ mA & 0.70 &  0.87 &  1.03  \\
$I=5$ mA & 2.32 &  2.90 & 3.49 \\
$I=8$ mA & 1.04 &  3.33 &  1.86 \\
$I=10$ mA &  1.33 &  1.98 &  1.39 \\
\hline
\end{tabular}
\label{table:maxe2} 
\end{table}

\clearpage
\subsubsection{Final Energy}
\begin{figure}[h!]
\begin{center}
\includegraphics[width=0.47\linewidth,angle=-0]{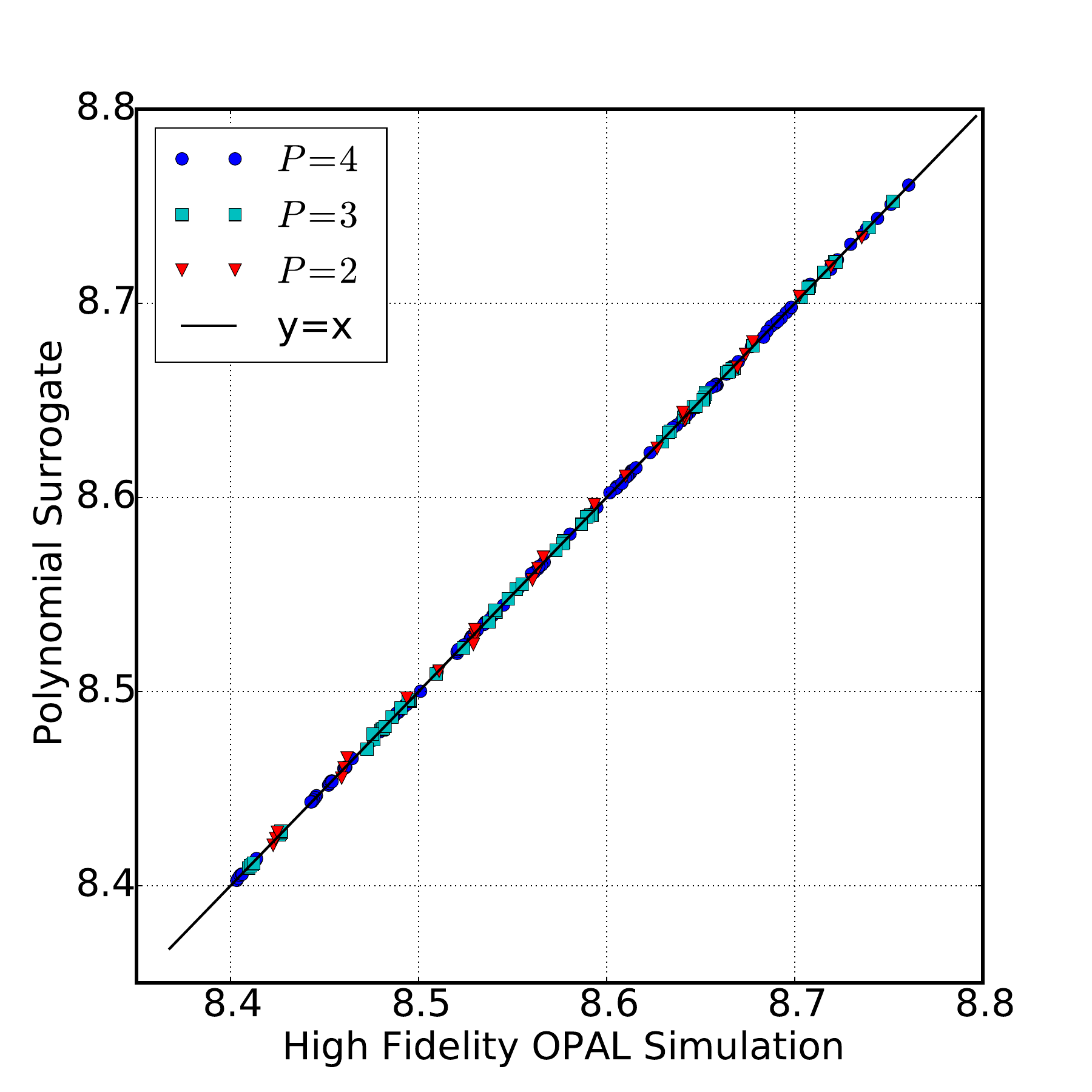}
\caption{Final Energy $E$ (MeV) for $I=10$ mA, and all experiments described in \tabref{table:designp2}.
\label{fig:surmodel3}}
\end{center}
\end{figure}

The energy dependence shown in \figref{fig:surmodel3} for 10 mA serves as an illustration of the expected behaviour for all other intensities.\  This is because of the 
small gain the third harmonic cavity is supposed to deliver (in the PSI Injector 2 we use the third harmonic cavity for acceleration).  For the given experiment only the last two turns are contributing. This fact is even better illustrated, when looking at the maximum training error, which is $\leq$ 0.07 \%,
as seen in \tabref{table:erre}.

\begin{table}[h!]
\caption{Maximum training error in \% between the high-fidelity and surrogate model for the final energy of the beam.}
\centering
\begin{tabular}[t]{ l  l l l}   
\hline \hline
	 & $P=4$ & $P=3$ & $P=2$ \\
\hline \hline
$I=1$ mA & 0.013 & 0.017 & 0.070 \\
$I=5$ mA & 0.013 & 0.036  & 0.066 \\
$I=8$ mA & 0.014  & 0.029  & 0.057 \\
$I=10$ mA & 0.010  & 0.027  & 0.056 \\
\hline
\end{tabular}
\label{table:erre} 
\end{table}

\clearpage

\subsubsection{Rms Energy Spread}
Despite the fact the rms energy spread is influenced by space charge, the collimation, and the change in phase, a very good agreement with absolute deviations $\le 5 \%$ was obtained. \tabref{table:designp3} shows details.
\begin{table}[h!]
\caption{Maximum training error in \% between the high fidelity and surrogate model for the energy spred $\Delta E$ of the beam.}
\centering
\begin{tabular}[t]{ l  l l l}   
\hline \hline
	 & $P=4$ & $P=3$ & $P=2$ \\
\hline \hline
$I=1$ mA & 0.97  &1.67 & 1.62 \\
$I=5$ mA & 2.56  & 1.04 & 1.29 \\
$I=8$ mA & 2.56  & 2.75 & 4.65 \\
$I=10$ mA & 3.00  & 3.70 &4.48\\
\hline
\end{tabular}
\label{table:designp3} 
\end{table}

\clearpage

\subsubsection{Halo Parameters}
The halo parameter was evaluated at turn 5 (\figref{fig:surmodel4}) and at turn 10  (\figref{fig:surmodel5}). As anticipated the halo grows and the surrogate model has a maximum absolute error
of $\le 5 \%$, again a very good accuracy.
\begin{figure}[h!]
\centering
\begin{tikzpicture}
\node[above right] (img) at (0,0) {\includegraphics[width=0.45\textwidth]{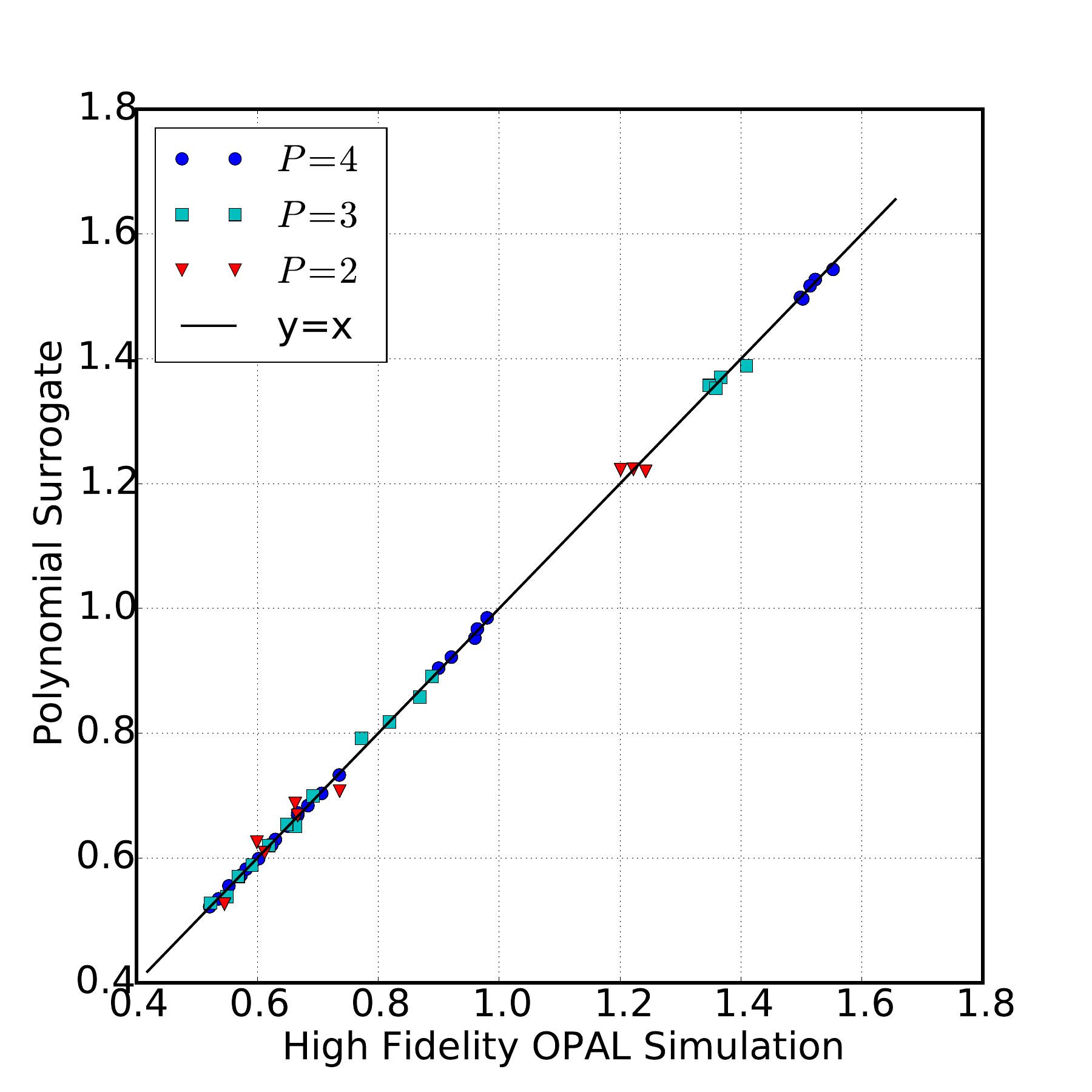}};
\node[below right] (img) at (0,0) {\includegraphics[width=0.45\textwidth]{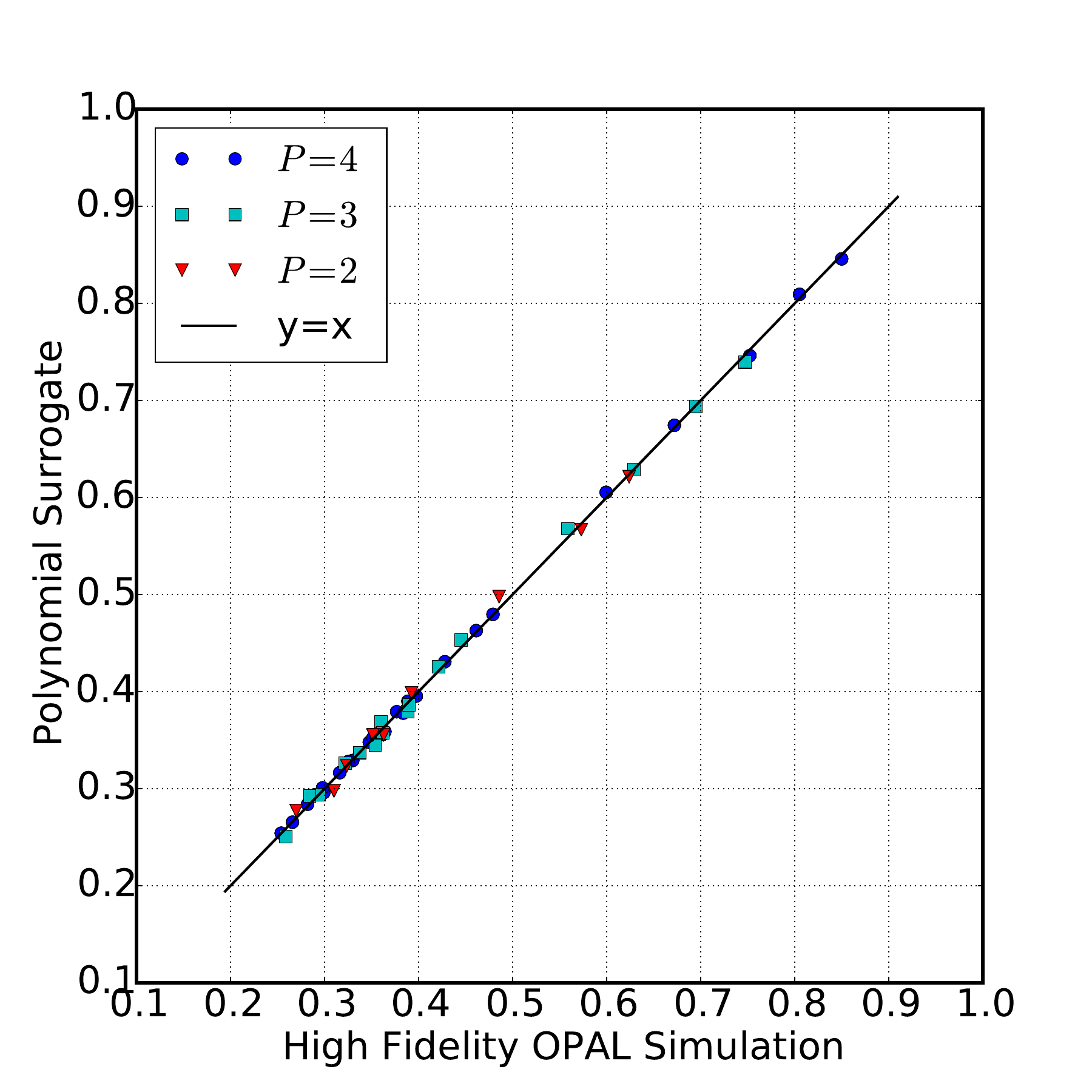}};
\node[above left] (img) at (0,0) {\includegraphics[width=0.45\textwidth]{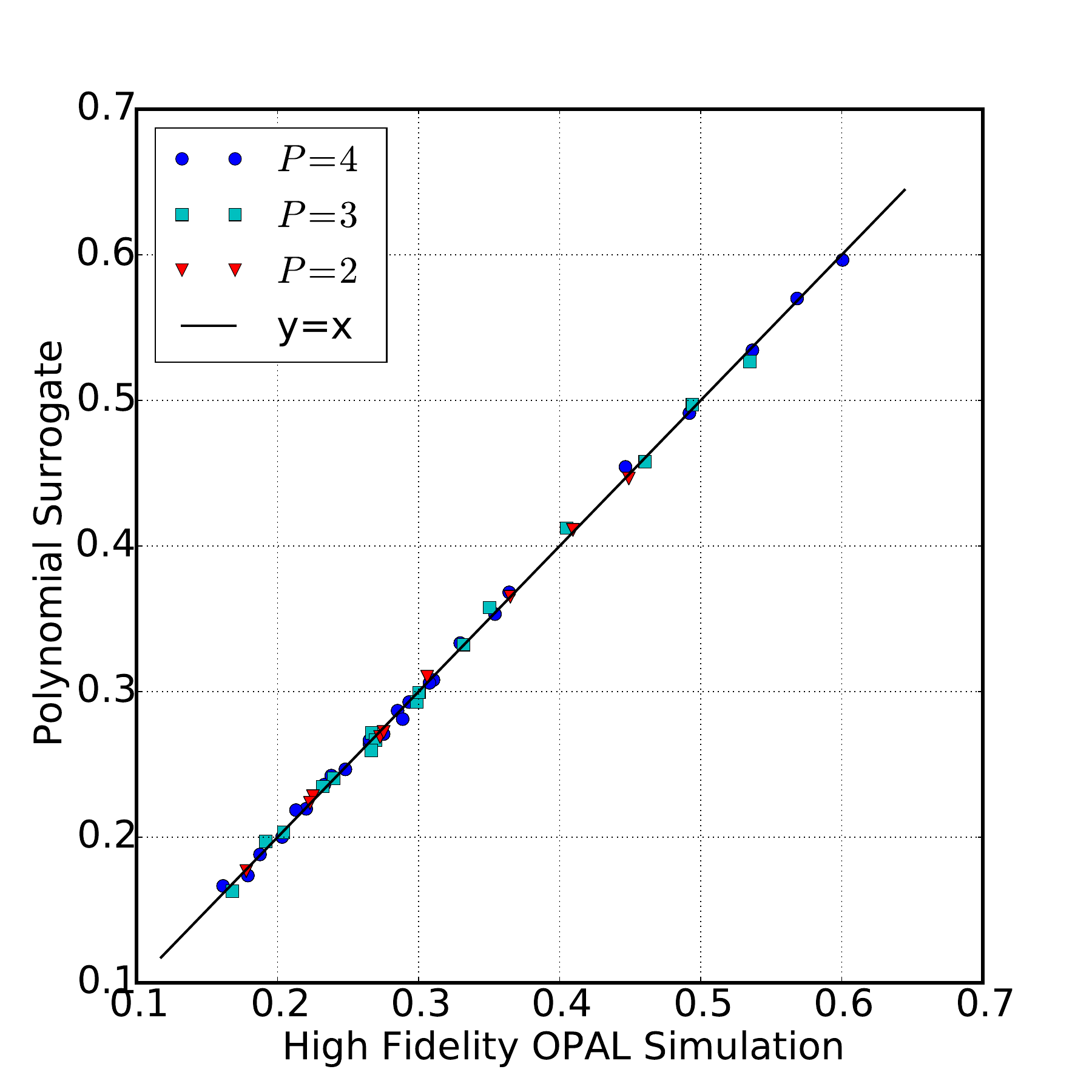}};
\node[below left] (img) at (0,0) {\includegraphics[width=0.45\textwidth]{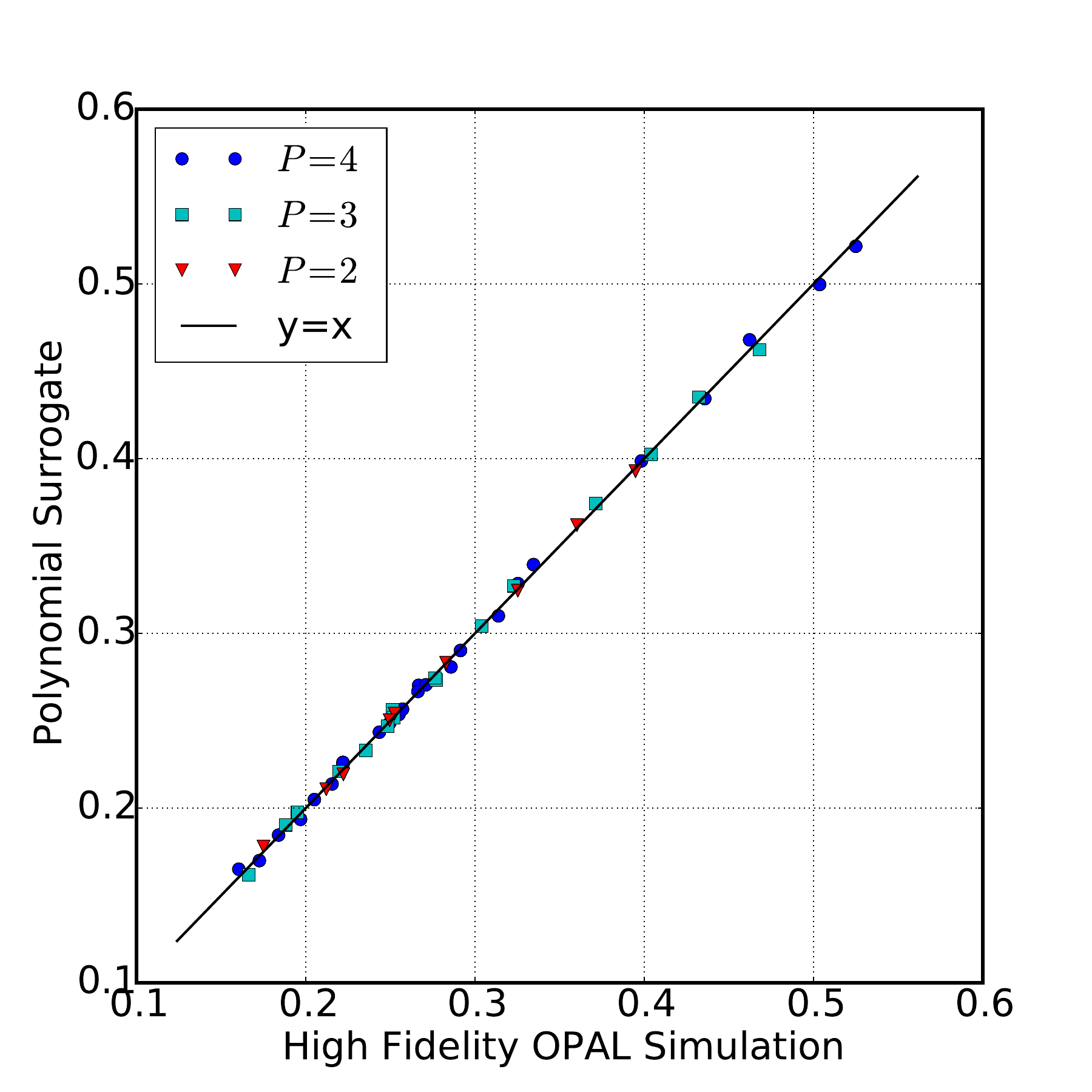}};
\node at (-100pt,190pt) {\footnotesize  $I=1$ mA};
\node at (100pt,190pt) {\footnotesize  $I=5$ mA};
\node at (-90pt,-15pt) {\footnotesize  $I=8$ mA};
\node at (90pt,-15pt) {\footnotesize $I=10$ mA};
\end{tikzpicture}
     \caption{The dimensionless halo parameter $h$ after turn 5 for all 3 experiments described in \tabref{table:designp2}. 
\label{fig:surmodel4}}
\end{figure}
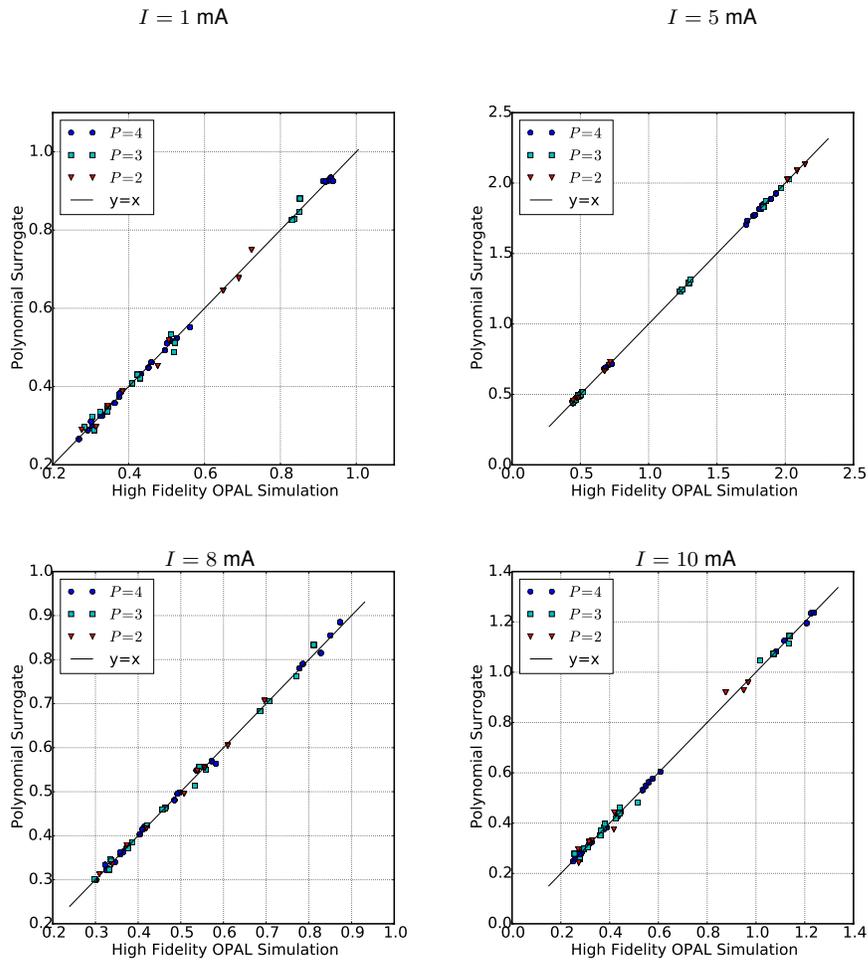
\begin{figure}[h!]
\centering
\begin{tikzpicture}
\node[above right] (img) at (0,0) {\includegraphics[width=0.45\textwidth]{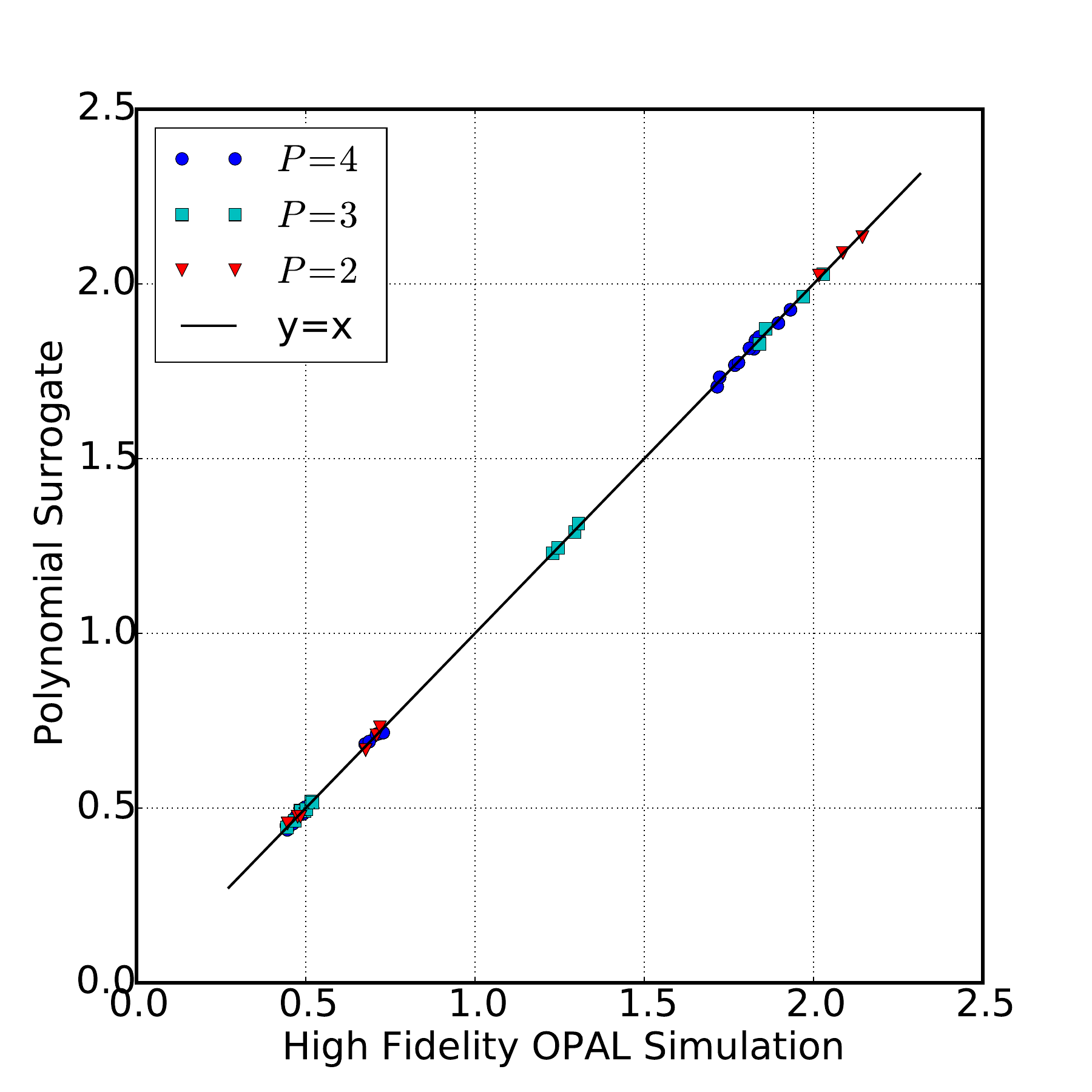}};
\node[below right] (img) at (0,0) {\includegraphics[width=0.45\textwidth]{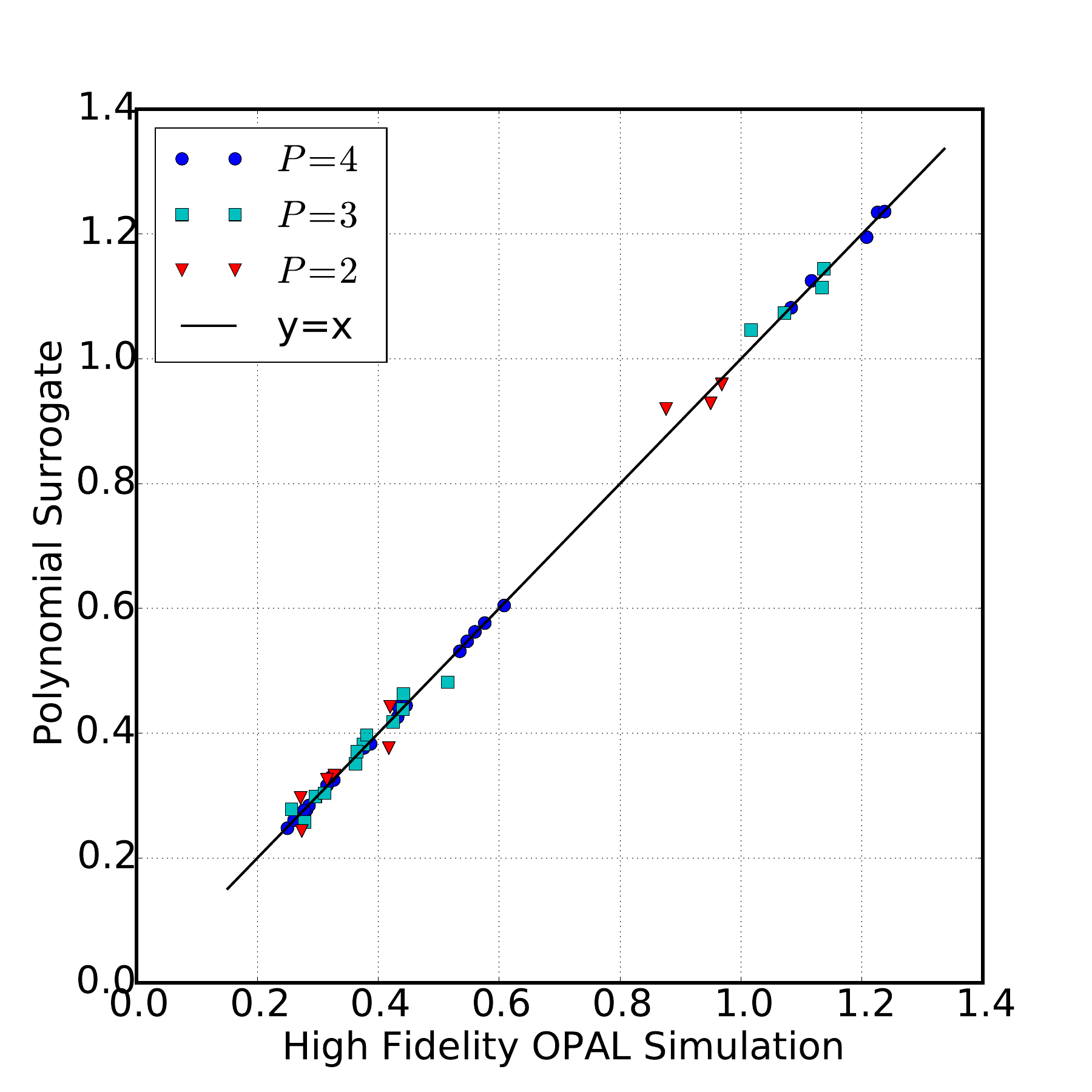}};
\node[above left] (img) at (0,0) {\includegraphics[width=0.45\textwidth]{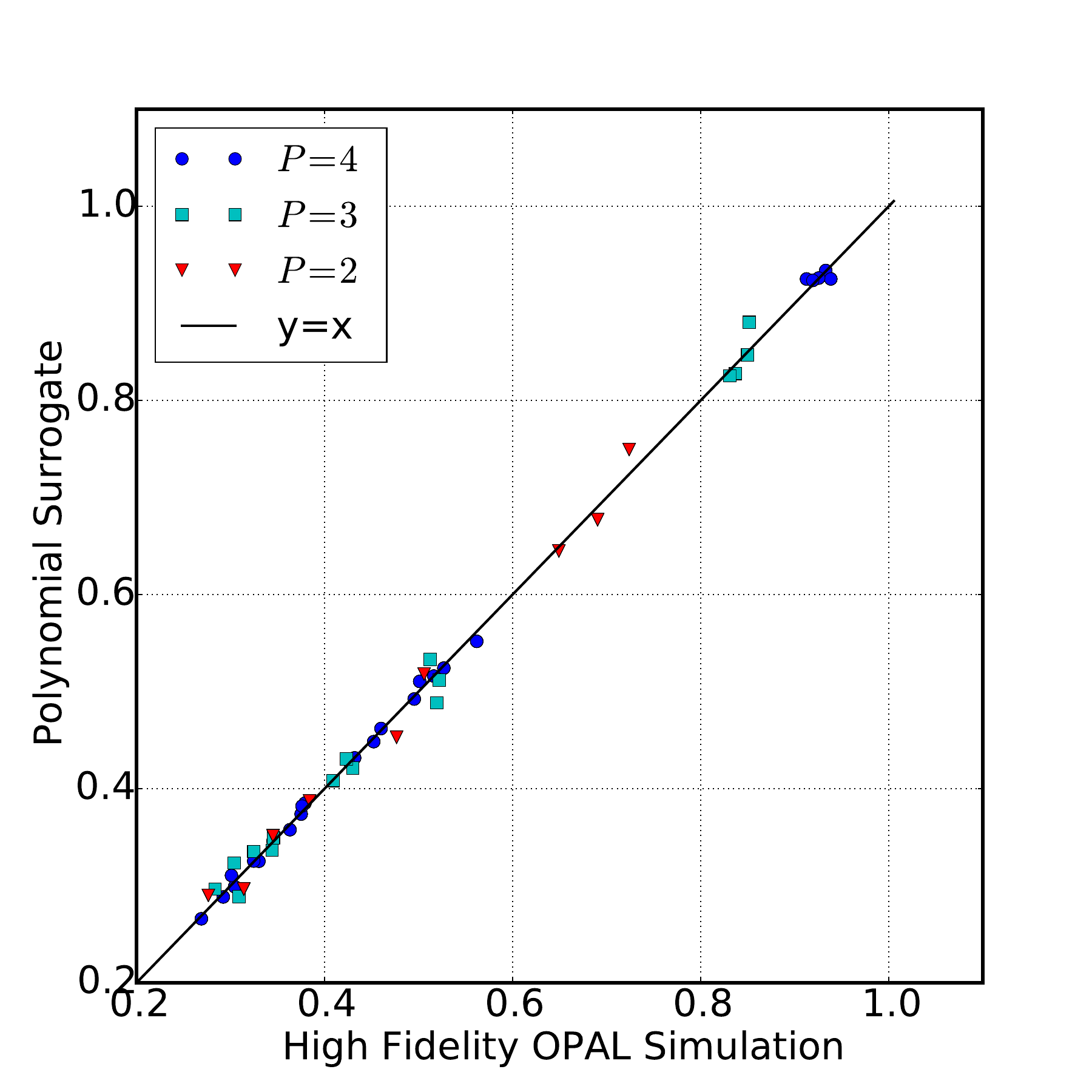}};
\node[below left] (img) at (0,0) {\includegraphics[width=0.45\textwidth]{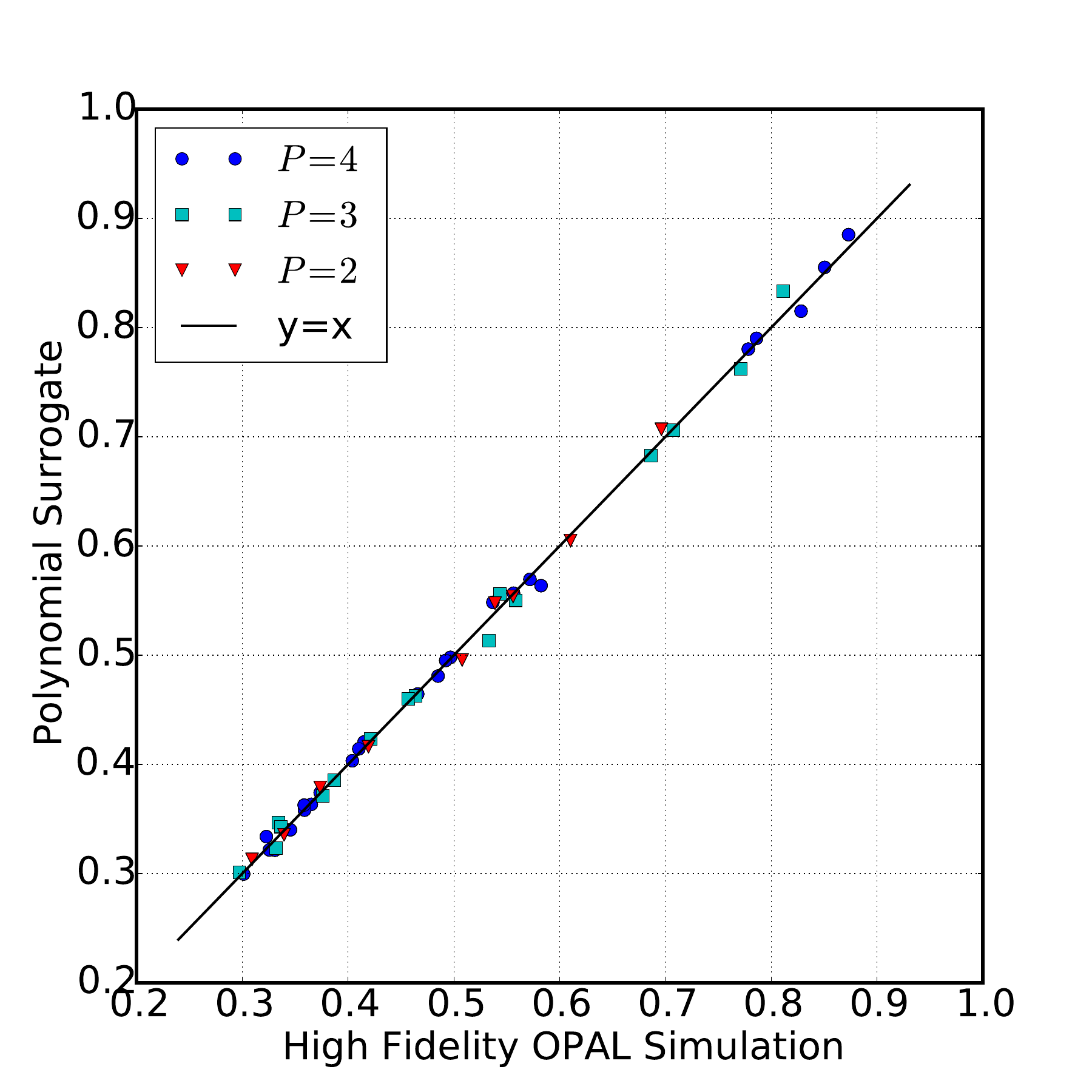}};
\node at (-100pt,190pt) {\footnotesize  $I=1$ mA};
\node at (100pt,190pt) {\footnotesize  $I=5$ mA};
\node at (-90pt,-15pt) {\footnotesize  $I=8$ mA};
\node at (90pt,-15pt) {\footnotesize $I=10$ mA};
\end{tikzpicture}
     \caption{The dimensionless halo parameter $h$ after turn 10 for all 3 experiments described in \tabref{table:designp2}. 
\label{fig:surmodel5}}
\end{figure}

\clearpage

\subsection{Sensitivity Analysis}
\label{sec:sensanal}
$S_{k}$ in \eqref{appeq:1sobol} can be interpreted as the fraction of the variance in model $\mathcal{M}$ that can be attributed to the i-th input parameter only. $S_{k}^{T}$ in \eqref{appeq:totsobol} measures the fractional contribution to the total variance due to the i-th parameter and its interactions with all other model parameters. In the sequel 
an analysis based on $S_{k}^{T}$ is shown for the model problem.

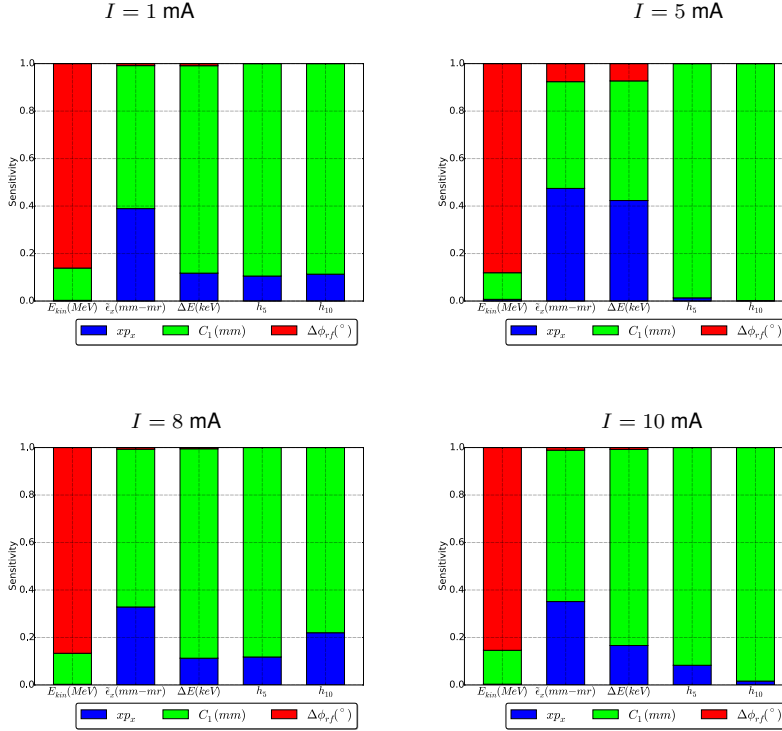
\begin{figure}[h!]
\begin{center}
\begin{tikzpicture}
\node[above right] (img) at (0,0) {\includegraphics[width=0.42\textwidth]{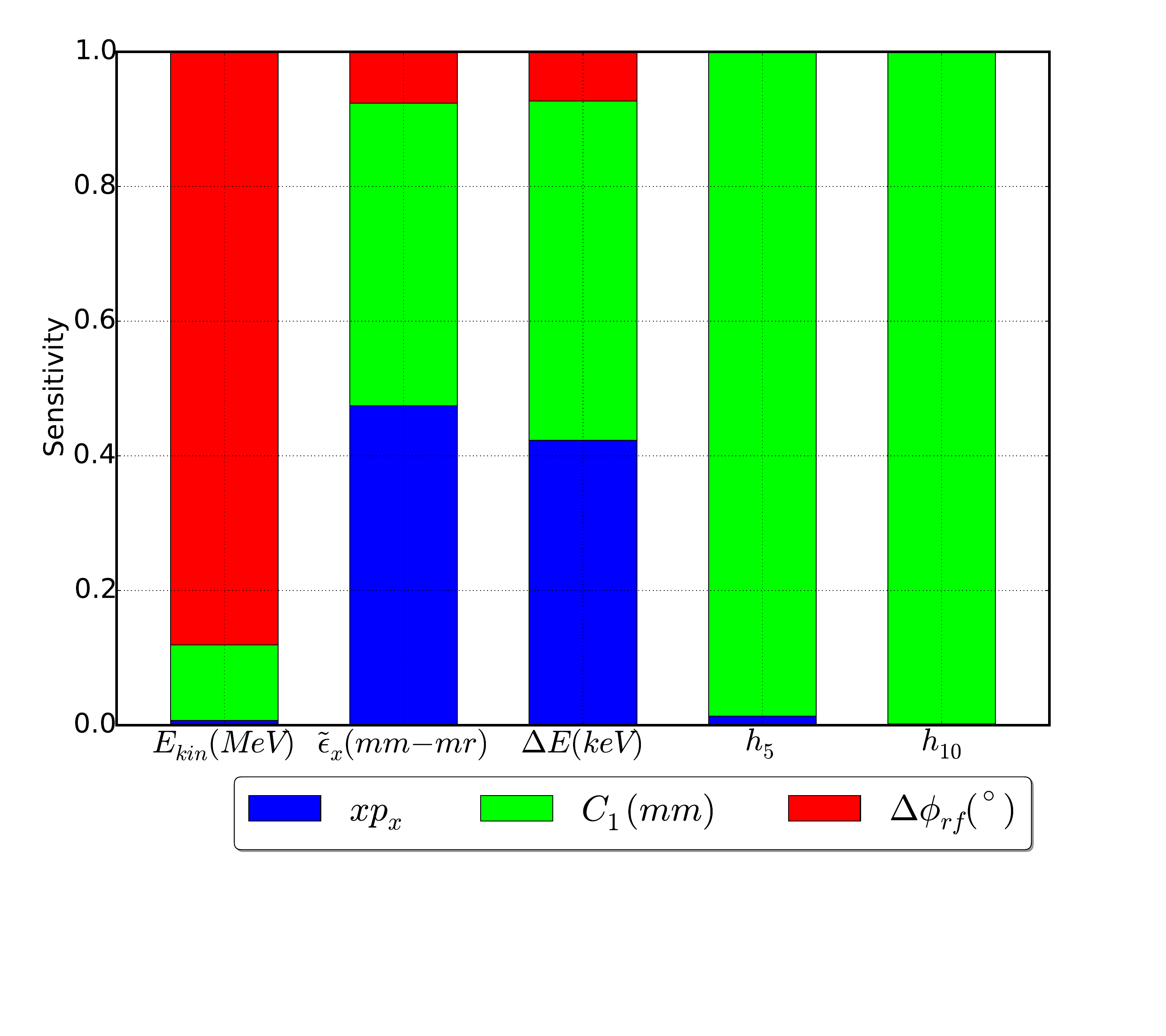}};
\node[below right] (img) at (0,0) {\includegraphics[width=0.42\textwidth]{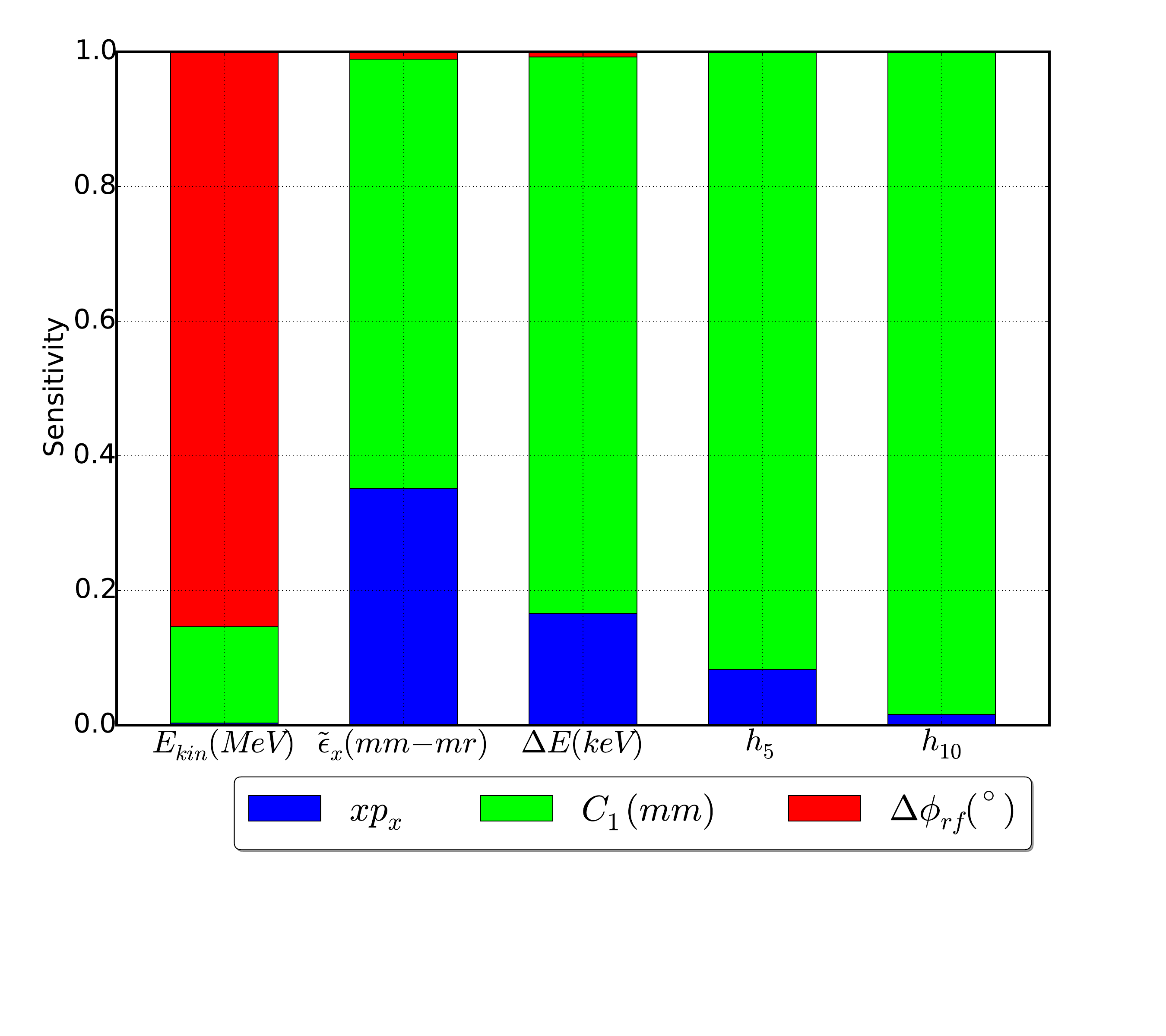}};
\node[above left] (img) at (0,0) {\includegraphics[width=0.42\textwidth]{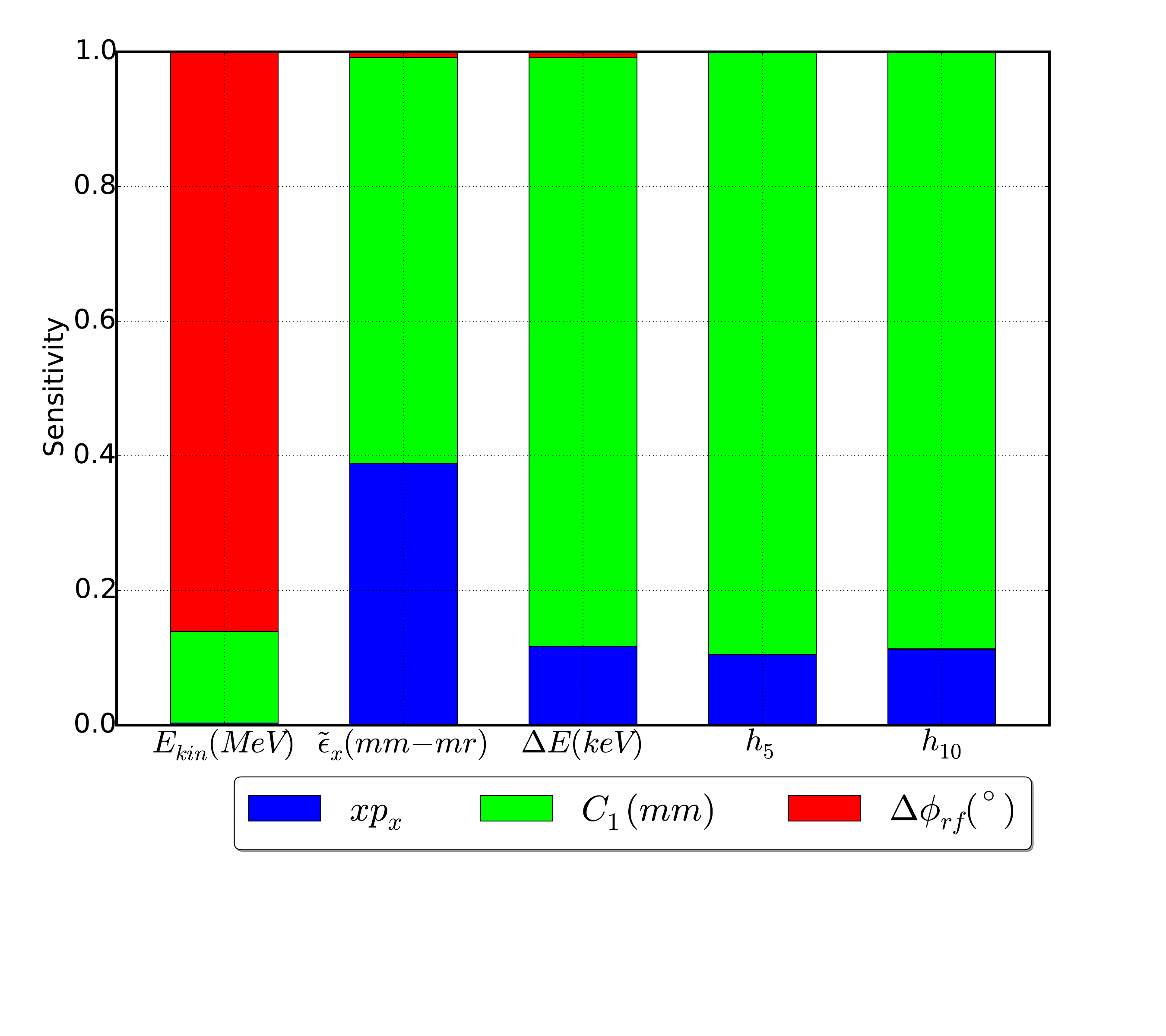}};
\node[below left] (img) at (0,0) {\includegraphics[width=0.42\textwidth]{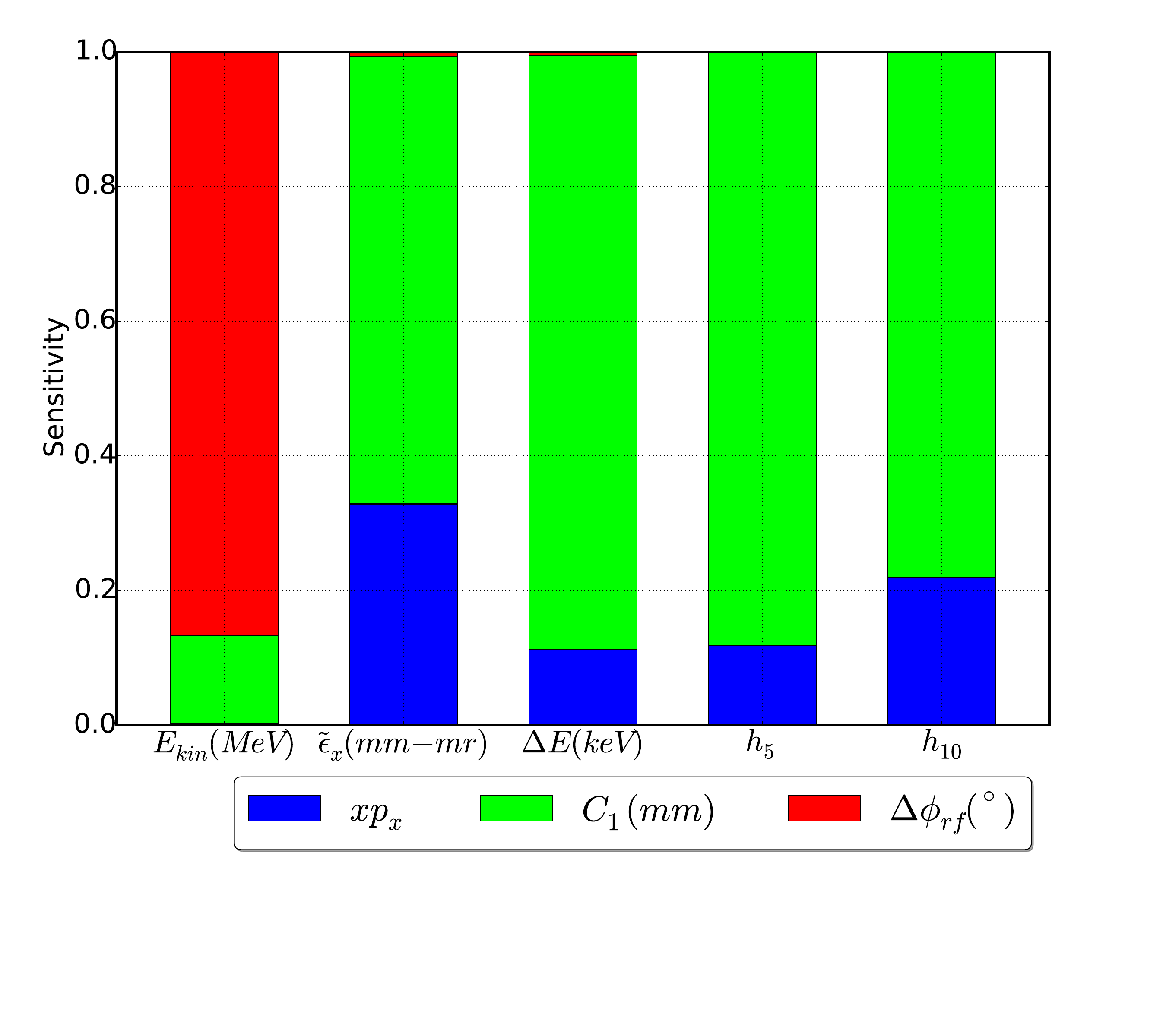}};
\node at (-100pt,155pt) {\footnotesize  $I=1$ mA};
\node at (100pt,155pt) {\footnotesize  $I=5$ mA};
\node at (-90pt,-0pt) {\footnotesize  $I=8$ mA};
\node at (90pt,-0pt) {\footnotesize $I=10$ mA};
\end{tikzpicture}
\vspace{-1.3cm}
\caption{Experiment 1: Global sensitivity analysis for intensities of 1,5,8 and 10 mA}
\label{fig:sensex1}
\end{center}
\end{figure}

\figref{fig:sensex1} shows, for a subset of the controllable parameter $I$, sensitivities of the QoI's with respect to the model parameters. The polynomial order is $p=4$, the similar correlations for other orders are not shown. 
Correlations, for example the insensitivity of the energy, and $x$, $p_x$ or the significant energy phase correlation,
are consistent with what is anticipated.\ A very mild dependence on
$x, p_x$ is observed and expected. There is a phase correlation appearing in the case of $I=5$ mA, which seems to be suppressed at other intensities, and the initial correlation of the distribution seems to become insignificant. A closer inspection of the phase space, beyond the scope
of this article, hints that the halo at this intensity has a minimum. This could explain the observed behaviour and is subject to a deeper 
investigation. 

These are very interesting findings that can guide new designs but also improve existing accelerators, and shows the quintessential merit and power 
of such a sensitivity analysis.


\subsection{Error Propagation and $L_{2}$ Error}\label{sec:l2err}
\begin{figure}[h!]
\begin{center}
\includegraphics[width=0.8\linewidth,angle=-0]{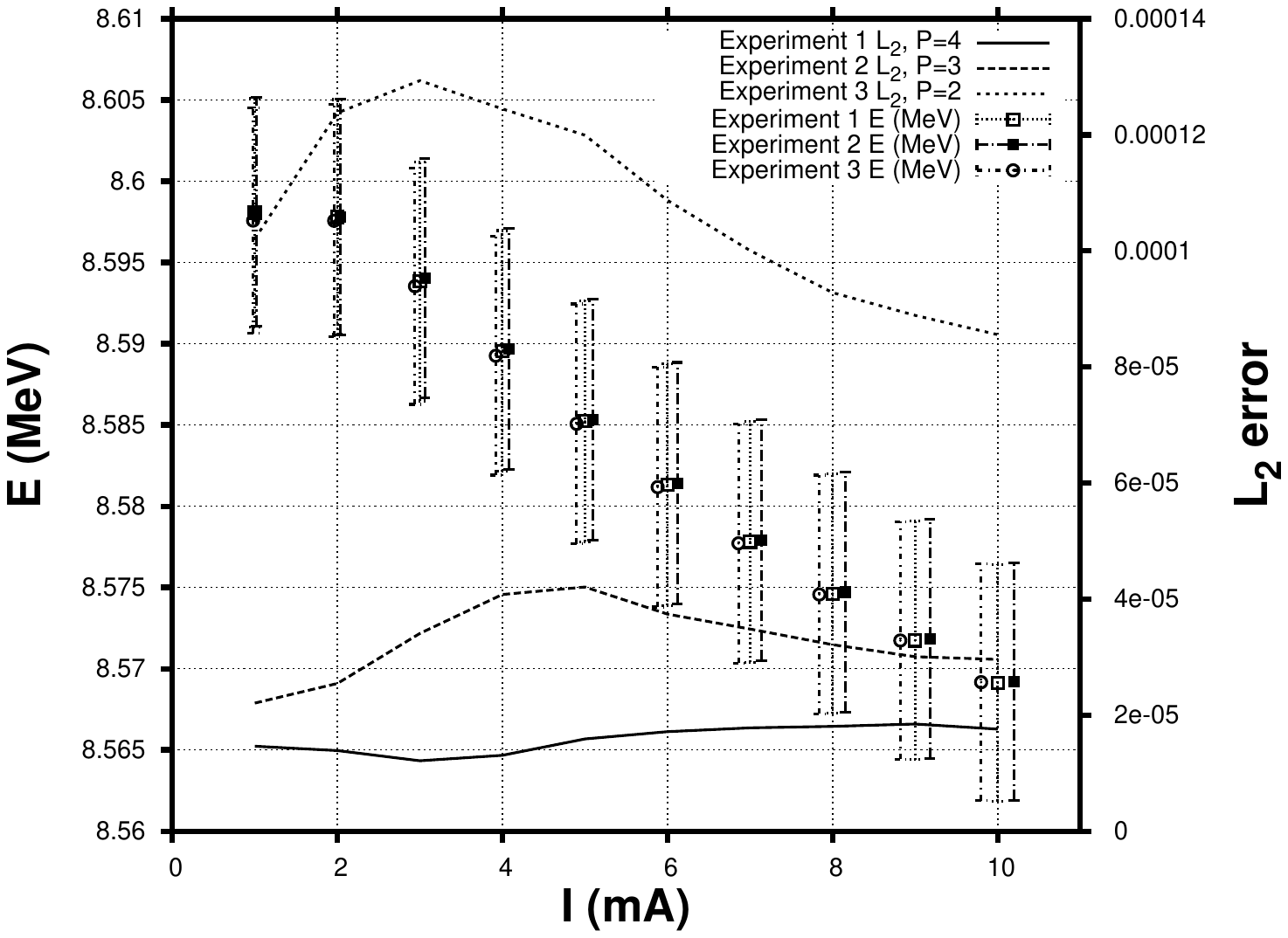}
\caption{Medium values, and variances are shown as dots and error bars, on the left y-axis for the extraction energy $E$. The global $L_{2}$ error (lines) between the high-fidelity and the surrogate model, for the final energy of the particle beam, is shown on the right y-axis.}
\label{fig:errprop1}
\end{center}
\end{figure}

In \figref{fig:errprop1}, the $L_2$ error 
\begin{displaymath}
L_2 = \frac{|| \hat{u} - u ||_2    } {|| \hat{u} ||_2  } 
\end{displaymath}
between the surrogate model $ \hat{u}$ and $u$, the high fidelity \opal\ model, is shown for $E$, the final energy of the particle beam and all values of the controllable parameter $I$.\ The mean value and variance are shown on the left y-axis.\ We can now precisely define the error and the dependence 
of the surrogate model on $P$.\ The expected convergence of the surrogate model as a function of $P$ is shown for one model parameter only, because of the similar behaviour in the other considered parameters.\ This clearly helps in choosing an appropriate order of the surrogate model.\  In addition, the accuracy was checked using a hold out model of $N_{rs} = 100$ uniform random samples over the model parameter domain $\vec{\lambda}$.

\subsection{Predictions}
\begin{figure}[h!]
\begin{center}
\includegraphics[width=0.99\linewidth,angle=-0]{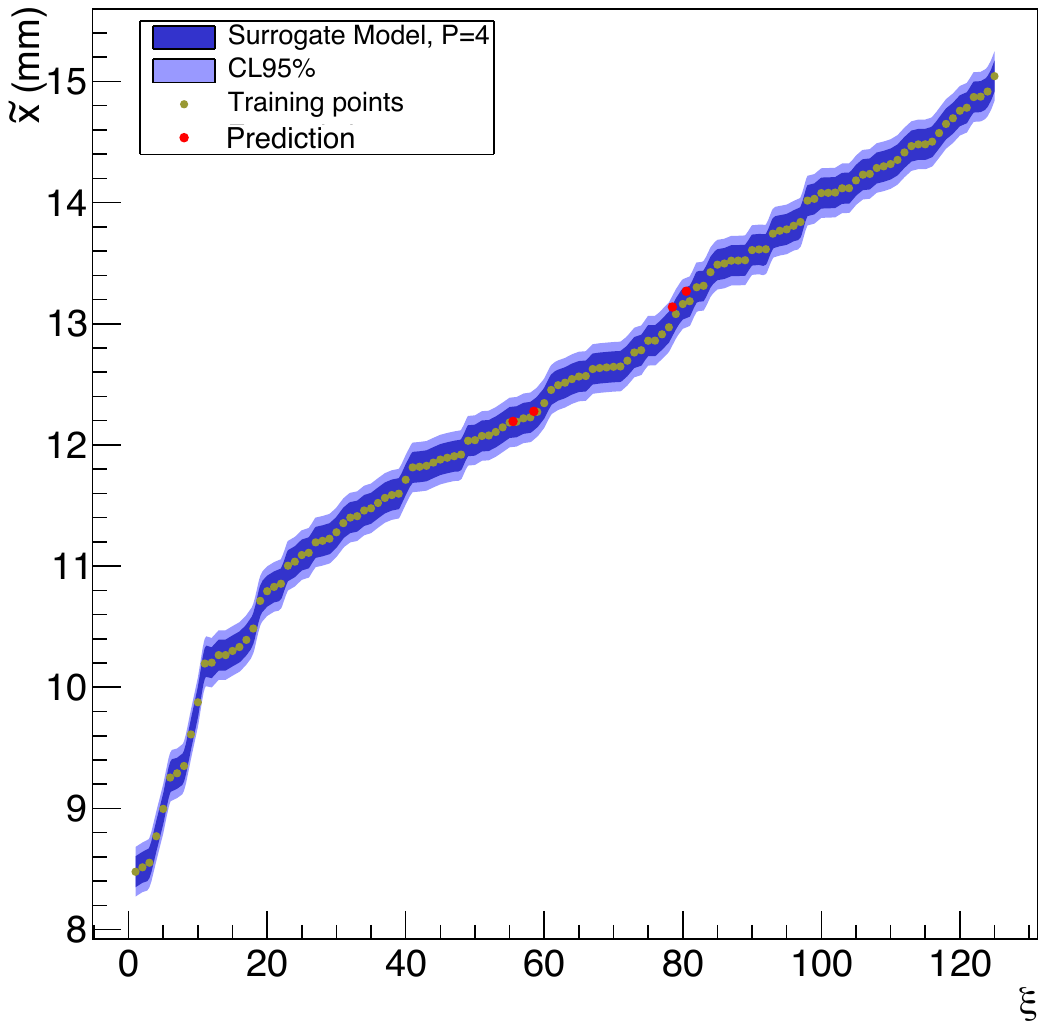}
\caption{The surrogate mode for $\tilde{x}$, together with training and prediction points. The 95\% CL of the model is also shown.}
\label{fig:errprop2}
\end{center}
\end{figure}

The surrogate model is constructed by selecting an appropriate number of training points in order to sample the input uncertainties of the design
parameter space.\ These finite number of training points are depicted as yellow points in \figref{fig:errprop2}.\ However, with the surrogate model we can choose
any point within the lower and upper bound specified ($a_i, b_i$ in \eqref{eq:lambdaj}) in order to obtain $\vec{\lambda}$ in \eqref{eq:pcin}.  In \figref{fig:errprop2} the red points are
arbitrarily chosen within the specified bounds and they are very well within the bounds of the surrogate model and the 95\% confidence level (CL) obtained by 
evaluating the Student-t test. The data presented in \figref{fig:errprop2} are only from experiment 3 in the case of 1 mA.

\subsection{Performance}
The presented surrogate model is the most simple, but gives, for the non-trivial model problem, statistically sound results. This fact and the remark
that the evaluation of the surrogate model is $\sim 800\times$ faster than the high-fidelity model ($0.5$ seconds v.s. $400$ seconds) opens up 
unprecedented possibilities in research areas such as on-line modelling and multi-objective \cite{iabc:12,yakb:13} optimization of charged particle accelerators.

\subsection{Conclusions for the Model Problem}
For a representative and at the same time non trivial model problem, an accurate and fast to evaluate surrogate model is presented.
From the sensitivity analysis, a phase correlation, in the case of $I=5$ mA, could be observed. A surrogate model for the halo parameter with
high fidelity is constructed. due to the low computational cost of the surrogate mode, future optimisation, minimisation of the halo, is conceivable.\  
This model problem should be understood as  "show case" demonstrating the applicability of this approach in a generalized accelerator setting.

\section{CONTRIBUTION TO THE DAED$\delta$ALUS/IsoDAR ACCELERATOR DESIGN EFFORT}
\label{sec:realwordprob}
\begin{figure}[t!]
	\centering
		\includegraphics[height=0.19\textheight]{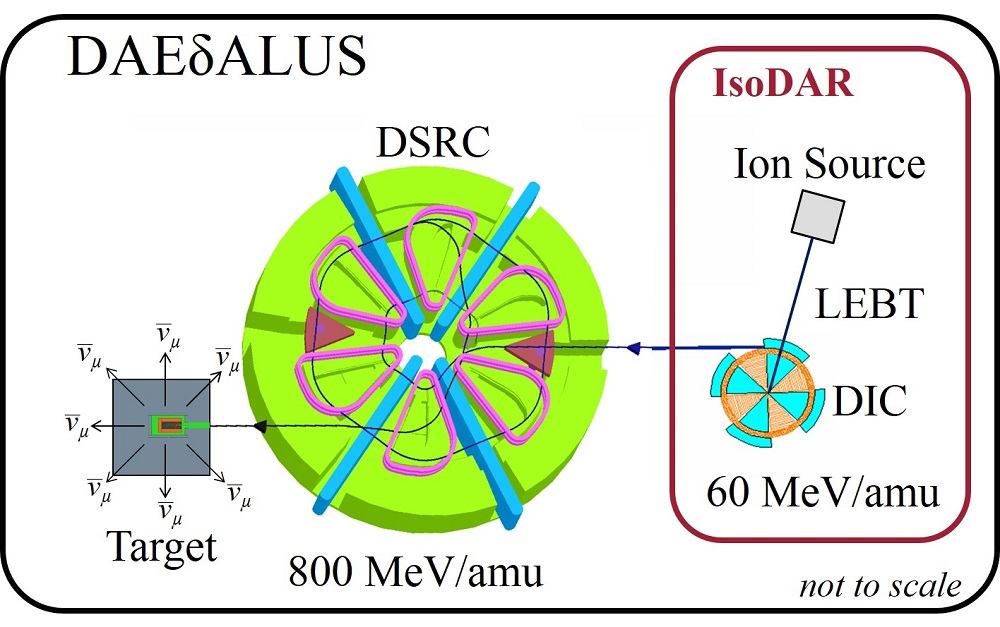}
		\caption{Cartoon picture of one single \DAD module with the 
				 injector part that can be used for IsoDAR highlighted
				 on the right.
	             \label{fig:dd_isodar}}
\end{figure}

The Decay-At-rest Experiment for $\delta_{\textrm{CP}}$ violation At a Laboratory for Underground Science (\DAD) \cite{Abs:2012wp} and the Isotope Decay-At-Rest experiment (IsoDAR)  \cite{PhysRevLett.109.141802} are 
proposed experiments to search for CP violation in the neutrino sector, and ``sterile'' neutrinos, respectively. In order to be decisive within 5 years, the neutrino flux and, consequently, the driver beam current, produced by a chain of cyclotrons cf.\ \figref{fig:dd_isodar}, must be high, higher than achieved today. 

\subsection{Physics Motivation}
The standard model of particle physics includes three so-called ``flavors'' of neutrinos:
\nue, \numu, and \nutau, and their respective anti-particles. These particles can change 
flavor (neutrino oscillations), a process that can be described using a mixing matrix.
This means that neutrinos must have a small mass \cite{Olive:1753419}. In addition,
some experiments aimed at measuring these oscillations in more detail have shown 
anomalies that led to the postulation of ``sterile'' neutrinos which would 
take part in the oscillation, but, contrary to
the three known flavors, do not interact through the weak force \cite{COLLIN2016354}. 
Another important question is whether the three neutrino model can give rise to a CP-violating phase
$\delta_{\textrm{CP}}$ \cite{PhysRevD.70.093011}, 
which might explain the matter-antimatter 
asymmetry in the universe today.

The main challenge, from the accelerator point of view, is the handling of the high intensity beams. Of utmost importance is the minimisation of particle losses, hence the understanding and mitigation of particle halo. A second, and related task is the optimization of the exit path out of the cyclotron. Here the  separation of the last two turns in the cyclotron has to be maximised.\
The conducted research by the \DAD/IsoDAR collaboration over the last couple of years, suggest that it is feasible, albeit challenging, to accelerate \SI{5}{mA} of \htp to \SI{60}{MeV/amu} in a compact cyclotron and boost it to \SI{800}{MeV/amu} in the DSRC ({\DAD} Superconducting Ring Cyclotron)  with clean extraction in both cases.\ 

The following surrogate model construction and sensitivity analysis of the IsoDAR cyclotron is research in progress, i.e.\ far from complete, but should illustrate the 
potential of the introduced methods on an ongoing design effort. 

\subsection{Initial Conditions for maximal Energy and Turn Separation} 
In order to run the physics experiment with the highest efficiency, a target energy of \SI{60}{MeV/amu} and lowest particle losses have to be reached.

\begin{figure}[h!]
	\centering
		\includegraphics[width=0.3\textwidth,angle=0]{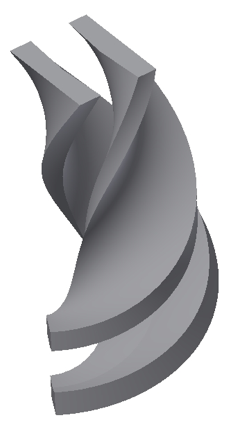} \includegraphics[width=0.3\textwidth,angle=0]{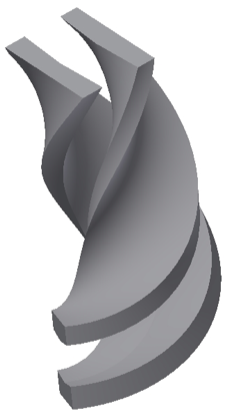}
		\caption{Two different spiral inflector position, with azimuth $\theta=135$ (left) and 
		 $\theta=140$ (right) degrees. 	             \label{fig:sensIsodar1}}
\end{figure}

\begin{figure}[h!]
\centering
\includegraphics[width=0.67\textwidth,angle=90]{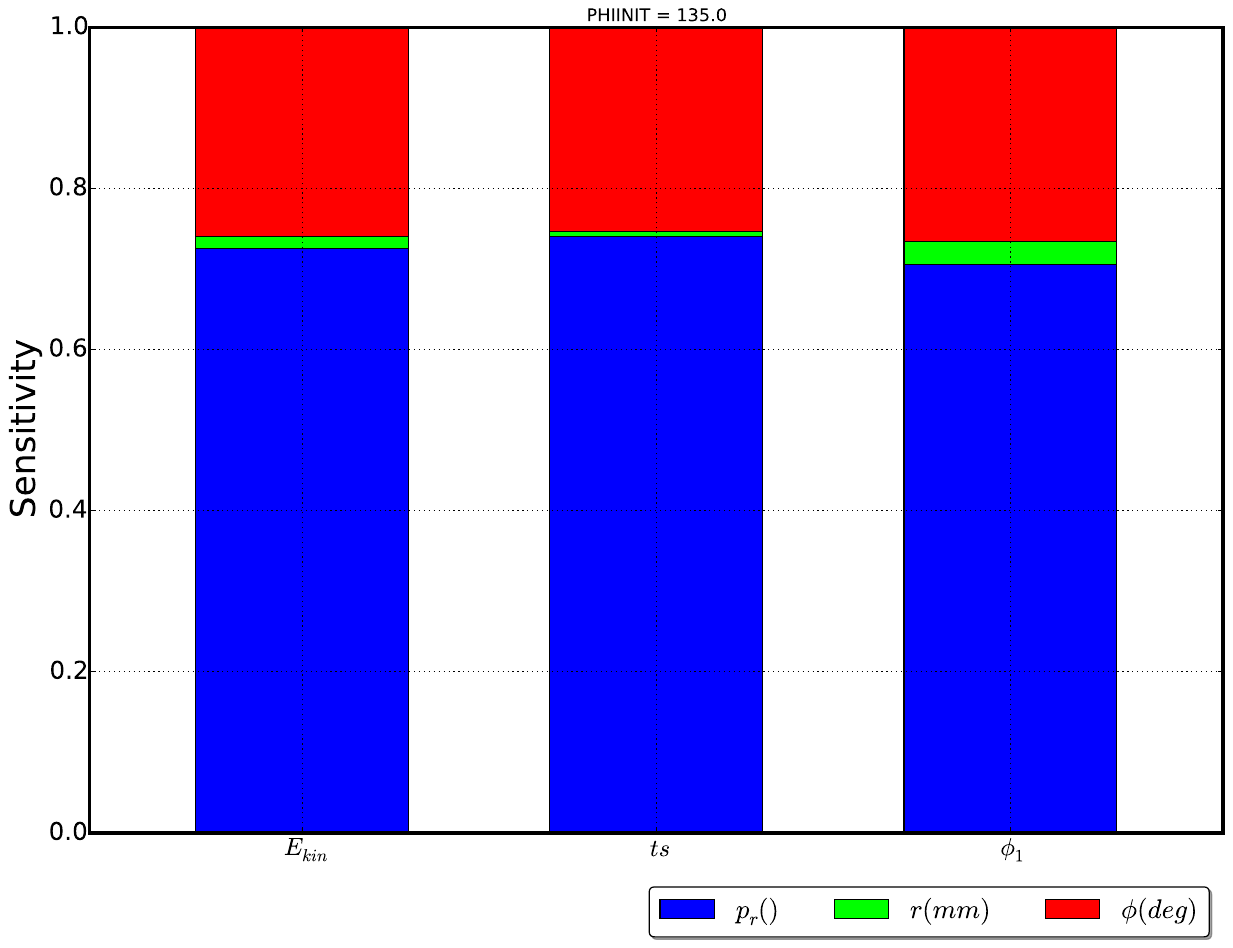} 
\includegraphics[width=0.67\textwidth,angle=90]{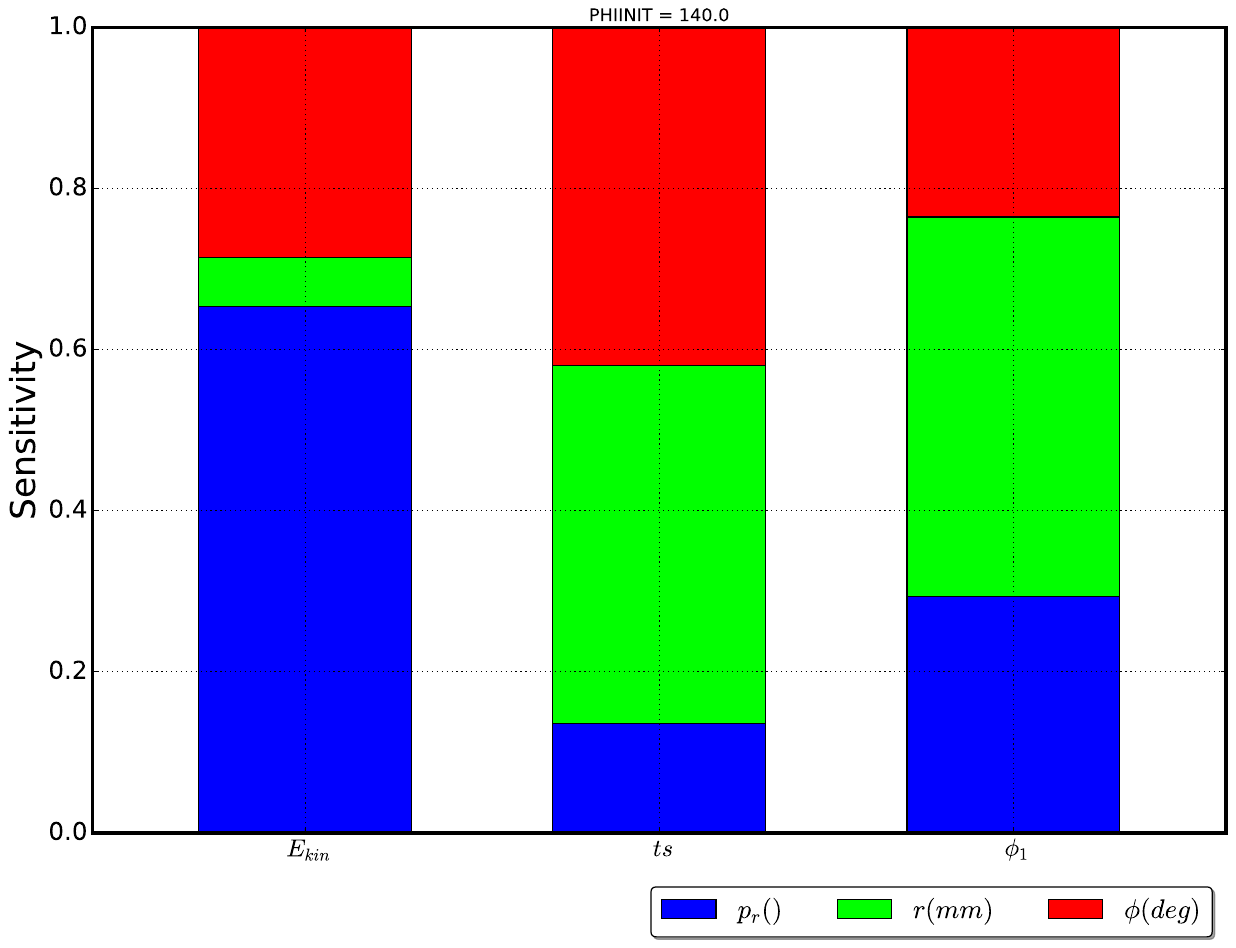}
\caption{Global sensitivity analysis for finding the most favoured spiral inflector position\label{fig:sensIsodar2}.\ On the left side, the sensitivities for a spiral inflector position of $\theta = 135 \deg$ is shown, while on the right side
a more favorable position of $\theta = 140 \deg$ reducing the influence of $p_r$ on the turn separation $ts$.}
\end{figure}

A large turn separation between the extracted turn $n$ and the turn $n-1$ allows the insertion of a septum to change the sign of curvature of the $n$th
orbit, hence facilitate clean (lossless) extraction of the beam. Detailed initial conditions for the 
cyclotron simulation are obtained from a 3D spiral inflector model, as shown schematically in \figref{fig:sensIsodar1}
\cite{PhysRevAccelBeams.20.124201}. From the exit of the spiral inflector, we need to find optimal initial 
conditions for the full cyclotron favorable. We restrict the number of parameters to 3 model parameters describing the beam initial conditions which are injection radius $r$, radial momenta $p_r$ (c.f.\ \figref{fig:sensIsodar0}) and the phase $\phi$ of the radio 
frequency of the acceleration cavities (not shown in \figref{fig:sensIsodar0}). The controllable parameter is the angle $\theta$ of the spiral inflector, which brings the beam from the vertical direction into the mid-plane. We varied the azimuthal angle, $\theta$, of the spiral inflector over a
range of $5 \deg$ as sown in \figref{fig:sensIsodar1}.
With the guidance of the sensitivity analysis c.f. \figref{fig:sensIsodar2}  we selected the most favorable case of $140$ degrees for the inflector angle to minimise the impact of $p_r$ on the 
turn separation (ts). The radial position $r$ of the beam and the phase $\phi(r)$ of the cavities are directly accessible to control while the radial momenta $p_r$ is not directly accessible for control, hence a low sensitivity of the QoI w.r.t.\ $p_r$ is desired.

The influence of the spiral inflector position was known to have an impact on the extraction efficiency, a direct
quantification of this fact was, to our knowledge, never described.

Having fixed the spiral inflector position, a surrogate model  was constructed to estimate the final energy $E$ and turn separation. We 
concentrate on the model for the energy and remark that the performance of the turn separations is very similar. A detailed discussion about the turn separation as well as the influence of the spiral inflector position will be given  in a forthcoming physics paper. 

\begin{figure} \centering
\begin{tikzpicture}[every node/.style={anchor=south west,inner sep=0pt},x=1mm, y=1mm,]   
     \node (fig1) at (0,0)
       {\includegraphics[scale=0.45]{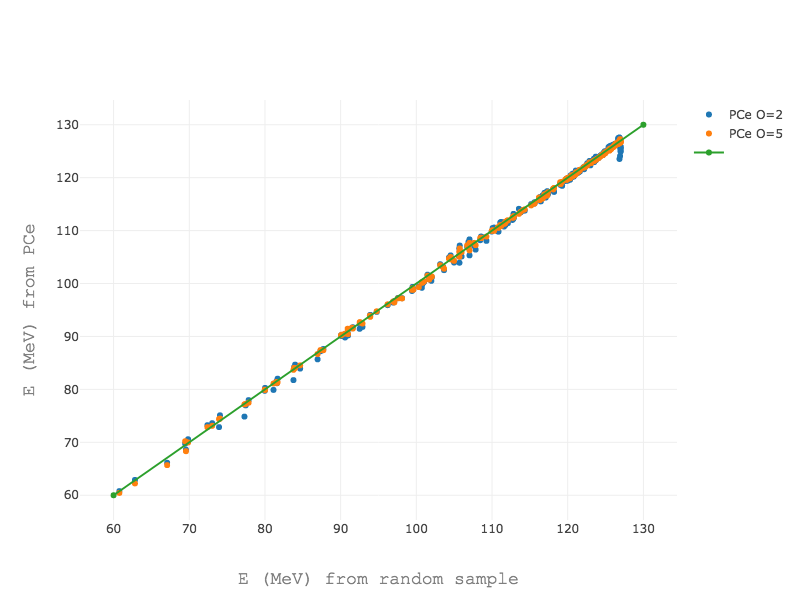}};
        \node (txt2) at (10,80)
      {a)};
     \node (fig2) at (65,10)
       {\includegraphics[scale=0.25]{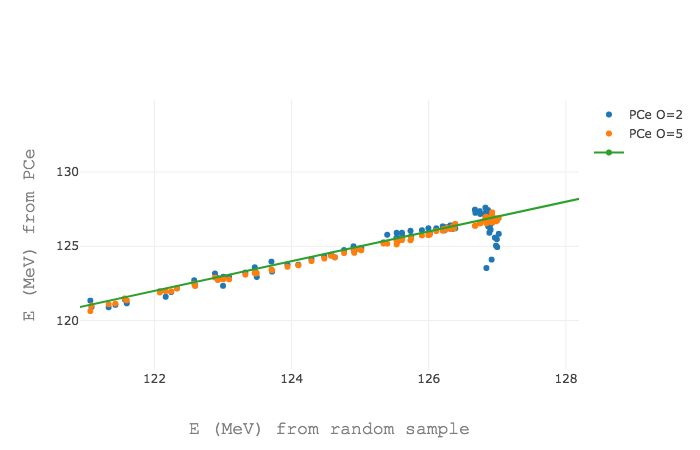}};  
     \node (txt2) at (65,40)
      {b)};
\end{tikzpicture}
\caption{a) Comparison of the surrogate model with order 2 and order 5 over the full energy range.\ b) is showing the 
high energy part of the spectrum where the 5 order is necessary to recover the high fidelity model. \label{fig:sensIsodar3}}
\end{figure}

In \figref{fig:sensIsodar3} a random sample N$_{rs} = 100$ is used to compare the high fidelity model to the surrogate model with orders 2 and 5.\ The second order model is behaving very well until, at high energies, 
non-linearities from the curvature of the radio frequency sine wave are present. In \figref{fig:sensIsodar3} b)
the performance of the 5th order model, in the high energy sector, is visible, in Table \ref{table:l2energy} the L$_2$ error is given.  

\begin{table}[h!]
\caption{L$_2$ error in energy, as function of the order of the surrogate model} 
\centering
\begin{tabular}[t]{ l  l }   
\hline \hline
Order	 & L$_2 \times 10^{-5}$, N$_{rs} = 100$   \\
\hline \hline
$2$ & 4.097  \\
$3$ & 1.846  \\
$4$ & 1.415  \\
$5$ & 1.020 \\
\hline
\end{tabular}
\label{table:l2energy} 
\end{table}

\subsection{Maximal Transmission}
Inspection of \figref{fig:sensIsodar0} reveals the fact that particles will terminate at some location very early in the machine. Particles with wrong dynamical properties need to be removed from the ensemble at low energies,
otherwise we would loose them at higher energies and activate and/or damage the machine. For this purpose
collimators are inserted, just after injection (not shown in the figures). These collimators can be spatially adjusted and will deliberately 
remove particles with wrong dynamical properties, such as large vertical momenta.  

The  following multi objective optimisation problem needs to be solved: given a range of target emittance at extraction (indicate the quality of the beam), maximise the transmission (minimise losses).\ We remark that in order to solve this problem, many time consuming particle-in-cell simulations have to be conducted.
Hence, an accurate surrogate model could have a substantial impact on the time to solution.

In order to construct such a model,  we consider 4 collimators as model parameter $\lambda$ and search for a 
surrogate model for the transmission $Q=N_{inj}/N_{ext} \times 100\%$, with  $N_{inj}$ the injected number of
particles and  $N_{ext}$ the surviving (to be extracted) number of particles.

\begin{figure}[h!]
\centering
\includegraphics[width=0.9\textwidth,angle=0]{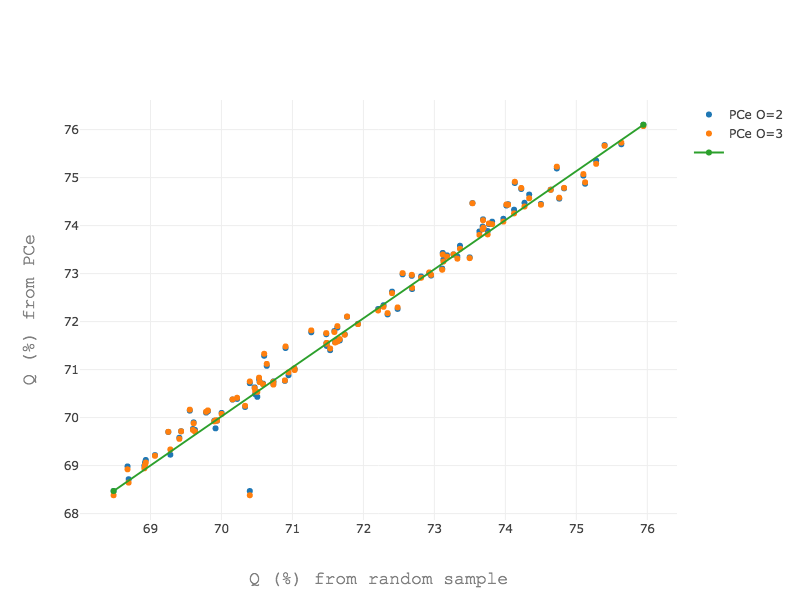} 
\caption{Surrogate model for the transmission $Q$, compared with random sampling\label{fig:suroQ!}.}
\end{figure}

The second QoI are the emittances defined in \secref{sec:qoidef}. As in the previous section, we use
a uniform random sample N$_{rs} = 100$ to evaluate the quality of the surrogate model.

\begin{figure}[h!] \centering
\begin{tikzpicture}[every node/.style={anchor=south west,inner sep=0pt},x=1mm, y=1mm,]   
     \node (fig1) at (-10,0)
       {\includegraphics[scale=0.25]{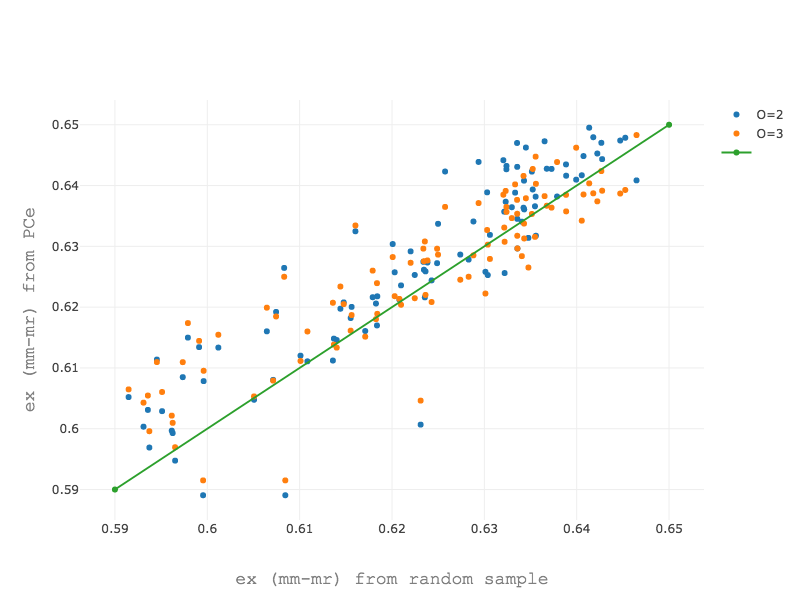}};
        \node (txt2) at (-10,45)
      {a)};
     \node (fig2) at (45,0)
       {\includegraphics[scale=0.25]{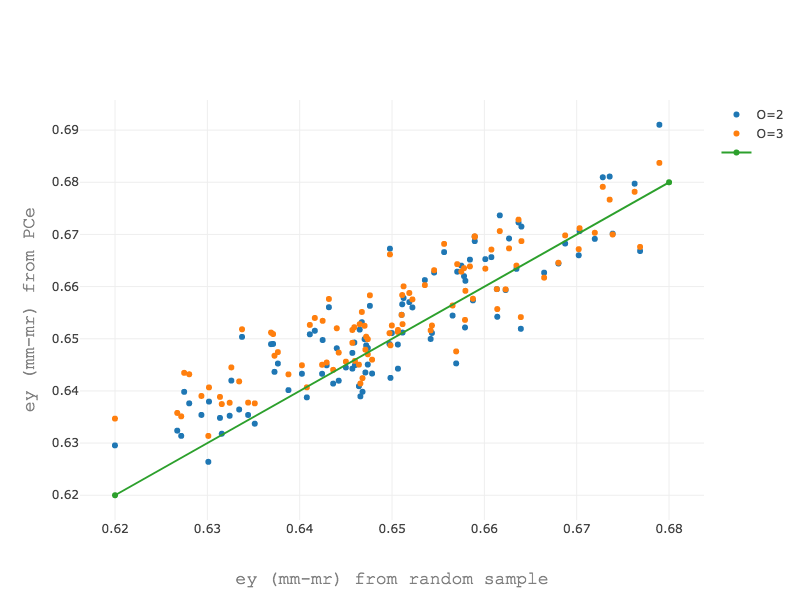}};  
     \node (txt2) at (45,45)
      {b)};
\end{tikzpicture}
\caption{Comparison of the surrogate model for $\epsilon_x$ and $\epsilon_y$ with order 2 and order 5\label{fig:sensIsodar5}.}
\end{figure}

In \figref{fig:suroQ!} we recognise a very good agreement between the random samples (from high fidelity \opal\ simulations) and the surrogate model. The same is true for the beam quality
shown in \figref{fig:sensIsodar5}. Hence we can conclude that for the four QoI's: energy $E$, transmission $Q$ and emittances $\epsilon_x$ and $\epsilon_y$ we can construct 
high fidelity surrogate models. 

One PIC simulation, used to train the model, runs for approximately $13090$ seconds on 8 cores (Intel KNL). The evaluation of the 4th order multivariate polynomial takes less than $0.002$ second using   
the UQTk software. This represents a speedup of $\approx 6.5 \times 10^6$ and allows to do large scale multi objective optimisation, using surrogate models as forward solver. These optimisations are part of ongoing research for the IsoDAR compact cyclotron design.

\section{CONCLUSIONS}
\label{sec:concl}

A sampling-based UQ approach is presented to study, for the first time, the effects of input uncertainties on the performance of particle accelerators.\ A particular, but complex, example in the form of a high-intensity cyclotron was
used to demonstrate the usefulness of the surrogate model as well as the global sensitivity analysis via computing the total \sobol\ indices.\ 
The presented physics problem is a model problem, with the aim of demonstrating the usefulness and applicability of the presented UQ approach.\ However, we claim to present a problem that can be recognised as a template for many high-intensity modelling attempts, and beyond.\

The proposed UQ approach is based on polynomial chaos expansion using the UQTk software.\ The goal is to achieve an accurate estimation of solution statistics using a minimal number of high-fidelity simulations.\ For several QoI's a surrogate model was constructed, the validity is proved by
comparing to a high-fidelity model.\ $L_2$ error norms show the expected convergence with regard to the degree of the polynomial chaos expansion.
For the rms beam size ($\tilde{x}$), holdout points, i.e.\ points that are not used in the training set, were evaluated and compared to the 
statistical expectations from the model. We found that the values are consistent with the surrogate model and clearly within the 95\% CL. 

The \sobol\ based global sensitivity analysis was in line with the expectation from the physics evaluation of the model problem.
optimization
A tremendous speedup of a factor of $800$ on the model problem and up to $\approx 6.5 \times 10^6$ was observed, comparing  the time to solution of the surrogate model to the high-fidelity model. This opens
up  possibilities for on-line modelling and multi-objective optimization of complex particle accelerators using surrogate models. 

Future research includes the continuation of the IsoDAR design effort by using the surrogate model in the parameter optimization, branching out into the field of proton therapy, with focus on understanding 
the uncertainty of accelerator parameters, in relation to the applied radiation dose to the patient.  An inverse problem to find initial particle distributions was solved in \cite{lee2006}. The presented Ansatz could be used to achieve similar goals.

In this article, conceptionally we followed the simplest approach towards UQ. Given the encouraging results, we plan to enhance this model by using Hermite chaos, and going to higher dimensions, which implies the use of sparse methods or latin hypercube sampling.

Particle accelerators in general create a vast amount of high-quality data, including the QoI's we have considered.\ Including such data into the the
model, or solving an inverse problem could be interesting research topics for the future.

\section{ACKNOWLEDGMENTS}
Drs.\ N. Neveu, N.\ Pogue and T.\ Schietinger for critical comments and English proof reading. Dr.\ V.\ Rizzoglio and Dr. D.\ Winklehner for many fruitful discussions and help with the art work.\ Dr.\ K.\ Sargsyan \& Dr. B. Debusschere for {\tt UQTk} and UQ related discussions.

\appendix

\section{Mathematical bases of Polynomial Chaos based UQ} \label{apptheory}

We briefly introduce the mathematical bases in the style and the notation of \cite{Smith2014,Ghanem91a, Xiu02,Xiu10a,Hadigol2015507}.\ Let $\left(\Omega,\mathcal{F},\mathcal{P} \right)$ be a complete probability space, where $\Omega$ is the sample set and $\mathcal{P}$ is a probability measure on $\mathcal{F}$, the $\sigma$-field (algebra) or Borel measure.\ Input uncertainties of the system have been discretised and approximated by the random vector $\bm{\xi} = \left(\xi_1,\cdots, \xi_d\right):\Omega \rightarrow \mathbb{R}^{d}$, $d \in \mathbb{N}$. The probability density function (pdf) of the random variable, $\xi_k$, is denoted by $\rho(\xi_k)$.\ Similarly, $\rho(\bm{\xi})$ represents the joint pdf of $\bm{\xi}$.\

Let  $\bm{i}$ be a  multi-index  ${\bm{i}} = (i_1, \dots , i_d) \in \mathcal{I}_{d,p}$ and the set of multi-indices $\mathcal{I}_{d,p}$ is defined by
\begin{equation} 
\mathcal{I}_{d,p} = \lbrace \bm i = (i_1, \dots , i_d)\in \mathbb{N}_0^d: \Vert\bm{i}\Vert_1 \leqslant p \rbrace,
\end{equation}
where $\Vert \cdot \Vert_1$ is the $l_1$ norm i.e.,\ $\Vert \cdot \Vert_1 = i_1+\dots + i_d$, and $p$ is the polynomial order.\

All square integrable, second-order random variables with finite variance output, $u(\bm{\xi}) \in L_2\left(\Omega,\mathcal{F},\mathcal{P} \right)$, can be written as
\begin{equation} 
\label{appeq:rvpce0}
u(\bm{\xi}) = \sum_{|\bm{i}| = 0}^\infty \alpha_{\bm{i}} \Psi_{\bm{i}}(\bm{\xi}).
\end{equation}
Hence $\alpha_{\bm{i}}$ denotes the deterministic coefficients and $\Psi_{\bm{i}}(\bm{\xi})$ are the multivariate PC basis 
functions \cite[10.1.1]{Smith2014} \cite{Ghanem91a}.\ Note that the uncertain QoI, $u$, is represented by a vector of deterministic parameters $\alpha_{\bm{i}}$.\

For the truncated PCE to order $p$ in $d$ dimensions of \eqref{appeq:rvpce0} we get
\begin{equation} 
\label{appeq:rvpce}
\hat{u}(\bm{\xi}) = \sum_{\bm{i} \in \mathcal{I}_{d,p}} \alpha_{\bm{i}} \Psi_{\bm{i}}(\bm{\xi}).\
\end{equation}

The basis functions $\Psi_{\bm i}(\bm\xi)$ in (\ref{appeq:rvpce}) are generated from
\begin{equation} 
\label{eqn:mvpol}
\Psi_{\bm{i}}(\bm{\xi}) = \prod_{k=1}^{d} \Psi_{{i}_k}(\xi_k), \ \ \ \ \bm{i} \in \mathcal{I}_{d,p},
\end{equation}
where $\Psi_{i_k}$ are univariate polynomials of degree ${i}_k  \in \mathbb{N}_0 := \mathbb{N}\cup \lbrace 0\rbrace$,
orthogonal with respect to $\rho_{k}(\bm{\xi})$ (see, e.g., Table \ref{table:Askey}), i.e.,
\begin{equation}
\label{appeq:1dpce}
\mathbb{E} [\Psi_{i_k} \Psi_{j_k}]  = \langle \Psi_{i_k} \Psi_{j_k} \rangle = \int \Psi_{i_k}(\xi_k) \Psi_{j_k}(\xi_k) \rho(\xi_k) \mathrm{d}\xi_k= \delta_{i_kj_k} \mathbb{E} [ \Psi_{i_k}^2 ].
\end{equation}
Here $\delta_{i_kj_k}$ denotes the Kronecker delta and $\mathbb{E}[\cdot]$ is the expectation operator. 

The number $K$ of PC basis functions of total order  $p$  in dimension $d$ can be calculated to 
\begin{displaymath} 
\label{appeq:pplo}
K=\vert \mathcal{I}_{d,p}\vert= \frac{(p+d)!}{p!d!}.
\end{displaymath}
The PC basis functions $\Psi_{\bm i}(\bm\xi)$ are orthogonal,  
\begin{equation}
\mathbb{E}[\Psi_{\bm i}\Psi_{\bm j}]=\delta_{\bm i,\bm j}\mathbb{E}[\Psi_{\bm i}^2],
\end{equation}
because of the orthogonality of $\Psi_{{i}_k}(\xi_k)$ and the independence of $\xi_k$. 
As $ p \rightarrow \infty$, the truncated PC expansion in (\ref{appeq:rvpce}) converges in the mean-square sense, if and only if  the following two conditions are fulfilled: 1)
$u(\bm\xi)$ has finite variance and 2) the coefficients $\alpha_{\bm{i}}$ are computed from the projection equation \citep{Xiu02}
\begin{equation}
\alpha_{\bm{i}} = \frac{\mathbb{E}[\hat{u} \Psi_{\bm i}]}{\mathbb{E}[\Psi_{\bm i}^2]}.
\end{equation}

\subsection{Global sensitivity analysis}
\label{appGSA}
The expensive, deterministic high-fidelity particle accelerator model, $\mathcal{M}$, is described by a function $\vec{u}=\mathcal{M}(\vec{x})$, where the input $\vec{x}$ is a point inside $\mathbf{D}$ (c.f. \figref {fig:modelred}) and $\vec{u}$ is a vector of QoI's.\ Finding correlations in these high dimensional spaces is nontrivial, however it is vital for a deep understanding of the underlying  physics.\ For example, reducing the search space is of great interest in the modelling and optimization process.\ In the spirit of Sobol' \citep{Sobol01}, let $\vec{u}^{*}=\mathcal{M}(\vec{x}^{*})$ be the sought  (true) solution.\ The local sensitivity of the solution $\vec{u}^{*}$ with respect to $x_{k}$ is estimated by $(\partial \vec{u} /\partial x_{k})_{\vec{x}=\vec{x}^{*}}$.\ On the contrary, the global sensitivity approach does not specify the input $\vec{x}=\vec{u}^{*}$, it only considers the model $\mathcal{M}(\vec{x})$.\ Therefore, global sensitivity analysis should be regarded as a tool for studying the mathematical model rather than a specific solution ($\vec{x}=\vec{x}^{*}$).\

Following \citep{Sobol01}, the problems that can be studied, in our context, with global sensitivity analysis can be categorised the following way:
\begin{enumerate}
  \item ranking of variables in $\vec{u} = \mathcal{M}(x_{1}, x_{2}, \ldots, x_{n})$,
  \item identifying variables with low impact on $\vec{u}$.\
\end{enumerate}
In this article, we use the \sobol\ indices \citep{Sobol01}, which are widely used due to their generality.\ Results can be found in \secref{sec:sensanal}.\ 

The first order PC-based \sobol\ index, $S_k$, represents the individual effects of the random input $\xi_k$ on the variability of $u(\bm{\xi})$, and is given by
\begin{equation}
\label{appeq:1sobol}
S_k = \frac{1}{\mathrm{Var}[u(\bm{\xi})]} \sum_{\bm{i} \in \mathcal{I}_{k}} \alpha_{\bm{i}}^2 ~\mathbb{E}[\Psi^2({\xivec}_{\bm{i}})], \quad \mathcal{I}_{k} = \lbrace \bm i \in \mathbb{N}_0^d: {i}_k > 0,  {i}_{m \neq k} = 0 \rbrace.\
\end{equation}
To compute $S_k$, all random inputs except $\xi_k$ are fixed.\ As a consequence, $S_k$ does not include effects arising from the interactions between $\xi_k$ and the other random inputs.\ This also means that $\mathcal{I}_{k}$ includes only the dimension $k$.\

The fractional contribution to the total variability of $u(\bm{\xi})$ due to parameter $\xi_k$, considering all other model parameters, is given by
\begin{equation}
\label{appeq:totsobol}
S_k^T =  \frac{1}{\mathrm{Var}[u(\bm{\xi})]}  \sum_{\bm{i} \in \mathcal{I}_{k}^T} \alpha_{\bm{i}}^2 ~\mathbb{E}[\Psi^2({\xivec}_{\bm{i}})] \quad \mathcal{I}_{k}^T = \lbrace \bm i \in \mathbb{N}_0^d: {i}_k > 0 \rbrace. 
\end{equation}
The set of multi indices $\mathcal{I}_{k}^T$ includes dimension $k$ among others.

Now we are in a position to rank the importance of the variables.\ The smaller $S_k^T$ is, the less important the random input, $\xi_k$, becomes.\ We note, for the extreme case $S_k^T \ll 1$, the variable $\xi_k$ is considered to be insignificant.\ In such a case, the variable can be replaced by its mean value without considerable effects on the variability of $u(\bm{\xi})$. We will make use
of this fact when discussing the model problem and use $S_k^T$ as a measure to identify the most important random inputs of the model.\

If one is interested in the fraction of the variance that is due to the joint contribution of the $i\text{-th}$ and  $j\text{-th}$ input parameter, we can easily compute

\begin{equation}
S_{i,j} = \frac{1}{\mathrm{Var}[u(\bm{\xi})]}  \sum_{\bm{i} \in \mathcal{I}_{i,j}} \alpha_{\bm{i}}^2 ~\mathbb{E}[\Psi^2({\xivec}_{\bm{i}})] \quad \mathcal{I}_{i,j} = \lbrace \bm i \in \mathbb{N}_0^d: {i}_i > 0, {i}_j > 0 \rbrace. 
\end{equation}
which describes this quantity.\ The set $\mathcal{I}_{i,j}$ of multi-indices includes dimensions $i$ and $j$, among others.

As an example to category 1 from above, consider a problem where $x_{i}$ and $x_{j}$ are two entries in the matrix of the second-order moments of the 
initial particle distribution within a simulation.\ We then find that $S_{i}$ and $S_{j}$ are both much smaller than $S_{i,j}$.\ Such a
situation will indicate that other entries in the matrix of second order moments significantly contribute.\
For category 2, refer to  \citep[Section 7.]{Sobol01}, where an approximation of $S$ is proven, when not considering all elements of $\vec{x}$.
\clearpage

\section{Wiener-Askey PC} \label{app1}
\begin{table}[h!]
\caption{The correspondence of Wiener-Askey PC and the pdf of the random variables \citep{Xiu02}.} 
\centering
\begin{tabular}[t]{ c   c  c }   
\hline \hline
$\rho(\xi_k)$ & Polynomial & Support \\  
\hline \hline
Beta & Jacobi & [a,b] \\  
\hline
Uniform & Legendre & [a,b] \\  
\hline
Gaussian & Hermite & (-$\infty$,+$\infty$) \\  
\hline
Gamma & Laguerre & (0,+$\infty$) \\  
\hline
\end{tabular}
\label{table:Askey} 
\end{table}

\section{Legendre polynomials}
The Legendre polynomials, or Legendre functions of the first kind \eqref{eq:leg}, \cite[p. 302]{whittaker_watson_1996}, are solutions to the Legendre differential equation, a second-order ordinary differential equation
\begin{equation}\label{eq:leg}
(1-x^2)\frac{d^2 y}{dx^2} - 2x \frac{dy}{dx} + l(l+1)y = 0.\
\end{equation}
In case of $l \in \mathcal{N}$, the solutions are  polynomials $P_n$. The first few polynomials
relevant to this paper are shown in \eqref{eq:P14}.
\begin{eqnarray} \label{eq:P14}
P_0(x)	&=&	1 \nonumber\\	
P_1(x)	&=&	x \nonumber\\	
P_2(x)	&=&	1/2(3x^2-1) \\	
P_3(x)	&=&	1/2(5x^3-3x) \nonumber\\	
P_4(x)	&=&	1/8(35x^4-30x^2+3) \nonumber\\
\ldots\nonumber
\end{eqnarray}

%
%

\bibliographystyle{unsrt}
{\bibliography{mybib}}

\end{document}